\documentclass[fleqn,usenatbib]{mnras}

\usepackage{newtxtext,newtxmath}
\usepackage[T1]{fontenc}

\usepackage{graphicx}	% Including figure files
\usepackage{amsmath}	% Advanced maths commands
\usepackage{xcolor}     % Allows to write coloured text
\usepackage{url}        % Allows to put url in references
\usepackage{hyperref}
\usepackage{subcaption} % Allows to use subfigures

\title[Clustering effects on dark sirens]{Clustering effects on the Dark Siren determination of $H_0$: A simulation study}

\author[Kalomenopoulos, Barbieri et al.]{
Marios Kalomenopoulos$^{1, 3}$\thanks{E-mail: v1mkalom@ed.ac.uk} \thanks{E-mail: marios.kalomenopoulos@[gmail.com][unlv.edu]} \href{https://orcid.org/0000-0001-6677-949X}{\includegraphics[scale=0.6]{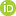}},
Riccardo Barbieri$^{2}$,
Sadegh Khochfar$^{1}$,
Jonathan Gair$^{2}$,\newauthor
Robert J. McGibbon$^{1, 4}$
\\
$^{1}$Institute for Astronomy, University of Edinburgh, Royal Observatory, Edinburgh EH9 3HJ, UK\\
$^{2}$Max Planck Institute for Gravitational Physics, Potsdam Science Park, Am Muhlenberg 1, D-14476 Potsdam, Germany\\
$^{3}$Department of Physics and Astronomy, University of Nevada, Las Vegas, NV 89154, USA\\
$^{4}$Leiden Observatory, Leiden University, PO Box 9513, 2300 RA Leiden, The Netherlands
}

\date{Accepted XXX. Received YYY; in original form ZZZ}

\pubyear{2023}

\begin{document}
\label{firstpage}
\pagerange{\pageref{firstpage}--\pageref{lastpage}}
\maketitle

\begin{abstract}
We investigate the cosmological power of gravitational wave (GWs) events without electromagnetic counterparts using high-resolution, cosmological simulations that include realistic clustering. 
We assume a near-future detector network (LVK + LIGO India in O5), with sky localisation and distance uncertainty capabilities for binary black holes, of the order of $60$ deg. squared and $10\%$ respectively, and with a range of about $1600$ Mpc/h.
We find that clustering leads to increased convergence of the $H_0$ posteriors, with clear recovery of the input value as early as $N_{\rm gw}=40$ events (a very conservative estimate for O5, which expects thousands of BBHs/year). In contrast, even with $N_{\rm gw}=100$ events in uniform catalogues, the posterior remains largely unconstrained. Additionally, we quantify the role of catalogue completeness and show that levels as low as $f=25\%$ can be competitive with fully complete catalogues, confirming the impact of clustering. Completeness levels of $f=50\%$ perform statistically similar to complete catalogues with as few as $N_{\rm gw}=40$ events. We conclude that improving GWs detection capabilities yields better results than more complete galaxy catalogues.Finally, we find that 1) physical weights do not provide significant additional constraining power on top of clustering, 2) luminosity distance errors contribute more for small cone openings, 3) the $H_0$ prior range does not introduce any biases in the $H_0$ posterior, 4) closer events ($\lesssim 100$ Mpc/h) yield about two times better precision than more distant events ($\lesssim 1600$ Mpc/h), and 5) our method is robust to different modelling choices.
\end{abstract}

%%%


%%%%%%%%%%%%%%%%%%%%%%%%%%%%%%%%%%%%%%%%%%%%%%%%%%

%%%%%%%%%%%%%%%%% BODY OF PAPER %%%%%%%%%%%%%%%%%%

\section{Introduction}

One of the basic parameters of the standard cosmological model is the Hubble-Lema\^{i}tre parameter  \citep{Hubble_1929, Lemaitre_1931}, which describes the expansion speed of our Universe \citep{Friedman_1922}. There is a long history of dedicated observations that try to measure it, and more specifically its value at the present time $H_0$ \citep{Freedman_Madore_2010_Ho, Jackson_2015, Steer_2020_Ho, Freedman_2021_Hubble_constant}. However, although the improvement of measurements have constrained $H_0$ to around $70\ \rm{km/s/Mpc}$, the parallel shrinkage of observational errors has led to statistically significant discrepancies among different probes, especially between ``early universe'' values of $H_0 \simeq 67\ \rm{km/s/Mpc}$ \citep{BAO_Ho_2019, Planck_Cosmo_param_2018} and ``late universe'' ones with $H_0 \simeq 73\ \rm{km/s/Mpc}$ \citep{Holicow_Ho_2020, SN_Ho_2019, SN_Ho_2021, H0_local_2025, Cosmoverse_2025}. 

Even though many possible solutions have been proposed (for a review of measurements and proposed solutions to the current tensions see \citet{Di_Valentino_et_al_2021_HoSolutions}), it is certain that more, independent, and complementary measurements need to be made to disentangle between the need of new physics and the potential of missing underlying systematic effects. A very promising and independent technique to measure the Hubble parameter is the use of gravitational wave observations \citep{Schutz_1986, Holz_Hughes_2005, Palmese_Mastrogiovanni_2025_Ho_cosmo_review, Poggiani_2025_Ho_cosmo_review}. Contrary to observations in the electromagnetic (EM) spectrum, GWs give direct information about the distance of the source. At the same time, a measurement of the redshift of the source is not guaranteed, and hard to obtain for most events. This is especially true for binary black holes (BBHs), which are not expected to have an identifiable EM counterpart, hence preventing a direct measurement of the source redshift.

The cleanest technique, a ``bright siren'', requires an EM counterpart and was performed for the first time in \citet{Abbott_et_al_2017_Standard_Siren_Ho} using the binary neutron star (BNS) merger event GW170817 \citep{GW170817_Obs_2017}, giving $H_0 = 70^{+12}_{-8}\ \rm{km/s/Mpc}$\footnote{Errors indicate everywhere the $68.3 \%$ credible interval, unless stated otherwise.}. Improved constraints of the inclination of the binary \citep{Hotokezaka_et_al_2019} reduced the errors to $H_0 = 68.9^{+4.7}_{-4.6}\ \rm{km/s/Mpc}$. Of course, with only one GW observation with an EM counterpart, the uncertainties are still large. At the same time, the detection of a counterpart has its own difficulties (e.g., host-galaxy misidentification, photometric redshift errors, peculiar velocity systematics, catalogue incompleteness etc.), and for the majority of GW events none is expected\footnote{For BNS, tidal effects can permit a direct redshift measurement \citep{Messenger_Read_NSz_2012, Messenger_et_al_2014}. These effects could be measurable on future GWs detectors, like the Einstein Telescope \citep{Einstein_Telescope}. For the case of a NSBH pair, an EM counterpart is possible, but depends crucially on the mass ratio of the two objects \citep{NSBH_mergers_EM_2019}. For the most populous channel of binary black holes (BBHs) events, in general, an associated EM signal is not expected, apart from peculiar cases, like in an active galactic nucleus environment \citep{Bahcall_Wolf_1976, AGN_BBH_EM_counter_2019}}.. Furthermore, cosmological model-dependent biases can pose further complications \citep{Shafieloo_Keeley_Linder_2020}. For this reason, a number of other methods have been proposed to get independent information about the redshift.

One is based on exploiting prior knowledge about the population of the sources \citep{Taylor_et_al_NSmass_2012}. The GW signals contain information about the redshifted ``chirp mass'' $\mathcal{M}_c = (1+z) M_c$, where $M_c = (m_1 m_2)^{3/5}/(m_1+m_2)^{1/5}$ is the chirp mass of the system in the source frame and $m_1, m_2$ the masses of the members of the binary. Including prior information about the source $M_c$ and measuring $\mathcal{M}_c$ can statistically put constraints on the redshift and hence on $H_0$. For more details on this technique, we refer to \citet{Mastrogiovanni_et_al_pop_2021}. Based on the GW observations until the end of O3, the third LIGO-Virgo-KAGRA (LVK) observing run \citep{AdvLIGO_O3_2019, AdvVirgo_O3_2019, AdvLIGO_O3_2020, AdvVirgo_DetChar_2023, GWTC3b_2021, GWTC3a_2021}, this method gave a value of the Hubble parameter $H_0 = 50^{+37.0}_{-30.0}\ \rm{km/s/Mpc}$~\citep{GWTC3-cosmic-expansion}, while after the first part of the fourth observing run (GWTC-4.0) \citep{AdvLIGO_2015, AdvVirgo_2015, KAGRA_2021, AdvLIGO_O4a_2025, LVK_DetChar_O4a_2025, GWTC4_Catalogue_2025}, the ``spectral siren'' result was $H_0 = 112.7^{+51.0}_{-35.9}\ \rm{km/s/Mpc}$~\citep{GWTC4-cosmic-expansion}. Both results correspond to the choice of a POWER LAW + PEAK model for the underlying mass distribution. A more realistic mass model, FULLPOP-4.0 \citep{GWTC4_Populations_2025}, yields $H_0 = 76.4^{+23.0}_{-18.1}\ \rm{km/s/Mpc}$ for GWTC-4.0 \citep{GWTC4-cosmic-expansion}.

Finally, another method to get information about the redshifts is the ``statistical dark sirens'' technique, where with ``dark'' we refer to GWs observations without an EM counterpart (see \citet{Hitchhiker_2022} for a pedagogical introduction to the method). It was first proposed in \citet{Schutz_1986} and the basic idea consists of using all the galaxies in the GW sky localisation volume as potential hosts, and statistically compute the $H_0$ parameter. 

Although this method can also be used for individual observations: GW170817 \citep{Fishbach_et_2019_Dark_siren}, GW170814 \citep{Soares-Santos_et_al_2019}, GW190814 \citep{Abbott_et_al_2020_HoGW190814, Palmese_et_al_2020_HoGW190814}, GW190412 \citep{Ballard_et_al_DESI_DarkSiren_2023} its real strength comes from combining multiple events, assumed to be independent from each other. 
 
Statistical determinations of $H_0$ have been performed with GW observations of the first three observing runs and the first part of the fourth observing run: using observations up until the end of O2 $\&$ GW170817 \citep{GWTC1_2019, GWTC2_2021, Abbott_et_al_2021_Ho_O2, Abbott_et_al_2021_Ho_O2_Erratum}, giving $H_0 = 68.7^{+17}_{-7.8}\ \rm{km/s/Mpc}$, using only observations up until the end of O2 \citep{Finke_et_al_DS_2021}, giving $H_0 = 67.3^{+27.6}_{-17.9}\ \rm{km/s/Mpc}$, using events observed up until the end of O3 \citep{GWTC3b_2021}, together with a BBH population model and GLADE+ galaxy catalogue \citep{GLADE_2018, GLADE+_2012} giving $H_0 = 67.0^{+13.0}_{-12.0}\ \rm{km/s/Mpc}$ \citep{GWTC3-cosmic-expansion}, $H_0 = 79.8^{+19.1}_{-12.8}\ \rm{km/s/Mpc}$ \citep{Palmese_et_al_DS_2021} using the Dark Energy Spectroscopic Instrument (DESI) \citep{DESI_Survey} and the Dark Energy Survey (DES) \citep{DES_Survey}, or $H_0 = 76.0^{+17.6}_{-13.5}\ \rm{km/s/Mpc}$ using the DECam Local Volume Exploration Survey (DELVE) \citep{Alfradique_et_al_DarkSirens_DELVE_2024}, and finally, till the first part of O4, giving $H_0 = 81.6^{+21.5}_{-15.9}\ \rm{km/s/Mpc}$, using the updated FULLPOP-4.0 population model and GLADE+ \citep{GWTC4-cosmic-expansion}, and $H_0 = 70.4^{+13.6}_{-11.7}\ \rm{km/s/Mpc}$ using the Legacy Survey catalogues \citep{Bom_et_al_DarkSiren_O4_2024}.

Still, this technique is in its infancy, with only a few GW sources statistically informative for a ``statistical dark sirens'' $H_0$ study \footnote{``Dark sirens'' techniques find also other cosmological applications. For example, they can constrain the cosmic dipole \citep{Chen_2025_dipole}, they can be used with galaxy clusters instead of individual galaxies \citep{Beirnaert_Ghosh_Dalya_2025}, or without any catalogue, but assuming knowledge of the redshift distribution of the merger rates of binary black holes \citep{Ding_et_al_2019, Leandro_Marra_Sturani_2022}. Finally, the possibility of extremely well localised ``dark sirens'' has been investigated in the context of future, ground based \citep{Bright_Sirens_Ground_2025}, and space based \citep{Bright_Sirens_Space_2025} detectors.}. For this reason, the power of the method is under active research, with improved codes developed that combine different techniques and include information from the BBH population \citep{Mastrogiovanni_et_al_2023_LVKcode1, Gray_et_al_2023_LVKcode2}. Investigations of the constraining power of ``statistical dark sirens'' predicted a determination of $H_0$ with a fractional uncertainty of the order of $\sim 10\%$, based on a number of $16^{+19}_{-14}$ well localised (sky volumes less than $10,000$ Mpc$^3$) BBHs and a $\sim 6\%$ accuracy from around $100$ BNS with no observed counterpart from the future runs of current detectors \citep{Chen_et_al_2018}. These results assumed a uniform distribution of galaxies in comoving volume and complete galaxy catalogues. These two assumptions have been investigated further in the literature. We discuss the main results below.

The existence of clustering is expected to improve the convergence of the method, since the possible hosts would be closer together. In that case, even if the true host of the GW source is missing from the galaxy survey, the presence of its observed neighbours can rectify its absence. \citet{Chen_et_al_2018} estimated that the effect of clustering can improve convergence by a factor of $\sim 2.5$. Realistic clustering was investigated for the first time in \citet{MacLeod_et_al_2008}, who used mock galaxy samples from SDSS DR6. They combined simple, cubical GW localisation areas between redshifts $z=0.1$ and $z=0.5$ representing observations of extreme-mass-ratio inspirals (EMRIs) with the future space GW detector, the Laser Interferometer Space Antenna (LISA) \citep{LISA_2017}. They found that the cross-correlation of 20 GW events with survey data within the LISA sky localisation, will yield a $1 \%$ determination of $H_0$. An updated study, but with specific patches of SDSS and sky localisation area (max $\Delta \Omega = 30$ degrees squared) corresponding to current and future ground-based detectors was performed in \citep{Nair_et_al_2018}, yielding $H_0 = 70.0^{+2.7}_{-2.8}\ \rm{km/s/Mpc}$, hence an error of the order of $4 \%$ with $N_{\rm gw} = 100$ GWs events. Finally, other studies \citep{Bera_et_al_2020_incompleteness, Mukherjee_Wandelt_Silk_2020, Mukherjee_et_al_2021, Ghosh_et_al_2024, Ferri_et_al_2025} investigated a statistical method to infer $H_0$ exploiting the clustering information of the GWs sources and nearby galaxies, either by calculating their angular cross-correlation or by combining information from the specific power-spectra respectively. Using this technique with GW events without electromagnetic counterparts from the third GW transient catalogue (GWTC-3.0), results to a median value of $H_0 = 82.4^{+23.0}_{-27.0}\ \rm{km/s/Mpc}$ \citep{Mukherjee_et_al_2024_crosscorr_GWTC3}, while forecasts for the LVK and DESI can reach a measurement of $H_0$ with precision of $\sim 7 \%$, while future, ground based detectors, like Cosmic Explorer \citep{Cosmic_Explorer_2019, Cosmic_Explorer_2021} and Einstein Telescope \citep{Einstein_Telescope, Einstein_Telescope_2023} could reduce this to $\sim 1\%$ \citep{Afroz_Mukherjee_2024_forecast}. The effects of GW bias in the GWs-galaxies cross-correlation method was explored in \citep{Cigarran_Mukherjee_2022_cross_corr_DESI_SphereX, Dehghani_et_al_2025_Bias_cross_corr}, showing that it can be constrained together with cosmological parameters and that is connected with astrophysical properties.

The effects of incompleteness, i.e., galaxies missing from a survey due to observational limitations, were analysed in more detail in \citep{Gray_et_al_2020}, where different fractions of galaxies were missing (fractions of galaxy completeness of 25\%, 50\%, 75\% and 100\%). Their mock catalogues were uniform in comoving volume. They quantified the less precise determination of $H_0$ with increased catalogue incompleteness and concluded that a $4.4 \%$ precision on $H_0$ based on galaxy catalogues is achievable with current detectors, assuming a uniform distribution of sources and $50 \%$ completeness from $249$ BNS events. Moreover, \citet{Turski_et_al_2025} studied the influence of the galaxy luminosity function \citep{Schechter_1976}, when dealing with incomplete catalogues, emphasising the need for an evolving Schechter function as this impacts the results on $H_0$. \citet{Finke_et_al_DS_2021} investigated the inference of $H_0$ from ``statistical dark sirens'' in \citep{GWTC2_2021} together with the GLADE galaxy catalogue \citep{GLADE_2018}, where they improved on the statistical analysis of various selection effects, but also in the completeness methodology. For the latter, apart from the uniform completion, they proposed a more realistic model of multiplicative completion, which takes into account the level of clustering of the relevant region compared to the mean density of galaxies in the Universe. More recently, \cite{Dalang_Baker_2023, Dalang_Fiorini_Baker_2024} developed a method that takes into account clustering information - specifically the variance of galaxy numbers - when dealing with surveys with missing galaxies, while \cite{Cosmic_cartography_2024, Cosmic_cartography_2025} used a Bayesian method for galaxy density and magnitude reconstruction based on large-scale structure information. Lastly, \cite{VanWyngarden_et_al_2025_cat_completeness} investigated how astrophysical information could alleviate the effects of catalogue incompleteness, and \citet{Perna_Mastrogiovanni_Ricciardone_2025} argued that a mismodeling of the former could lead to biases on $H_0$.

In this work, we are interested in investigating further the two scenarios discussed above - effects of clustering \& catalogue completeness - within the context of the ``statistical dark sirens'' statistical technique. We perform this task by exploiting for the first time cosmological N-body simulations combined with galaxy properties provided from machine learning (ML) techniques \citep{Rob_ML_2021}. This approach will ensure self-consistent clustering information as well as galaxy properties. We investigate the consequences of introducing physical weights, like masses and magnitudes, to individual galaxies when calculating the final posteriors. Our work aims to clarify some of the important intricacies of the ``statitical dark sirens'' method and contribute to the observational strategies of future GWs observations. 

Our paper is structured as follows: In section \ref{sec:methods} we introduce our simulations and present our methodology for calculating $H_0$, in section \ref{sec:results} we comment on our results on the $H_0$ posteriors in a clustered environment, with complete and incomplete catalogues and finally, in section \ref{sec:conclusions} we discuss our conclusions and possible extensions.

\section{Methods}\label{sec:methods}

\subsection{Simulations}\label{sec:simulations}

To calculate realistic density anisotropies, we  rely on cosmological dark matter (DM) only simulations, run with \emph{Gadget-4} \citep{Gadget_2, Gadget_4} for the LEGACY project \citep{Meriot_et_al_2022}.

The latter is composed by two primary volumes of $1600$ Mpc/h and $100$ Mpc/h box sizes run down to $z=0$ with $2048^3$ resolution elements, as well as a set of zoom-in simulations on the larger box, with size $83$ Mpc/h, and an effective resolution of $32768^3\ (1700^3)$. The parameter $h$ is defined as $h=H_0/100$. The latter have been designed to sample regions with density $-2, -1, 0, +1$, and $+2$ $\sigma$ of the mean density of the Universe, as well as extremely high (cluster) and very low (void) density regions and are therefore ideal to study different environmental effects. We use ROCKSTAR \citep{Rockstar_2013} to construct halo catalogues for each snapshot.

For our analysis, we use the data from four different boxes, at redshift $z=0$, with box size of: $L_{\rm box} = 1600, 703.125, 100$ Mpc/h. Three of them are high resolution simulations, while one low resolution simulation is used to examine the impact of resolution in the results. The details are shown in Table \ref{table:simulation_table}.

\begin{table}
\centering
\begin{tabular}{|c c c c c|}
     \hline
      $L_{\rm box}$ [Mpc/h] & $1600$ & $703.125$  & $100$ & $100$ \\ [0.5ex] 
     \hline\hline
     $N_{\rm eff}$ & $2048^3$ & $4096^3$ & $2048^3$ & $256^3$ \\ 
     \hline
     $M_{\rm res}$ [$M_{\sun}$] & $5.43 \times 10^{10}$  & $6.78 \times 10^{9}$ & $1.32 \times 10^{7}$ & $6.78 \times 10^{9}$ \\ 
     \hline
\end{tabular}
\caption{Simulations' properties: $L_{\rm box}$ - comoving box length in Mpc/h, $N_{\rm eff}$ - effective particle number, $M_{\rm res}$ - mass resolution in $M_{\sun}$.}
\label{table:simulation_table}
\end{table}

\begin{figure}
	\includegraphics[width=\columnwidth]{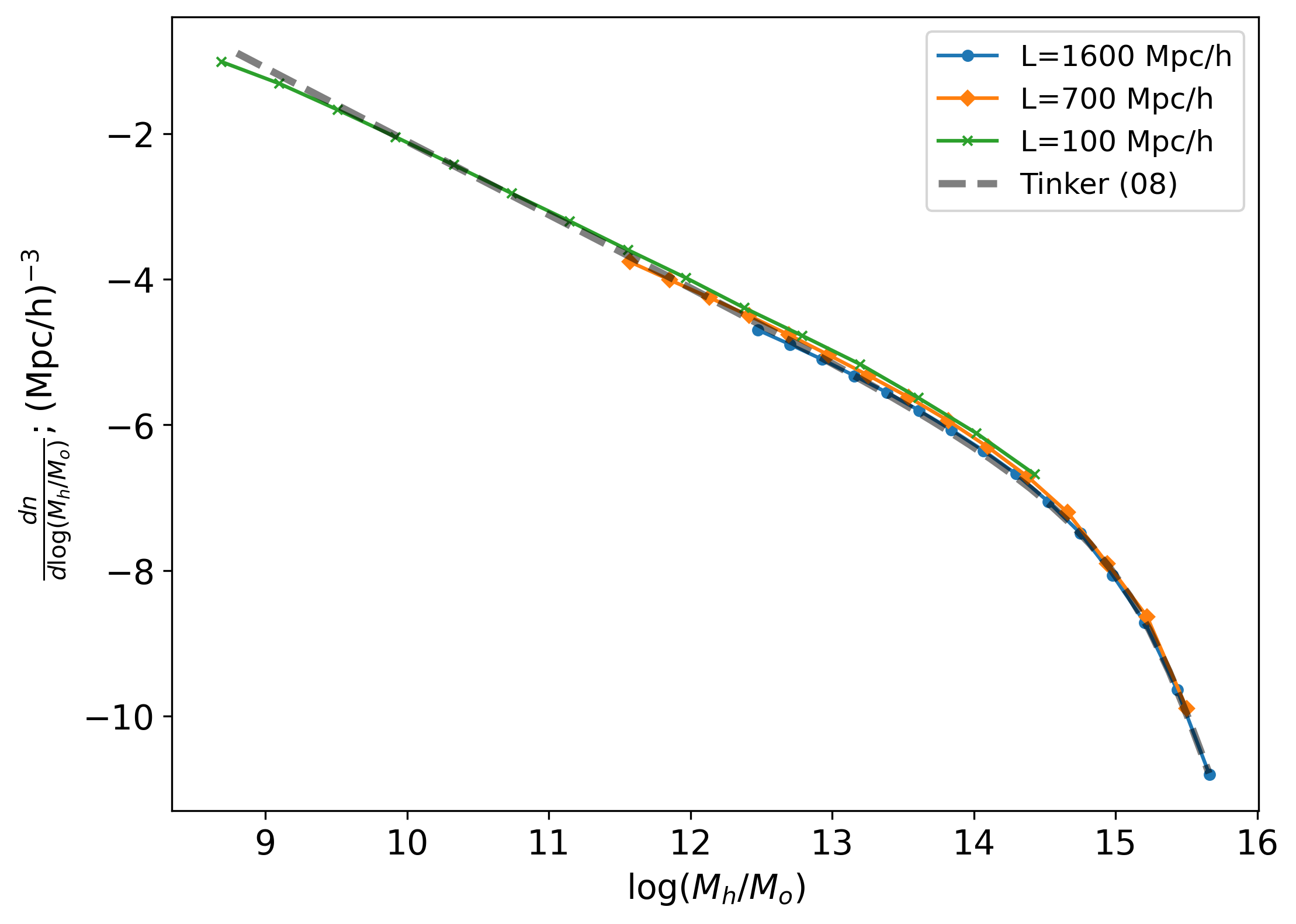}
    \caption{DM halo mass function for three of the large particle number LEGACY boxes that are used in the main analysis of this work. (Green solid) $L=100$ Mpc/h, (Orange solid) $L\simeq700$ Mpc/h zoom-in simulation, and the $L=1600$ Mpc/h box (Blue solid). Their mass functions are shown in the same order from left to right (as expected, only the largest box can form the more massive haloes). The grey dashed line provides the fit based on \citet{Tinker_et_al_2008}.}
    \label{fig:halo_mass_functions}
\end{figure}

All simulations have an accompanying halo catalogue, with their halo mass function shown in Figure \ref{fig:halo_mass_functions}. The mass functions are consistent with the fit proposed in \citet{Tinker_et_al_2008}. The effects of different boxes are discussed in section \ref{sec:Box_size_posteriors} and appendix \ref{ap:resolution_effects}.

The cosmological parameters of the simulations, are taken from WMAP9 \citep{Hinshaw_et_al_2013_WMAP}, with ($\Omega_m,\ \Omega_{\Lambda},\ \Omega_k,\ H_0) = (0.285,\ 0.715,\ 0,\ 69.5$ km/s/Mpc).

\subsubsection{Stellar-to-Halo mass connection}\label{sec:stellar_Halo_mass}

Since the simulations only follow DM, it is necessary to assign galaxy masses to individual haloes via models. Here we follow the standard parameterisation from \citep{Yang_et_al_2003, Moster_et_al_2010}, which relates halo mass to stellar mass of galaxies being hosted by the halo, with the fit from \citet{Girelli_et_al_2020}:
\begin{equation}\label{eq:SHM}
\frac{M_*}{M_h}(z) = 2A(z)\left[(M_h/M_A(z))^{-\beta(z)}+(M_h/M_A(z))^{\gamma(z)}\right]^{-1},
\end{equation}
where the different parameters are given by (for $z=0$, which is our redshift of interest):
\begin{itemize}
    \item $\log M_A(z) = \log M_A(z=0)$ 
    \item $A(z) = A_0$
    \item $\gamma(z) = \gamma_0$
    \item $\beta(z) = \beta_0$.
\end{itemize}
 
The fit\footnote{The above fit, based on the halo mass functions in \citep{Despali_et_al_2016}}, is qualitatively very similar to other fits in the literature, \citep[e.g.][]{Moster_et_al_2013, Behroozi_et_al_2013}. We have confirmed that the specific selection is not influencing our results. We note also that the masses only play a role in the case of physical weights - sections \ref{sec:Geometric_physical_weights}, \ref{sec:weights_results} - but are irrelevant for the other analyses. Even in this case, the weighting scheme of the galaxies is expected to be more important than the stellar-to-halo mass mapping. We adopt from \citet{Girelli_et_al_2020} the values ($\log M_{A,0}, A_0, \gamma_0, \beta_0$) = ($11.77, 0.0465, 0.702, 1$) which are based on the subhalo abundance matching technique, where there is a one-to-one connection between halo (sub)structures and galaxies  \citep[see][for an investigation of this assumption]{Neistein_et_al_2011}.

In our simulations we assume the equivalent mapping of our haloes to galaxies and we use the two terms interchangeably from now on. Note that we ignore any scatter in the above relation, since for the level of complexity of this work, we are mainly interested in the general trend of galaxies' masses. Physical weights are not used for the majority of the analyses, except from section \ref{sec:weights_results}.

Finally we use ML to assign each galaxy an \emph{absolute magnitude} in band $B$ ($M_B$) and \emph{star formation rate} (SFR) in $M_{\odot}/{\rm yr}$. We choose these properties for consistency with \citet{Gray_et_al_2020}, and because the B-band acts as a proxy for SFR \citep{Singer_et_al_2016, Bands_Ducoin_et_al_2020}. The ML model is built using the ``extremely randomised tree'' algorithm and makes a correspondence between our DM haloes and their physical characteristics. This is done in a self-consistent way for all our simulation boxes, and for this reason, we choose this methodology instead of semi-analytic prescriptions that may depend on different assumptions for each parameter. Nevertheless, as mentioned above, the specific mapping approach is less important that the weighting scheme used in the cases of physical weights (see also section \ref{sec:Geometric_physical_weights}). Exploiting information across many redshift snapshots for increased accuracy, the ML model predicts magnitudes and star formation rates. To learn this mapping, the ML model is trained using the IllustrisTNG300 cosmological run \citep{IllustrisTNG_Springel_2018, IllustrisTNG_Marinacci_2018, IllustrisTNG_Naiman_2018, IllustrisTNG_Nelson_2018, IllustrisTNG_Pillepich_2018} and takes subhalo spin, maximum rotational velocity, velocity dispersion and DM halo mass as input features. The reliability of the model was tested by comparing baryonic properties predicted using the DM information of the IllustrisTNG simulations versus the original values provided by them \citep[see discussion in section $3.2$ of][]{Rob_ML_2021}. A complete description of this method, as well as its validation can be found in \citep{Rob_ML_2021, quotas}.

In summary, for each galaxy we have the following properties: 1) Their $3D$ coordinates, 2) their stellar mass, 3) their absolute magnitudes in the B band, and 4) their star formation rate.

\subsection{Cosmological inference}\label{sec:cosmo_inference}

\begin{figure}
     \centering
     \includegraphics[width=1\columnwidth]{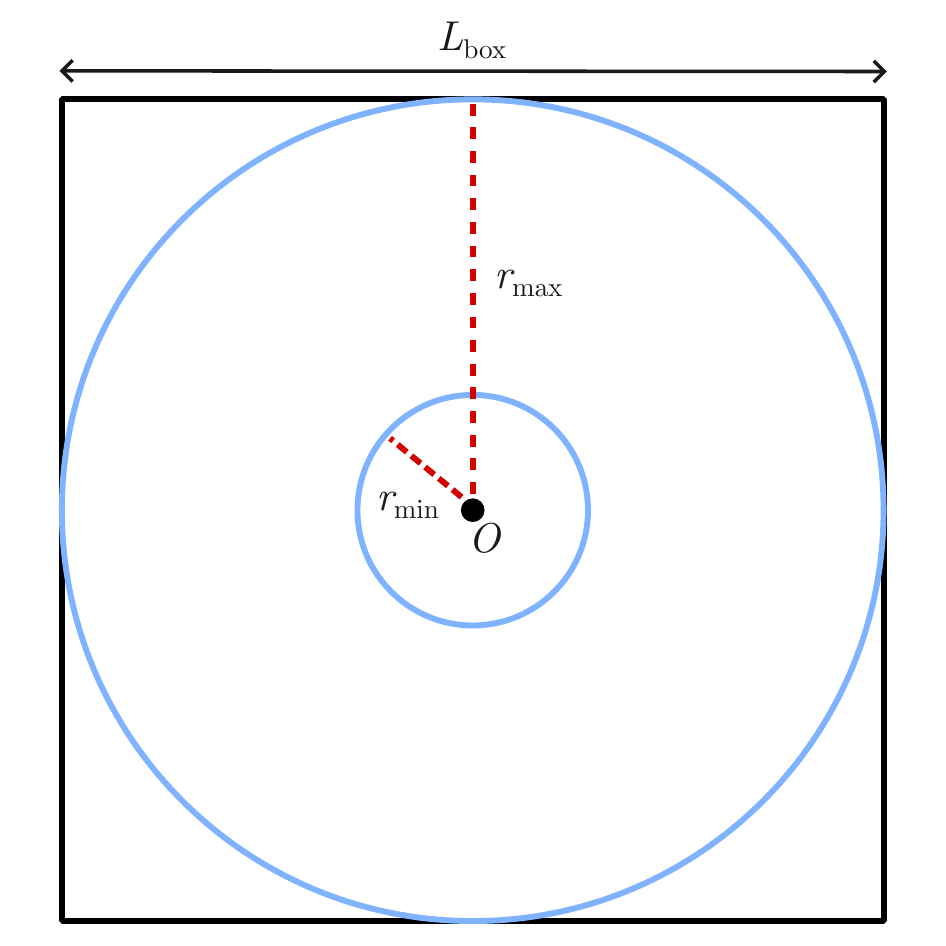}
     \caption{For each simulation we define an ``observable spherical cell'' in $3D$. A $2D$ plane projection is shown here, with $[r_{\rm min}=10\ {\rm Mpc}, r_{\rm max} = L_{\rm box}/2]$, as described in the main text.}
     \label{fig:observable_sphere}
\end{figure}

In our analysis, we use halo catalogues from the different simulation boxes and estimate the effects of galaxy clustering and survey completeness on the determination of $H_0$. Our general methodology tries to mimic as closely as possible the observational studies and consists of the following steps:

\begin{itemize}
    \item \emph{Step 1}: We place an observer at the centre of the box. We limit the maximum distance range to a sphere of $L_{\rm box}/2$, which corresponds to about $1200$, $506$, and $72$ Mpc radii, for the different boxes\footnote{Note that here we include $h$ in the calculation. As a reminder, the original box sizes were reported in units of Mpc/h.}.
       
    We also define a minimum distance threshold of $10$ Mpc. This means that for each case we only consider the galaxies that are within a spherical cell with range $[r_{\rm min}=10\ {\rm Mpc}, r_{\rm max} = L_{\rm box}/2]$ In redshifts, the largest range covered is $z \in [0.005, 0.23]$, assuming the simulation cosmology (see also Figure \ref{fig:observable_sphere}).
    
    \item \emph{Step 2}: We choose randomly a halo from the simulation catalogue: this is our `true' GW source, with coordinates $(d^{\rm src}, \theta^{\rm src}, \phi^{\rm src})$.
    
    \item \emph{Step 3}: To avoid centering artificially our observational region at the `true' source, we perturb the centre of our localisation cone. To find the new position, we construct a $2D$ Gaussian distribution centered at the `true' source for the angular perturbation and a $1D$ Gaussian for the distance one, and draw a new position $(d^{\rm gw}, \theta^c, \phi^c)$, $c$ for `cone centre'. The latter will be the central point, around which we construct our localisation cone. The widths of this Gaussian distribution and the final cone depend on the observational errors we choose. These correspond to the distance uncertainty $\sigma_{d_L}$ and the sky localisation errors $\sigma_\theta^2$ of our detections. Section \ref{sec:Error_priors_source_selection} provides the details of this procedure. 
        
    \item \emph{Step 4}: For all the haloes inside the cone, we find their distance to the observer. We also calculate the haloes' projected angles on the sky as seen by the observer (see also Figure \ref{fig:galaxy_opening_angle}).
    
    \item \emph{Step 5}: For all the haloes inside the cone, we calculate $H_{0,i}$, using their distance information, corresponding to the EM $d_L$, and the GWs distance $d^{\rm gw}$ (see section \ref{sec:Ho_calculation}). 
    
    \item \emph{Step 6}: Finally, depending on the method (see section \ref{sec:Ho_posteriors}), we calculate the individual $H_0$ posterior for the cone.
    
    \item \emph{Step 7}: For multiple events (cones), we repeat the procedure above and compute the final $H_0$ posterior distribution, by combining observations. The maximum number of GW events $N_{\rm gw}$ is specified for each case, with a fiducial number $N_{\rm gw, max} = 100$, consistent with previous studies. 
    
    Although this number is of the order of magnitude of the observed number of BBH mergers, $N_{\rm BBH} \sim 70$ from the first three observing runs \citep{GWTC1_2019, GWTC2_2021, GWTC3a_2021, GWTC3b_2021}, and $N_{\rm BBH} = 120$ from GWTC-4.0 \citep{GWTC4_Catalogue_2025}, as we discuss at the end of section \ref{sec:Error_priors_source_selection}, our error configuration is too optimistic for current capabilities, but more relevant and conservative for a near future detector network which would be expected to reach these numbers of BBH detections within months, and to detect thousands of BBHs per year. Measurements using different numbers of GWs events are studied in section \ref{sec:convergence_studies}.
\end{itemize}{}

\begin{figure*}
	%\addtocounter{figure}{-1}
	\centering
	\begin{subfigure}[b]{\textwidth}
		\centering
		\includegraphics[width=\textwidth]{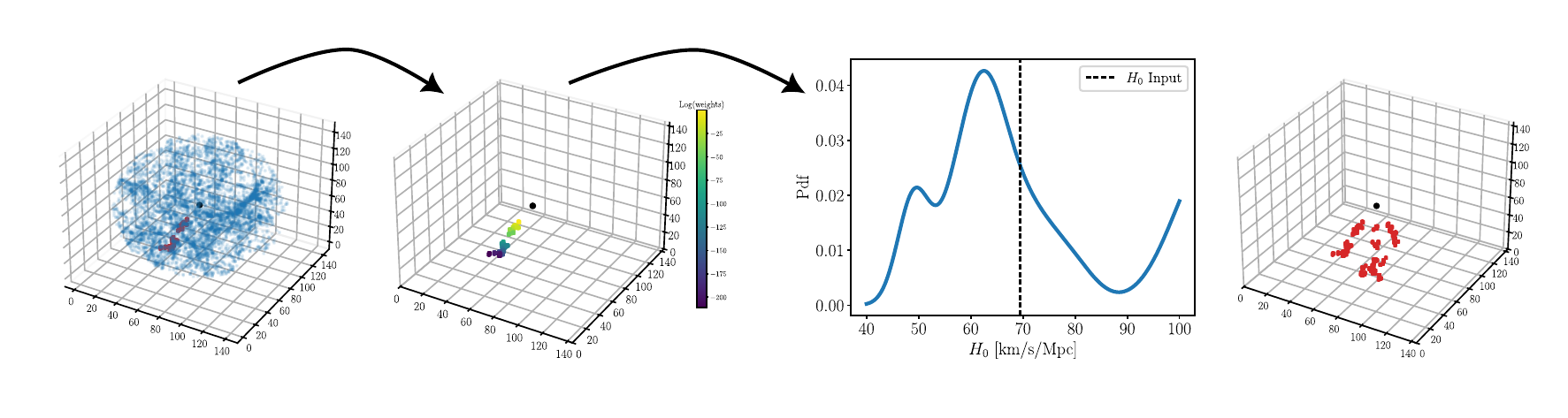}
	\end{subfigure}
	\caption{\emph{Graphical summary of our analysis}: a) We cut an ``observational sphere'' from the test simulation box. A GW event is modeled as a 3D cone (red dots); b) the observer (black dot) is put at the centre of the ``observational'' sphere and the galaxies in the cone are assigned weights (geometrical or physical); c) each event produces an $H_0$ posterior; d) this procedure is repeated for multiple events and their $H_0$ posterior pdfs are combined. The $L_{\rm box}=100$ Mpc/h box, with fewer galaxies, has been chosen for visualisation purposes.}
	\label{fig:Graphical_summary}
\end{figure*}

We will describe some of these steps in more detail in the following sections. For a graphical summary, see Figure \ref{fig:Graphical_summary}.

\subsection{Error priors and choosing the GW source}\label{sec:Error_priors_source_selection}

To select our GW sources and construct our conical regions (section \ref{sec:Cone_construction}), we mimic the observational situation of real GW events \citep{Soares-Santos_et_al_2019}. More specifically, we follow a procedure similar to \citet{Muttoni_et_al_2022}, who studied the multiband detection of massive BHs ($M_\bullet > 120 M_\odot$) by a combination of future generation, ground and space based detectors, LISA \citep{LISA_2017} and Einstein Telescope \citep{Einstein_Telescope, Einstein_Telescope_2023}. 

\citet{Muttoni_et_al_2022} used a Fisher matrix approach to estimate observational errors and to position their GWs sources. In our case, since we want to investigate the effects of different cone openings and sizes, we use a simplified, custom prescription to insert observational errors, described in more detail below and in section \ref{sec:Cone_construction}, but we position our GW sources in a similar way.

More specifically, there are two errors that are relevant in our case:  

\begin{itemize}
    \item A distance error $\sigma_{d_L}$, that corresponds to the error in the luminosity distance of the GW source. Here we model this error as a linear function of luminosity distance, i.e., $\sigma_{d_L} = A\cdot d^{\rm gw}$, where $A$ is a fixed proportion constant that we select in each simulated analysis.

    \item An angle error $\sigma_{\theta}^2$ (or $\Delta \Omega$ below), that corresponds to the uncertainty of the sky localisation area (in units of squared degrees).
\end{itemize}

Usually, the errors quoted correspond to the $90 \%$ confidence interval\footnote{This means that the standard deviation is $\tilde{\sigma} = \sigma/1.645$. The tilde sign $(\sim)$ will be used to make this distinction clear, where needed.}.

Based on these errors we build a theoretical $3D$ Gaussian distribution around the GW source, given by the product $\mathcal{N}(d^{\rm  src}, \tilde{\sigma}_{d_L})$ for the luminosity distance uncertainty and $\mathcal{N}(\theta^{\rm src}, \tilde{\sigma}_{\theta})$ for the uncertainty in the two angles respectively. We consider the three errors as independent, and use them to draw the centres of our cones, by sampling them from these distributions. In other words, the original GW source and the centre of the cones (the GW events' 3D localisation) are in general, different.

In specific cases, described in section \ref{sec:Cone_construction}, the GW source needs to be discarded and a new random galaxy is to be selected. Then we repeat the procedure above, till all the criteria are satisfied.

\begin{figure*}
	%\addtocounter{figure}{-1}
	\centering
	\begin{subfigure}[b]{0.305\textwidth}
		\centering
		\includegraphics[width=\textwidth]{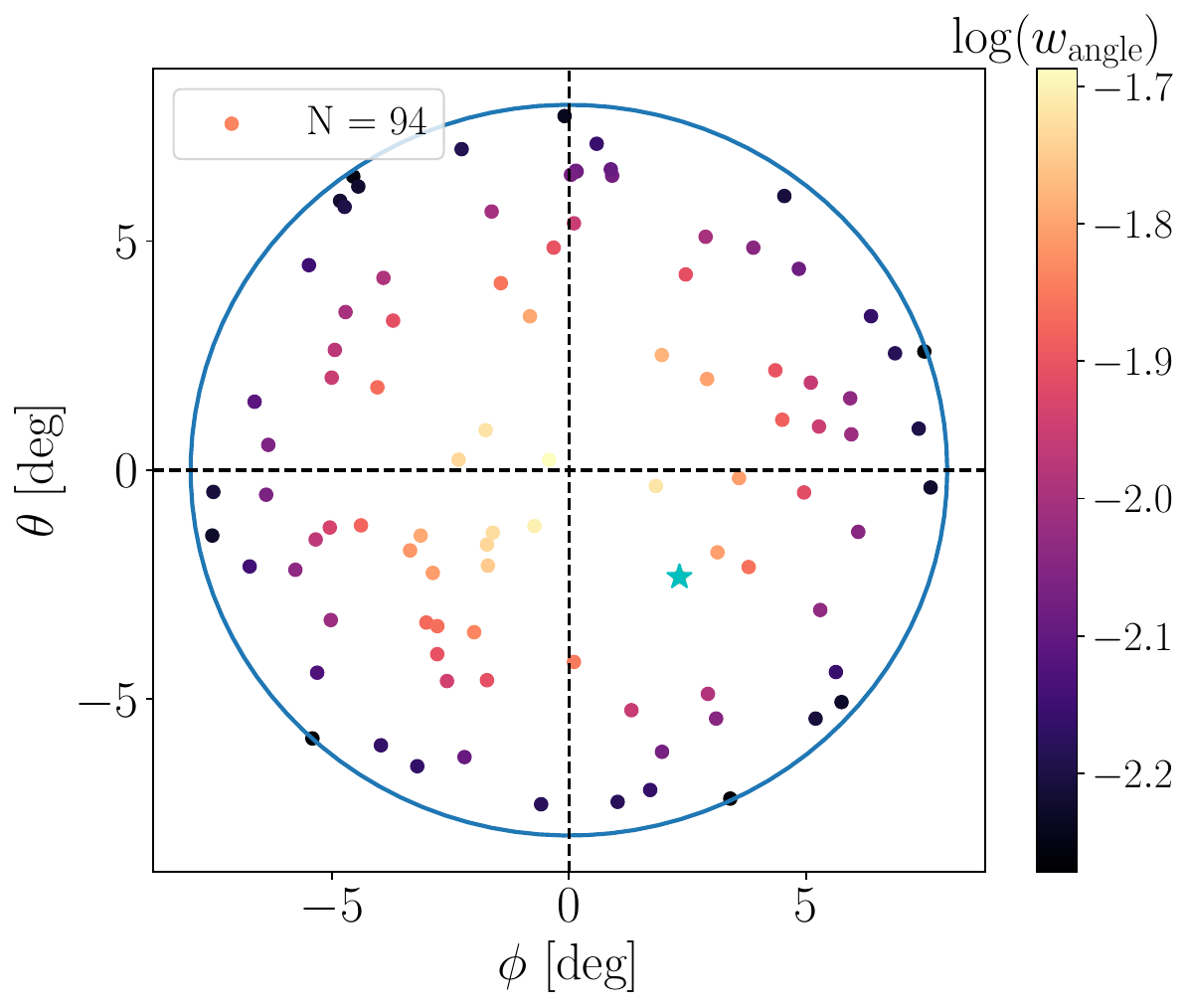}
	\end{subfigure}
	\hspace{0.5cm}
	\begin{subfigure}[b]{0.305\textwidth}
		\centering
		\includegraphics[width=\textwidth]{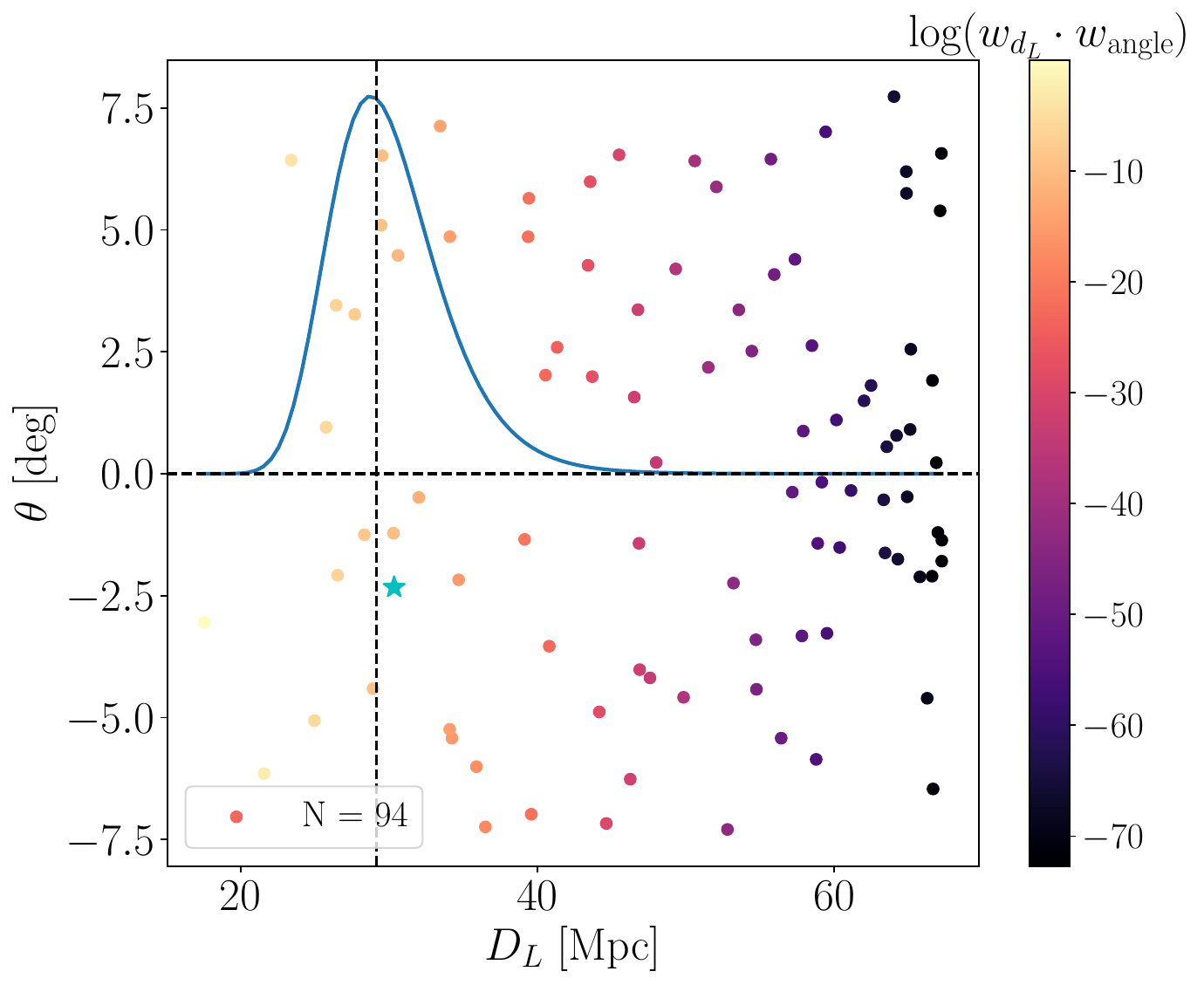}
	\end{subfigure}
	\hspace{0.5cm}
	\begin{subfigure}[b]{0.305\textwidth}
		\centering
		\includegraphics[width=\textwidth]{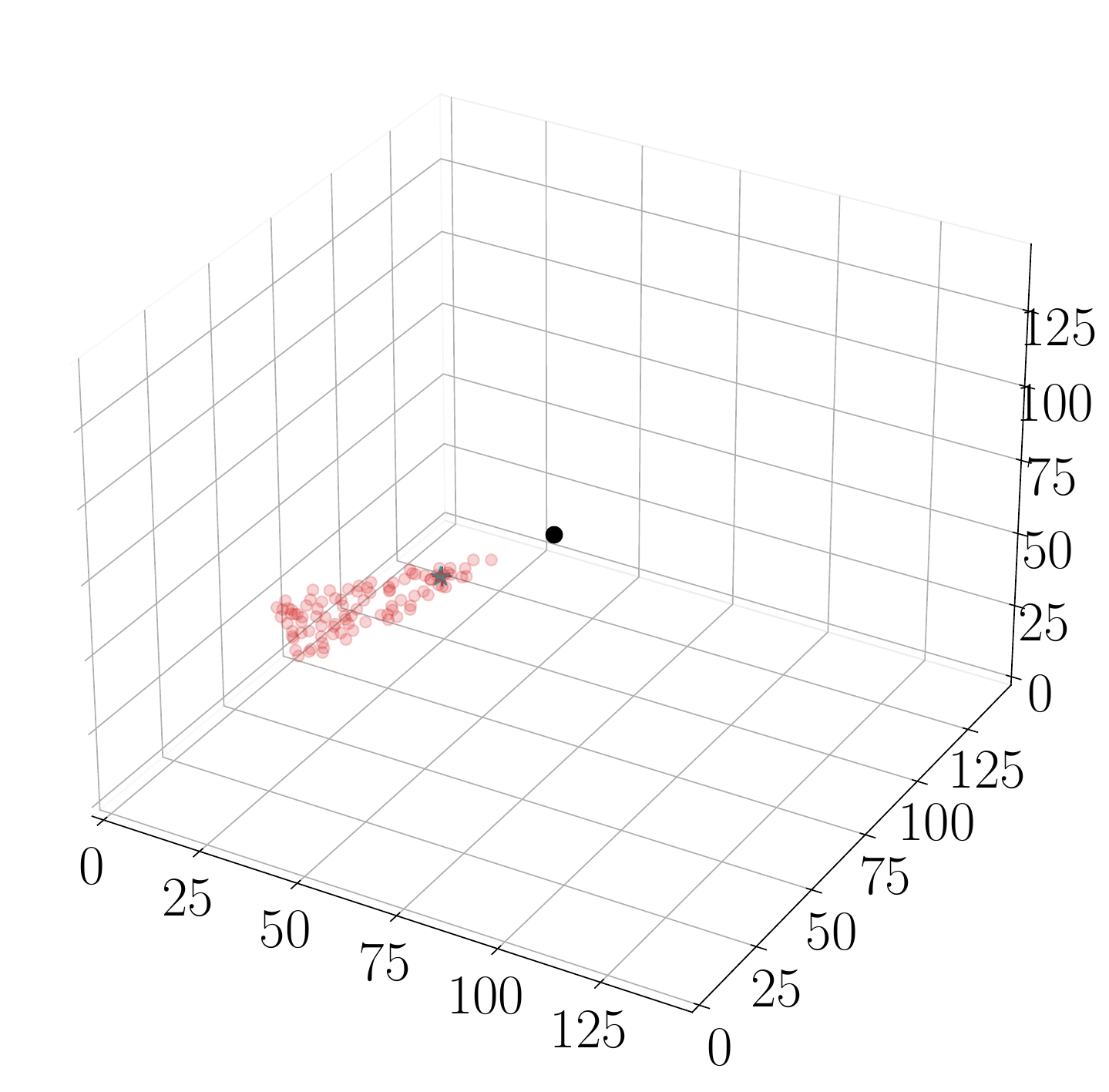}
	\end{subfigure}
	\caption{\emph{Example of cone construction and of geometric weights}: We show one example of cone construction and the weights assigned to each galaxy inside the cone. The black, dashed lines show the coordinates of the cone centre, the cyan star shows the position of the true GWs source, while the other dots show the galaxies that belong to this cone. \emph{Left}: We show the angular position of the galaxies ($\phi, \theta$), together with the geometric weights they are assigned with respect their position to the cone centre. \emph{Centre}: Same cone, but from the side, with the x-axis, showing the luminosity distance of the galaxies. The weights here correspond to the combination of angular and distance weights $w=w_{\rm \theta} \cdot w_{L}$ (see section \ref{sec:Geometric_physical_weights}). We also plot the distance likelihood, eq. (\ref{eq:distance_lhd}) - blue, solid line. Note that the maximum distance of a galaxy inside the cone is bigger than the $95\%$ range of the likelihood, and this is due to our consideration of the $H_0$ prior (recall section \ref{sec:Cone_construction}), i.e. for larger $H_0$ distances with the same redshift would be smaller, and as a result will be within the limits of our cone. \emph{Right}: 3D configuration of the cone. The black dot corresponds to the centre of the box, i.e. the observer. The $L_{\rm box}=100$ Mpc/h box, with fewer galaxies, has been chosen for visualisation purposes.}
	\label{fig:cone_profiles}
\end{figure*}

We find that this procedure is robust and avoids introducing a clear bias in our final results. Also, it is similar to observed signals, where the true source is not necessarily located exactly at the centre of the sky localisation region. In the \emph{physics implementation}, we introduce some extra weights $w_i$ before the selection of the GW source (section \ref{sec:Geometric_physical_weights}). Examples of the above procedure are shown in Figure \ref{fig:cone_profiles}.

These two errors ($A, \Delta \Omega$) are selected at the start of each simulated analysis, and are kept fixed for all events of the specific run. Below we study the impact of different error choices at the inference of $H_0$ (section \ref{sec:observational_errors_posteriors}), and we choose our `default' values as $(A, \Delta \Omega) = (0.1, 60)$. We note here that these observational errors are very optimistic for current GW detectors. For example, the large majority ($>85 \%$) of BBH detected in the LVK observing runs up to  O4a \citep{GWTC4-cosmic-expansion} have $\Delta \Omega > 100$ deg$^2$, and larger distance uncertainties. The main reason for this fiducial choice is the limitations from the size of our cosmological simulations. Nevertheless, in section \ref{sec:observational_errors_posteriors}, we investigate the effects of different observational errors, covering ranges $30 \ {\rm deg}^2\ \leq \Delta \Omega \leq 120\ {\rm deg}^2$ and $0.05 \leq A \leq 0.1$. The primary analysis here would be most relevant for a near-future detector network - LIGO Livingston, LIGO Hanford, LIGO India, Virgo and KAGRA (by the end of the O5 run \citep{Prospects_loc_O5}) - where thousands of BBHs events are expected to have a sky localisation and fractional distance uncertainty below $100$ deg$^2$ and $0.1$, per year, respectively \citep{Maggiore_et_al_future_loc_2025}. 

\subsection{Cone construction}\label{sec:Cone_construction}

Based on the error selection (section \ref{sec:Error_priors_source_selection}), we can construct our cones. Their geometry is fixed as following: The height of the cone depends on the distance error $\sigma_{d_L}$ and the $H_0$ prior range. More specifically, the limits of the cone would depend on how we model our distance likelihood (section \ref{sec:bayesian_analysis_method}). The default setting is a skewed Gaussian on the luminosity distance, following \citep{Hitchhiker_2022},
\begin{align}\label{eq:distance_lhd}
    \mathcal{L}(d^{\rm gw}|&d_L(z, H_0)) = \nonumber \\ 
    &= \frac{1}{\sqrt{2 \pi} \tilde{\sigma}_{d_L}}\exp{\left(-\frac{1}{2} \left[ \frac{d^{\rm gw} - d_L(z, H_0)}{\tilde{\sigma}_{d_L}}\right]^2\right)},
\end{align}
where $d^{\rm gw}$ is the distance of the `observed' GWs event (centre of the cone) and the dependence of the error in the denominator on $d_L$, i.e. $\tilde{\sigma} \sim A \cdot d_L$ breaks the symmetry of the Gaussian, allowing a bigger `tail' on larger distances.

\begin{figure}
    \centering
	\includegraphics[width=1.0\columnwidth]{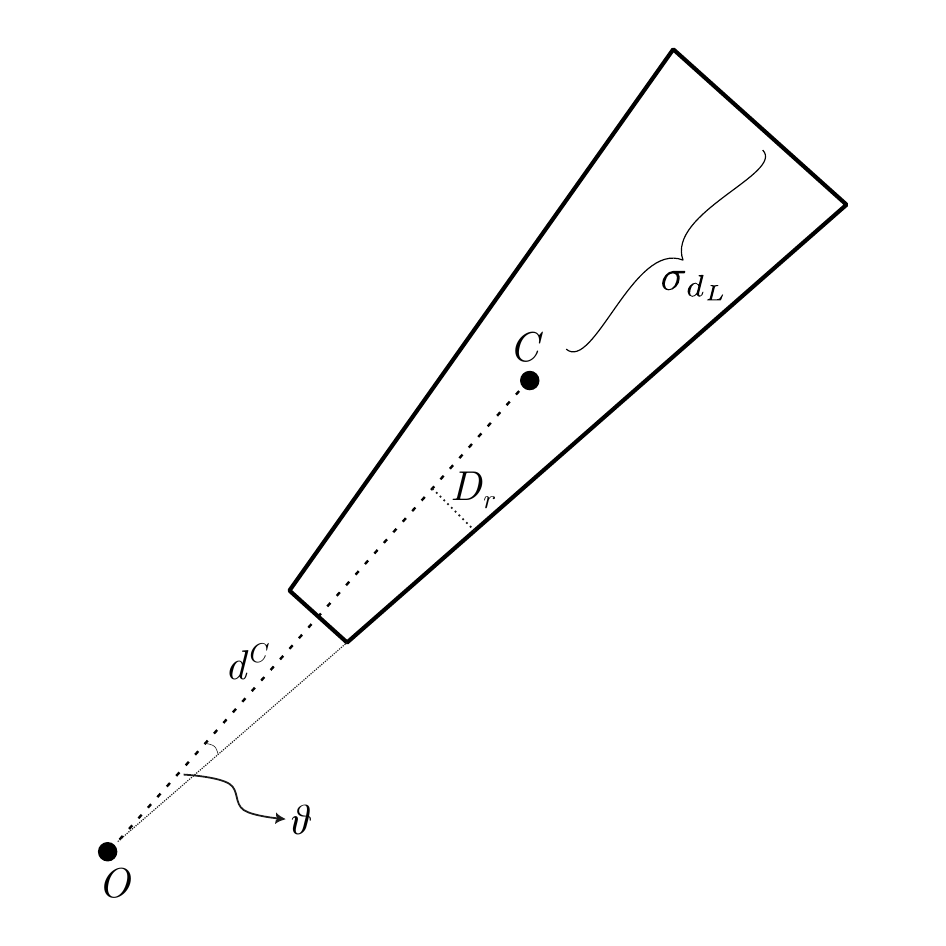}
    \caption{The most important distances in our cone construction. See discussion in section \ref{sec:Cone_construction} for details.}
    \label{fig:Cone_construction}
\end{figure}

\begin{figure}
     \centering
     \includegraphics[width=\columnwidth]{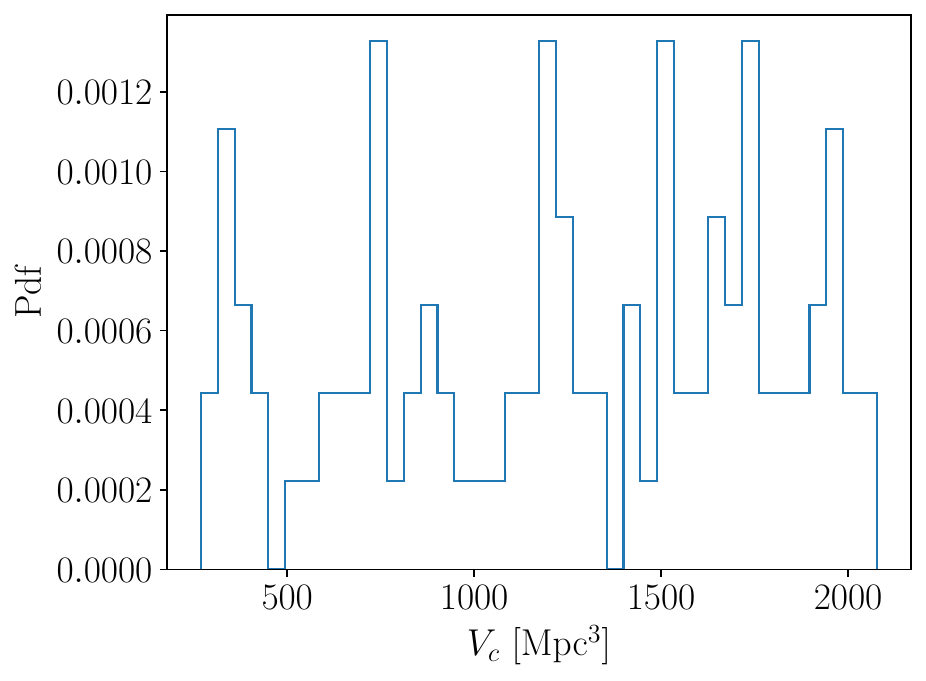}
     \caption{As discussed in section \ref{sec:Cone_construction}, the shape of our sky region should not, statistically speaking, affect the results, as long as our regions are large enough to contain large-scale structure information. We plot here a histogram of the volume of each cone, in a run with $N=100$ events in the $L_{\rm box}=100$ Mpc/h. The size of the cones, even in that case, is large enough, such that no bias is expected from the shape of our cuts.}
     \label{fig:cone_volumes}
\end{figure}

Using this distribution, we calculate the extent of the cone. We use the cumulative distribution function of the distance likelihood, to find the distance range that corresponds to the $90 \%$ probability of the sky localisation:
\begin{equation}
    p = \frac{1}{2} \left[1+{\rm erf}{\left( \frac{d_L-d^{\rm gw}}{\tilde{\sigma} \sqrt{2}} \right)} \right].
\end{equation}
Solving for $d$, for the cases of $p=5 \%$ and $p=95 \%$, we find the minimum and maximum distances along the line of sight, $d_L^{5}$ and $d_L^{95}$ respectively. Note that these limits would vary for a different underlying cosmology, so we take into account the $H_0$ prior range considered in this work. This will lead to a more extensive range of distances $[d_{\rm min}, d_{\rm max}]$:
\begin{align}
    d_{\rm max} &= d_L^{95} \times H_0^{\rm prior\ max}/H_0^{\rm sim} \\
    d_{\rm min} &= d_L^{5} \times H_0^{\rm prior\ min}/H_0^{\rm sim}.
\end{align}
The details of this scaling are described in section \ref{sec:Ho_calculation} and in eq. (\ref{eq:Ho_calculation}). The above constrain the line-of-sight extent of the cone. Using this distance range we perform the first galaxy cut, i.e. we keep only galaxies with distances compatible with $[d_{\rm min}, d_{\rm max}]$. Additionally, we need to consider its angular opening. Although the sky localisation area is rarely a circle, we idealise it as such with our cone structure\footnote{An asymmetry in the sky localisation, as in most real events, will introduce two opposite effects. On the one hand, it could improve determination of $H_0$, since some galaxies will be disfavoured more clearly, based on their position. On the other hand, it usually includes a higher number of galaxies, flattening the $H_0$ posterior. In the final analysis, the two approaches should be statistically equivalent, as long as the regions are big enough to include the large-scale structure effects of interest. We comment further on that in Figure \ref{fig:cone_volumes} and in section \ref{sec:results}.}. From the sky localisation area error $\Delta \Omega = \sigma_{\theta}^2$, we calculate the opening angle as $\theta = \sqrt{\sigma_{\theta}^2/\pi}$. The geometry can be seen in Figure \ref{fig:Cone_construction}. We calculate the radius of the cone for all distances as:
\begin{equation}\label{eq:maximum_cone_radius}
    D_r(d_p) = d_p \tan \theta,
\end{equation}
where $d_p$ is the projected line-of-sight distance between a halo in the cone and the observer. Implementing this second constraint, we keep only the galaxies that lie within the cone.

Finally, when constructing the cones, we take into account some additional restrictions: since our observational sphere has a specific maximum limit (Figure \ref{fig:observable_sphere}), periodic boundary conditions should be imposed, or we need to constrain the regions where the cones can be constructed to avoid ``overflowing'' outside the box. We follow the latter option, i.e., we avoid placing our cones' centres close to the boundaries. We note that this procedure could produce an ``edge of the sphere'' effect, where a hard cut-off on the presence of galaxies outside the sphere biases the $H_0$ inference. However, in our case we avoid this problem by restricting the cone positions so they do not intersect the boundary of the box. This is mildly inconsistent, but is analogous to assuming that the statistical distribution of galaxies away from the boundaries of the sphere, i.e. closer-by galaxies, is representative of the distribution across the whole sky, which is a valid assumption for the volumes we use.

More specifically, we implement the following procedure:
\begin{itemize}
    \item We choose a spherical geometry roughly consistent with the ``horizon distance'' of current, ground-based GW detectors.
    \item We impose a maximum distance of allowed galaxies' positions $d_L^{\rm max} = L_{\rm box}/2$, to avoid `overflowing' outside the box, both for data generation and the $H_0$ inference. 
    \item We impose a minimum distance of allowed galaxies' positions $d_L^{\rm min}=10$ Mpc, to avoid the observer being close to the cone, both for data generation and the $H_0$ inference. 
    \item If a cone is created, such that a galaxy inside has a distance with $d_L>d_L^{\rm max}$ or $d_L<d_L^{\rm min}$, the cone is drawn again by selecting a new GWs source and centre - refer to section \ref{sec:Error_priors_source_selection}.
\end{itemize}

This procedure creates cones (events) which lie within our ``observable spherical cell'' (Figure \ref{fig:observable_sphere}).

\subsection{\texorpdfstring{$H_0$}{Ho} calculation}\label{sec:Ho_calculation}

The standard method to calculate the Hubble parameter at low redshifts ($z \ll 1$) is to use the approximation \citep{Weinberg_Cosmo}:
\begin{equation}
    H_0 = \frac{cz}{d_L}.
\end{equation}
This relation is cosmology independent, but using it at higher redshifts can lead to a serious bias - towards higher values - in the measurement of $H_0$, so extra care is needed \citep{Redshifts_DavisScrimgeour_2014}. In practice, for higher redshifts we cannot avoid the cosmological dependence \citep{Redshifts_DavisScrimgeour_2014}, so we have to either keep extra redshift terms in the luminosity distance expansion or fit the standard luminosity distance for a flat, $\Lambda$CDM universe:
\begin{equation}\label{eq:distance_LCDM}
    d_L (z) = (1+z)\frac{c}{H_0} \int_{(1+z)^{-1}}^1 \frac{da}{\sqrt{\Omega_m a + (1-\Omega_m)a^4}} = \frac{K(z)}{H_0},
\end{equation}
where we have neglected the contribution of cosmic radiation and have introduced the scale-factor $a$, with $a_0=1$. In our scenario, the GW observation provides a direct luminosity distance measurement, i.e., $d_L = d^{\rm gw}$, while the redshift is obtained through the photometric or spectroscopic information of the galaxies in the catalogue. Assuming that each galaxy in the localisation region acts as a potential host, one can marginalise over luminosity distance and redshift information, to infer the different cosmological parameters.

In a real observational scenario, one would have direct access to the galaxies' redshifts. In our case, the cosmological simulations provide us with information about the 3D positions of the galaxies, so one first has to translate these into redshifts, using the relevant input cosmology, and then proceed as above. Since this step would only make the analysis more time consuming, without affecting the final inference, we choose to circumvent it using a ``distance rescaling'' (more details on this below). This general procedure does not use any approximation for the expression of the luminosity distance and consequently the general conclusions are valid for future GW observations, which are expected to reach $z>1$ \citep{Chen_Ezquiaga_Gupta_2024_NextGen}. For this work, we fix all other cosmological parameters to the values of the simulations, and we only infer $H_0$.

To understand the validity of the rescaling, we start from equation (\ref{eq:distance_LCDM}). For a galaxy at redshift $z^g$, for a given $H_0$, this leads to a distance $d_L^g = K(z^g)/H_0$. If we assume that this galaxy is the source of the GW event, then $d_L^g=d^{\rm gw} \Rightarrow H_0^g = K(z^g)/d^{\rm gw}$.

As we mentioned above, in an idealised observational setting with precise knowledge of $d^{\rm gw}, z^g$ and of the other cosmological parameters, this relation would return an $H_0$ value for each galaxy. In contrast, the simulations provide the coordinates of the galaxies, and not their redshifts. Instead of performing a two step procedure: distances $\rightarrow z \rightarrow$ distances, we circumvent these steps by exploiting the easily obtained distance information with respect to an observer. To do this, we modify this relation to
\begin{equation}
    H_0^g = \frac{K \left[ z^g(d^g, H_0^{\rm sim}) \right]}{d^{\rm gw}}.
\end{equation}

We note that, from eq. (\ref{eq:distance_LCDM}), $K(z^g) = d_L^g \cdot H_0^{\rm sim}$, i.e. we use the simulation cosmology $H_0^{\rm sim}$ to connect a given distance with a redshift\footnote{As one could have done, if they decided to perform the analysis in redshift space.}, and hence we can write
\begin{equation}\label{eq:Ho_calculation}
    H_0^g = H_0^{\rm sim} \frac{d_L^g}{d^{\rm gw}}.
\end{equation}

The distances are calculated directly from the coordinate information given by the simulation\footnote{For the respective redshift ranges, we assume that comoving and physical distances are the same. However, this will not affect the final conclusions of our work, since we are interested in distance ratios.}. In eq. (\ref{eq:Ho_calculation}) we have: $H_0^g$, the inferred Hubble parameter for galaxy $g$, $H_0^{\rm sim}$ the injected Hubble parameter, i.e., the one used in our simulation, $d^{\rm gw}$ the distance inferred from the GW observation, i.e., the centre of the cone in our case and $d^g_L$ the distance of a galaxy within the conical region. This procedure gives us, for each cone, $N$ values of $H_0^g$, one for each galaxy $d^g_L$. Note that we first select the cone centre $d^{\rm gw}$, which is then kept fixed for the specific cone. 

The injected value of $H_0^{\rm sim}$, is taken from WMAP9 \citep{Hinshaw_et_al_2013_WMAP}, in order to be consistent with our cosmological simulations.

The preceding arguments indicate that the main contribution to the $H_0$ estimate will come from the relative position of each galaxy in the region compared to the estimated location of the GW source. This formulation makes clear the positive effect of clustering to the final statistics. When clustering is present, there will be a higher probability that the true source is grouped together with other galaxies in the region, and as a result, share similar distances.

Finally, we note that this procedure to calculate $H_0$ is used directly in section \ref{sec:Ho_histogram_stacking}, where we `stack' the $H_0$ values of all the galaxies in the sky localisation areas of the events. If one wants to use this procedure in a real observational scenario, they would need to work directly with redshifts, and could not use the rescaling procedure - eq. (\ref{eq:Ho_calculation}) - which exploits the fact that the input $H_0^{\rm sim}$ of the simulations is known. Instead, one would use directly eq. (\ref{eq:distance_LCDM}), to get $H_0^g$ from the ($K(z^g), d^{\rm gw}$) pair. In section \ref{sec:bayesian_analysis_method}, where we develop a Bayesian analysis to infer $H_0$ we use this procedure to update the distances of the galaxies, but not to calculate $H_0$ directly.

\subsection{\texorpdfstring{$H_0$}{Ho} posteriors}\label{sec:Ho_posteriors}

To calculate the posterior on $H_0$ we follow two different methods, of increasing complexity and similarity to observational techniques. We analyse them in turn.

\subsubsection{\texorpdfstring{$H_0$}{Ho} stacking}\label{sec:Ho_histogram_stacking}

This case extends the work of \citet{MacLeod_et_al_2008}. In their work, they consider:

\begin{itemize}
    \item Boxes, as the localisation volume, which avoid realistic geometry biases.
    \item Galaxies with equal weights, i.e., they are not weighted based on their true position in the box.
    \item Very precise distance determination, which would be more relevant for LISA \citep{LISA_2017}.
    \item GW events with equal weights, i.e., irrespective of the number of galaxies in the box/sky uncertainty.
\end{itemize}

In our implementation of this method, we aim to improve on these assumptions, making the following changes:

\begin{itemize}
    \item We implement a more realistic sky localisation, both in terms of capabilities of current ground-based GWs detectors, but more importantly in terms of the sky volume, i.e., using skewed cones, instead of square boxes.
    \item We include geometric and physical weights for each galaxy (section \ref{sec:Geometric_physical_weights}).
    \item We simulate a uniform catalogue, to estimate the importance of clustering, by comparing to a realistic catalogue from cosmological simulations. 
    \item We take into account the different quality of each GW event (section \ref{sec:Geometric_physical_weights}). 
\end{itemize}

For the simpler case of `$H_0$ stacking', or `histogram analysis', we follow equation (\ref{eq:Ho_calculation}). 

This means that for each cone we have $N_c$ values of $H_0^g$, where $N_c$ is the number of galaxies in the cone. When considering the final $H_0$ histogram, we stack together all these $H_0^g$ values from multiple cones/events. 

Finally, we note a couple of things:

\begin{enumerate}
    \item Firstly, the centre of the cone does not correspond to the coordinates of any specific galaxy, since the cone centre is perturbed away from the true host galaxy by the observational error - sections \ref{sec:Error_priors_source_selection} and \ref{sec:Cone_construction}. 
       
    This means that we do not induce an artificial peak towards the input value $H_0^{\rm sim}$, and more clearly demonstrates the power of the method, and in particular clustering, since we recover $H_0^{\rm sim}$ even without any knowledge of the `true' host galaxy of the source.

    \item  Secondly, unless a correction is made, this `stacking' method will have a clear bias to higher values of the Hubble-Lema\^itre parameter, because of geometric effects, i.e., more galaxies are in the far side of the cones. To make this clearer we return to equation (\ref{eq:Ho_calculation}). The cone centre distance $d^{\rm gw}$, is the geometric centre of the cone. However, due to the cone geometry, with consecutive rings covering more volume at larger distances, this leads to an increased number of galaxies with $d_L^g>d^{\rm gw}$, and from equation (\ref{eq:Ho_calculation}), $H_0^g>H_0^{\rm sim}$. 
    
    A consistent resolution to these ``volume'' effects is described in section \ref{sec:bayesian_analysis_method}, where the full likelihood is calculated. To address these in the simplified ``stacking'' method, we introduce the GWs measurement errors as weights and build a ``weighted histogram'', where each galaxy's contribution is weighted by its position in the cone, using the weights from section \ref{sec:Geometric_physical_weights}. 
\end{enumerate}

The results of this method are discussed in section \ref{sec:stacking_results}.

\subsubsection{Geometric and Physical Weights}\label{sec:Geometric_physical_weights}

\begin{figure}
    \centering
	\includegraphics[width=1.0\columnwidth]{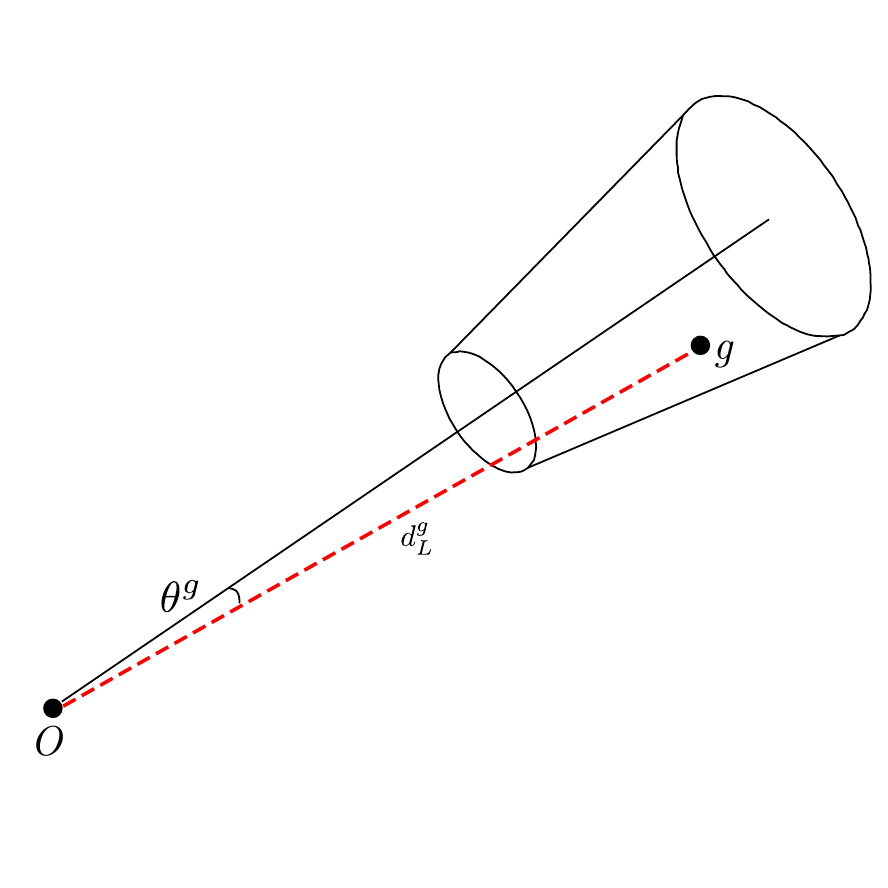}
    \caption{To calculate the geometric weights for each galaxy \emph{g}, we use its angle $\theta^g$ with respect to the centre of our observational cone, and its luminosity distance $d_L^g$. See discussion in section \ref{sec:Geometric_physical_weights} for details.}
    \label{fig:galaxy_opening_angle}
\end{figure}

For each galaxy we have information about its position inside the cone, both in terms of distance from the observer, but also its sky coordinates. So, we have a luminosity distance between the galaxy and observer $d^g_L$ and a pair of angles ($\phi, \theta$) depending on sky position. Hence, we can use the observational errors to assign geometric weights to the galaxies. In the full analysis - section \ref{sec:bayesian_analysis_method} - these weights are part of the likelihood construction: a priori each galaxy is considered equally likely to be the host of the GW event. However, in the presence of GW data, specific sky locations and distances are disfavoured as hosts of the true source by the GW likelihood. For the `$H_0$ stacking' method - section \ref{sec:Ho_histogram_stacking} - these are included as additional weights when constructing the histograms. The weighting is based on the galaxies' coordinates versus the coordinates of the `observed' GW, at the centre of the cone. For the luminosity distance:
\begin{equation}\label{eq:los_errors}
    w_L \sim \exp{\left[-\frac{1}{2}\left(\frac{d^g_L - d^{\rm gw}}{\tilde{\sigma}_{d_L}}\right)^2\right]},
\end{equation}
where we recall that $\tilde{\sigma}_{d_L} = \sigma_{d_L}/1.645$, since $\sigma_{d_L}$ corresponds to the $90 \%$ confidence interval. For the angles, we exploit the symmetry of the cones and define their angular position with respect to the line-of-sight between the observer and the cone centre (Figure \ref{fig:galaxy_opening_angle}). 

The weight is then given by:
\begin{equation}\label{eq:angle_errors}
    w_{\theta} \sim \exp{\left[-\frac{1}{2}\left(\frac{\theta^g-\theta^{\rm gw}}{\tilde{\sigma}_{\theta}}\right)^2\right]}.    
\end{equation}

Apart from the above geometric weights, we also include, in some runs, weights based on the physical properties of the galaxies - magnitude, star formation rate or stellar mass - as a simple multiplicative factor. For example, for masses, we calculate a weight for each galaxy $w_M^g = M^g/\sum_i M^i$. The weights can be used to penalise the contribution of individual galaxies to the final $H_0$ posterior, the selection of hosts for GW sources or both. To avoid any statistical bias due to low number of galaxies inside each cone, we calculate the physical weights based on the initial catalogue, with all the galaxies present, not including only the cone galaxies for each event. 

Finally, we note that the physical weights are used in two places in our analysis. One, when a detection threshold is chosen, in order to construct incomplete catalogues (section \ref{sec:completeness_inv}), and two, when the impact of physical information on the $H_0$ posteriors is investigated  (section \ref{sec:weights_results}). Therefore, the specific calculation of the physical properties - in our work based on the ML techniques described in section \ref{sec:stellar_Halo_mass} - should be negligible to the final results for two main reasons: (1) in calculating the incompleteness fraction, we are only interested in a threshold that allows a specific percentage of galaxies to ``survive''. Since the physical properties are estimated in a uniform way between our simulation boxes, they allow a consistent calculation of incompleteness. The completed catalogues are then investigated on their properties (e.g., clustering power, distribution of galaxies) and give reasonable results (see also Figure \ref{fig:sphere_completion} and Appendix \ref{ap:resolution_effects}). (2) In the case of $H_0$ posteriors, the greatest impact originates from the modelling of the weighting. In our simple, linear model, we have checked that the actual values of the physical parameters (stellar masses, magnitudes, star formation rates) are sensible, and since they are calculated consistently among runs, they are not expected to bias our results (section \ref{sec:weights_results} discusses further the implication of physical weights on the $H_0$ inference).

\subsubsection{Bayesian analysis}\label{sec:bayesian_analysis_method}

We now develop a Bayesian analysis to calculate the posteriors of $H_0$. We follow the approach in \citet{Del_Pozzo_2012, Chen_et_al_2018, Muttoni_et_al_2022, Hitchhiker_2022}. From Bayes' theorem, we have:
\begin{equation}
    P(H_0|D) = \frac{\mathcal{L}(D|H_0) \pi(H_0)}{\mathcal{Z}},
\end{equation}
where the different terms are:

\begin{itemize}
    \item $\pi(H_0)$: Prior degree of belief on the $H_0$ value. We assume a flat distribution with $H_0\ \epsilon\ [40, 100]$ km/s/Mpc. Recall our input value from the simulations is in the middle of this range, with $H_0^{\rm sim} = 69.5$ km/s/Mpc. 
        
    \item $\mathcal{L}(D_c|H_0)$: Marginalised likelihood, given an $H_0$ value and a model.
    
    Data $D_c$ here are the observed distance and sky localisation of the GW sources. We call this a ``marginalised likelihood'', since all other GW source parameters, such as masses and spins, are assumed to be marginalised over. The same is true for galaxies' positions and physical properties. We make this procedure explicit below. The cosmological parameters, other than the Hubble-Lema\^itre parameter, are fixed as described in section \ref{sec:simulations}.
    
    \item $\mathcal{Z}$: The evidence. Used for model selection, but acts as a normalizing factor for parameter estimation. We will ignore it, since we will normalise our posteriors at the end.
\end{itemize}

For each cone, we calculate a marginalised likelihood $\mathcal{L}_c(D_c|H_0)$. Since each cone corresponds to a single, independent event, the full (marginalised) likelihood for $N_c$ events is:
\begin{equation}
    \mathcal{L}(D_c|H_0) = \prod_c^{N_c} \mathcal{L}_c(D_c|H_0).
\end{equation}
Each cone/event includes the contributions of all the galaxies in the sky localisation volume. For a single cone, this likelihood $\mathcal{L}_c$ marginalises over a number of terms, which we analyse in order below. Hence the data for a single observation are the cone distance and position in the sky, coming from the GW observations, i.e., $D_c = (d^c_L, \theta^c)$. The likelihood is constructed by marginalising over the properties of the galaxies in the region, i.e., $D_g = (z^g, \theta^g, w^g)$, which are independent for each observation. The weights $w^g$ depend on each galaxy's physical characteristics, like mass, star formation rate, or B-mag, as described in section \ref{sec:simulations}. The single cone marginalised likelihood is then:
\begin{align}\label{eq:single_cone_LHD_init}
    \mathcal{L}_{\rm c}(D_c|H_0) &= \int \mathcal{L}_{\rm c}(D_c, D_g|H_0) \ {\rm d} D_g =  \nonumber \\
    &= \int \mathcal{L}_{\rm c}(d^c_L, \theta^c, z^g, \theta^g, w^g|H_0) \ {\rm d} D_g =  \nonumber \\
    &= \int p(d_L^c|d_L) p(\theta^c|\theta) \nonumber \\
    & \hspace{1cm}\times p(d_L|z, H_0) p(\theta|\theta^g) p(z|z^g) p(w|w^g) \nonumber \\
    & \hspace{1.5cm} \times \pi(z^g, \theta^g, w^g) \ {\rm d}d_L {\rm d}z^g {\rm d}\theta^g {\rm d}w^g ,
\end{align}
where we have assumed that the contributions from distances and angles are independent. $\pi(z^g, \theta^g, w^g)$ corresponds to the priors on the galaxy's parameters, which for ($z^g, \theta^g$) in practice restrict the integral to only the galaxies in the sky localisation region, while for $w^g$ it covers uniformly the full range of possible values in the catalogue. In the following, we describe the ingredients in detail. 

We start by specifying $p(d_L|z, H_0)$. The distance information of the galaxies comes from their redshifts and a set of cosmological parameters (in our case $H_0$), through eq. (\ref{eq:distance_LCDM}), $d_L = K(z)/H_0$. Hence, for a given cosmological model only one of $d_L, z$ is independent:

\begin{itemize}
    \item $p(d_L|z, H_0)$: how we infer our distances given the redshift and cosmological parameters:   
    \begin{equation}
        p(d_L|z, H_0) = \delta ( d_L-d_L(z, H_0) ).
    \end{equation}
\end{itemize}
This simplifies eq. (\ref{eq:single_cone_LHD_init}) to
\begin{align}\label{eq:single_cone_LHD_marg_dist}
    \mathcal{L}_{\rm c}(D_c, \theta^c|H_0) &=  \int p(d_L^c|d_L(z, H_0)) p(\theta^c|\theta) \nonumber \\
    & \hspace{1cm}p(\theta|\theta^g) p(z|z^g) p(w|w^g) \nonumber \\
    & \hspace{1.5cm} \pi(z^g, \theta^g, w^g)\ {\rm d}z^g {\rm d}\theta^g {\rm d}w^g.
\end{align}

The galaxies' redshifts and sky positions come from EM observations, with their own uncertainties, which introduces a redshift probability $p(z|z^g)$ and a sky location probability $p(\theta|\theta^g)$.
\begin{itemize}
    \item $p(z|z^g)$: characterizes how well we know the redshifts of the galaxies. Here, we assume perfect knowledge of the galaxies' redshifts, i.e. $\delta(z-z^g)$. Since our main emphasis is on the effects of clustering and incompleteness, in the following we use precise galaxy redshifts. A detailed analysis of redshift effects on $H_0$ inference with ``statistical dark sirens'' can be found in \citet{Turski_et_al_Redshifts_2023, Cross-Parkin_et_al_2025_redshifts}.
    \item $p(\theta|\theta^g)$: characterizes how well we know the galaxies' sky positions. Again, we assume here perfect knowledge of the galaxies' sky positions, i.e., $\delta(\theta-\theta^g)$.
    \item $p(w|w^g)$: characterizes how we assign probabilities of hosting a GWs source to galaxy properties. We model the extra weights assuming a linear relation to the specific physical parameter, i.e., $p(w|w^g) \propto w \cdot \delta(w-w^g)$. We choose the linear modelling for simplicity, and comparison with some works in the literature \citep[e.g.][]{Gray_et_al_2020}. We note that this assumption is not unique and more physically informed weights probabilities are possible. For example, \citet{Perna_Mastrogiovanni_Ricciardone_2025} introduce a weighting procedure motivated by the Schechter function \citep{Schechter_1976}.
\end{itemize}
Substituting $p(z|z^g)p(\theta|\theta^g)p(w|w^g) \propto w \cdot \delta(w-w^g)\delta(z-z^g)\delta(\theta-\theta^g)$ in eq. (\ref{eq:single_cone_LHD_marg_dist}) and performing the marginalisation integrals, taking into account that in each cone we have multiple galaxies, we get
\begin{equation}\label{eq:single_cone_LHD_final}
    \mathcal{L}_{\rm c}(D_c|H_0) \propto  \sum_g w^g \cdot p(d_L^c|d_L(z^g, H_0)) p(\theta^c|\theta^g).
\end{equation}

We should note that all the $z$ integrals above, range from $z_{\rm min}$ to $z_{\rm max}$, constructed to cover the full cone constructed by $d^{\rm gw} \pm {\sigma}_d$, transformed to redshifts taking additionally into account the $H_0$ prior range. In practice, this means that we consider galaxies in the distance range $[d_{\rm min}, d_{\rm max}]$, defined in section \ref{sec:Cone_construction}.

We model the remaining two distributions as (skewed) Gaussians:
\begin{itemize}
    \item $p(\theta^{\rm c}| \theta^g)$: how far in the sky is the position $\theta^g$ of each galaxy in the cone compared to the cone centre.
    \begin{equation}\label{eq:angles_pdfs}
        p(\theta^{\rm c}|\theta^g) \sim \exp{\left[-\frac{1}{2}\left(\frac{\theta^c- \theta^g}{\tilde{\sigma}_{\theta}}\right)^2\right]}. 
    \end{equation}

    \item $p(d_L^c|d_L^g)$: the likelihood of each luminosity distance compared to our GWs information. We use the form adopted in \citet{Hitchhiker_2022}:
    \begin{equation}\label{eq:distances_pdfs}
        p(d_L^c|d_L^g) \sim \frac{1}{\tilde{\sigma}_{d_L}\sqrt{2 \pi}} \cdot \exp \left[-\frac{1}{2}\left(\frac{d_L^g-d_L^{\rm c}}{\tilde{\sigma}_{d_L}}\right)^2 \right],
    \end{equation} 
    where $\tilde{\sigma}_{d_L} \sim A d_L(z, H_0)$, with $A$ the factor discussed in section \ref{sec:Error_priors_source_selection}.    
\end{itemize}
Equation (\ref{eq:single_cone_LHD_final}), together with pdfs (\ref{eq:angles_pdfs}) and (\ref{eq:distances_pdfs}), gives the final expression for the likelihood for a single cone.

\begin{equation}
    \mathcal{L}_c(d^c_L, \theta^c|H_0) \propto \sum_g^N p(d_L^c|d_L(z^g, H_0)) \mathcal{N}(\theta^c|\theta^g) w^g.
\end{equation}

In the majority of the analyses below, we ignore any physical weights, i.e., we assume that the GW host can be any of the galaxies in the cone. Hence, unless the \emph{physics implementation} is selected, we set $w^g=1$, otherwise we use normalised weights based on physical properties of the galaxies. For example, if we weight with the masses, we define $w^g = m_g/\sum m_g$, where the sum extends to all the galaxies in the catalogue.

To create the final posterior distribution, we calculate the likelihood above throughout the prior range of $H_0$.

Finally, it is important to take into account GW selection effects, i.e., the importance of each event based on how likely they are to be detected \citep{Chen_et_al_2018, Gray_et_al_2020, Hitchhiker_2022}. For this reason, we divide the likelihood of each cone with

\begin{equation}
    P_{\rm det}(H_0) = \int_{{\rm det_{\rm GW}}} \mathcal{L}(D|D_{\rm g}, H_0) {\rm d} D 
    %{\rm d} D_{\rm g},
\end{equation}
where the integrals are over all possible detectable GW signals. For GWs, in a real scenario, one needs to take into account all the GW parameters and the detector capabilities. There are also selection effects for EM observation, that involve a number of properties of the sources and of the telescopes. In our case, we use a simplified model, by considering only GW selection criteria\footnote{Our procedure involves completing all the cases of incomplete catalogues, i.e., catalogues where galaxies are missing due to a magnitude threshold, by putting galaxies back, but in a uniform way (section \ref{sec:completeness_inv}). Consequently, all these catalogues can be dealt with as if they are ``complete'' in terms of EM selection criteria, which are not investigated further here.} and specifically a distance threshold $d_L^{\rm thr}$ to which GW sources can be reliably detected. As a result we can ignore all other terms in the likelihood except for $p(d_L|d_L(z, H_0))$. Following \citet{Hitchhiker_2022} to perform the distance integral, the detection probability $P_{\rm det}(z, H_0)$ is given by

\begin{equation}
    P_{\rm det}(z, H_0) = \frac{1}{2} \Biggr[ 1 + {\rm erf} \left( \frac{d_L^{\rm thr}-d_L(z, H_0)}{\sqrt{2} A d_L(z, H_0)} \right) \Biggl],
\end{equation}
where $d_L^{\rm thr}$ corresponds to the threshold distance where we can observe GW events. Since we only allow GW events, where their relevant cones are within our observational sphere, the maximum distance allowed is smaller than $L_{\rm box}/2$ and depends on the observational errors we assume for each analysis. More specifically, it is calculated as
\begin{equation}\label{eq:dcone_centre_threshold}
    d_L^{\rm thr} = L_{\rm box}/2 \cdot H_0^{\rm min}/H_0^{\rm sim} \cdot \cos(\theta)/(1+A),
\end{equation}
with $A$ the proportionality constant of the luminosity distance error, as described in section \ref{sec:Error_priors_source_selection}, while $\theta$ correspond to the opening angle of the cone radius, as shown in Figure \ref{fig:Cone_construction}. The latter corrects for the fact that the maximum distance from the observer in a cone, is not the line-of-sight direction, but the point that corresponds at the edge of the cone.

Marginalising over the population of galaxies in the catalogue, we end up with
\begin{align}
    P_{\rm det}(H_0) &= \int P_{\rm det}(z, H_0) p(z|\{z^g\}) {\rm d} z \nonumber \\
    & = \sum_g^N \int P_{\rm det}(z, H_0) \delta(z-z^g) {\rm d} z \nonumber \\
    & = \sum_g^N P_{\rm det}(z^g, H_0),
\end{align}
where in the second line we have replaced the population prior by the catalogue's discrete galaxy distribution, assuming it represents accurately a discrete realisation of the underlying galaxy population. The selection effects for each box are calculated once at the beginning of each run. 

Finally, our likelihood for a single cone is updated to
\begin{equation}\label{eq:likelihood_final_selection_effects}
    \mathcal{L}_c(D^{\rm gw}|H_0) \propto \frac{\sum_g^N p(d_L^c|d_L^g(H_0)) \mathcal{N}(\theta^c|\theta^g) w_g}{P_{\rm det}(H_0)}.
\end{equation}

This is equivalent to eq. ($15$) in \citet{Hitchhiker_2022}, where a catalogue of galaxies with perfect redshifts is assumed. Here we also include a term for the angular positions of the galaxies, which was negligible in the configuration of \citep{Hitchhiker_2022}. We now have all the tools in order to infer $H_0$ from dark sirens in simulations!

\subsection{Investigating completeness}\label{sec:completeness_inv}

The power of the statistical method lies in our capability to cross-correlate the GW sky localisation volume with the galaxies in the region.

So far, we have assumed perfect knowledge of all the galaxies that reside inside our observational sphere. However, in realistic scenarios, not all galaxies inside the sky localisation volume are observable. This is due to observational limitations of the EM surveys which are likely to produce galaxy catalogues that are incomplete, i.e., with galaxies missing due to specific sensitivities in magnitude. 

In these cases, the standard procedure, that we also follow here, is to complete the catalogue, based on the mean density of the Universe for the specific volume, by uniformly adding galaxies back. Note that this method is similar, but not the same as in \citet{Chen_et_al_2018, Gray_et_al_2020}, where the contribution of the missing galaxies is calculated statistically. In our case, we add galaxies back in the catalogue and take their contributions into account as described in section \ref{sec:Ho_posteriors}. We note that for the added galaxies we have only geometrical information, and do not assign physical characteristics, i.e., in the analysis we only consider their positions in the sky. 
In future work (Barbieri et al., in prep.), we are expanding this framework, in order to complete the galaxy catalogues exploiting prior clustering information, but also by assigning physical properties, like magnitudes.

To investigate completeness effects, we include a ``completeness fraction'' $f$ in our data generation. This selects a percentage of the original galaxies in our observational sphere that remain in the incomplete catalogue, based on specific physical thresholds. For this reason, we adopt a selection procedure that tries to mimic observational studies by removing galaxies based on magnitude:

A fraction of the least ``bright'' galaxies is removed, which replicates the magnitude limitations in a survey. Following \citet{Gray_et_al_2020}, these depend on the distance of the galaxies from our observer and, in our case, their absolute ($B$-band) magnitude. So the relevant magnitude formula becomes: $m = M_B+5\log_{10}(d_L/{\rm Mpc})+25$.

The selection cuts are performed taking into account the properties of the galaxies of the observable sphere. For a concrete example, refer to section \ref{sec:complete_a_catalogue}. This leads to a completeness fraction estimated via $f=N^{\rm survived}/N^{\rm all}$. This is a ``global'' completeness level, hence note that there are expected to be significant variations in the completeness of individual cones. This is expected in a realistic scenario, since the magnitude cut introduces a distance scale, i.e., closer-by galaxies are most likely to be observed in all cases, hence closer-by GW events will have more complete cones. For our analysis, we choose to retain the global percentage cut, which more closely resembles an observational scenario.

Unless stated otherwise, we investigate four incompleteness fractions $f$, that correspond to $f=$ $75 \%$, $50 \%$, $25 \%$ and $5 \%$ survey completeness. In Table \ref{table:completeness_magnitude_table} we provide the correspondence of these fractions to the magnitude threshold used to remove galaxies accordingly, for all high resolution simulations. We note that this correspondence should be used with care: the magnitude thresholds we find are similar to the ones reported in \citep{Gray_et_al_2020}, but specific galaxy surveys take into account many more technical details, e.g., of the specific instrument, that have been ignored here. However, the completeness fractions, which can also be estimated from the galaxy surveys themselves, would make a more reliable comparison.

\begin{table}
\centering
\begin{tabular}{|c c c c c|}
     \hline
     $f [\%]$ & $5$ & $25$ & $50$ & $75$ \\ 
     \hline
     $m_{\rm B}^{\rm thr}$ [L1600] & $16.74$  & $18.12$ & $18.98$ & $19.60$ \\ 
     \hline
     $m_{\rm B}^{\rm thr}$ [L700] & $16.03$  & $17.40$ & $18.12$ & $18.82$ \\ 
     \hline
     $m_{\rm B}^{\rm thr}$ [L100] & $14.62$  & $16.41$ & $17.63$ & $20.21$ \\ 
\end{tabular}
\caption{Correspondence of completeness levels to apparent B magnitudes for all high-res simulations used in this work. As expected, we observe that lower completeness levels, i.e. more galaxies missing, correspond to a lower apparent magnitude threshold. Galaxies with magnitudes larger than the threshold will not be observed, based on our convention.}
\label{table:completeness_magnitude_table}
\end{table}

\begin{figure}
     \centering
     \subfloat[][Initial Sphere]{\includegraphics[width=0.48\columnwidth]{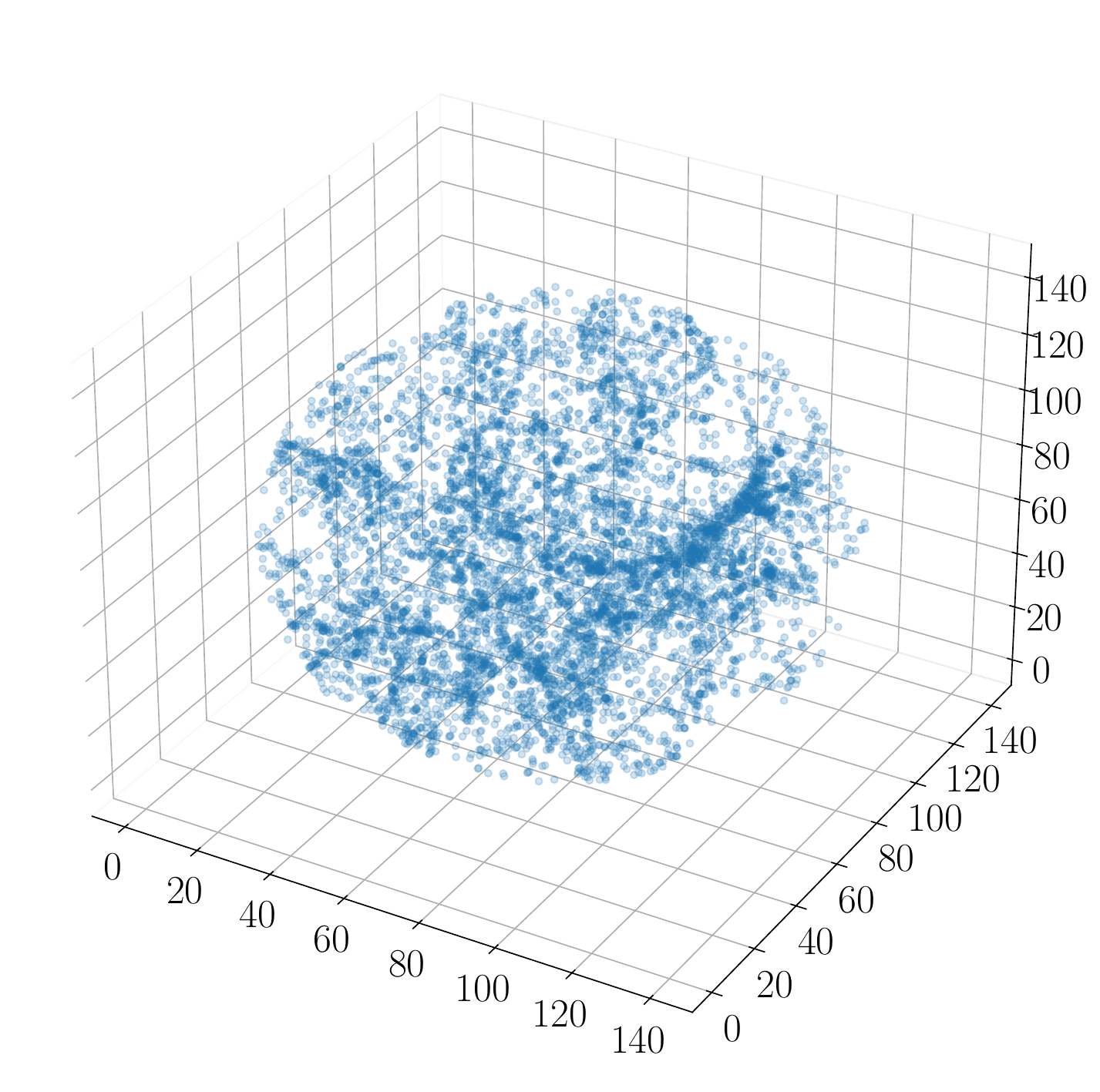}}\\

     \subfloat[][Sphere after a $50 \%$ cut]{\includegraphics[width=0.48\columnwidth]{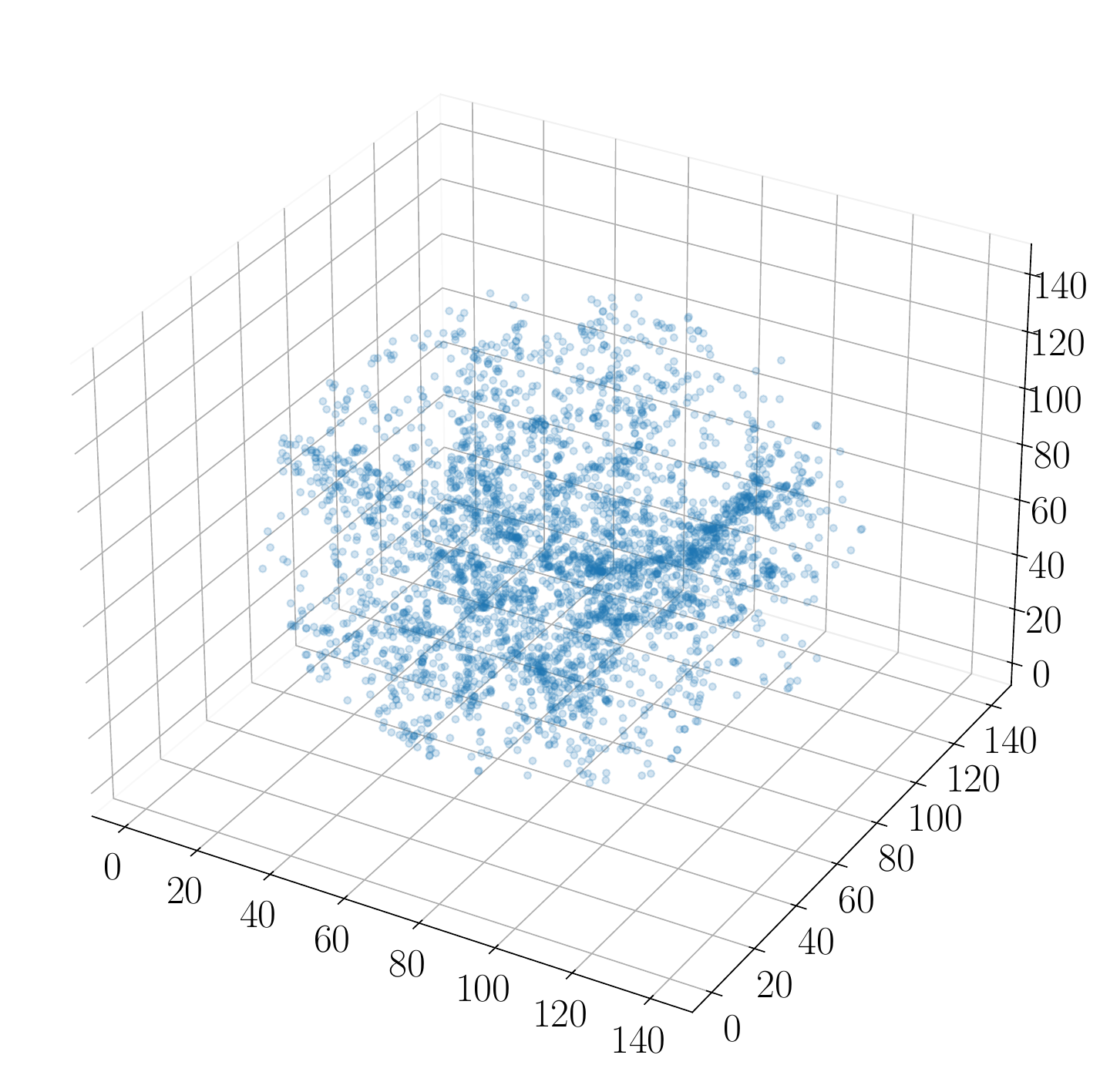}}\\
     
    \subfloat[][Uniformly Completed Sphere]{\includegraphics[width=0.48\columnwidth]{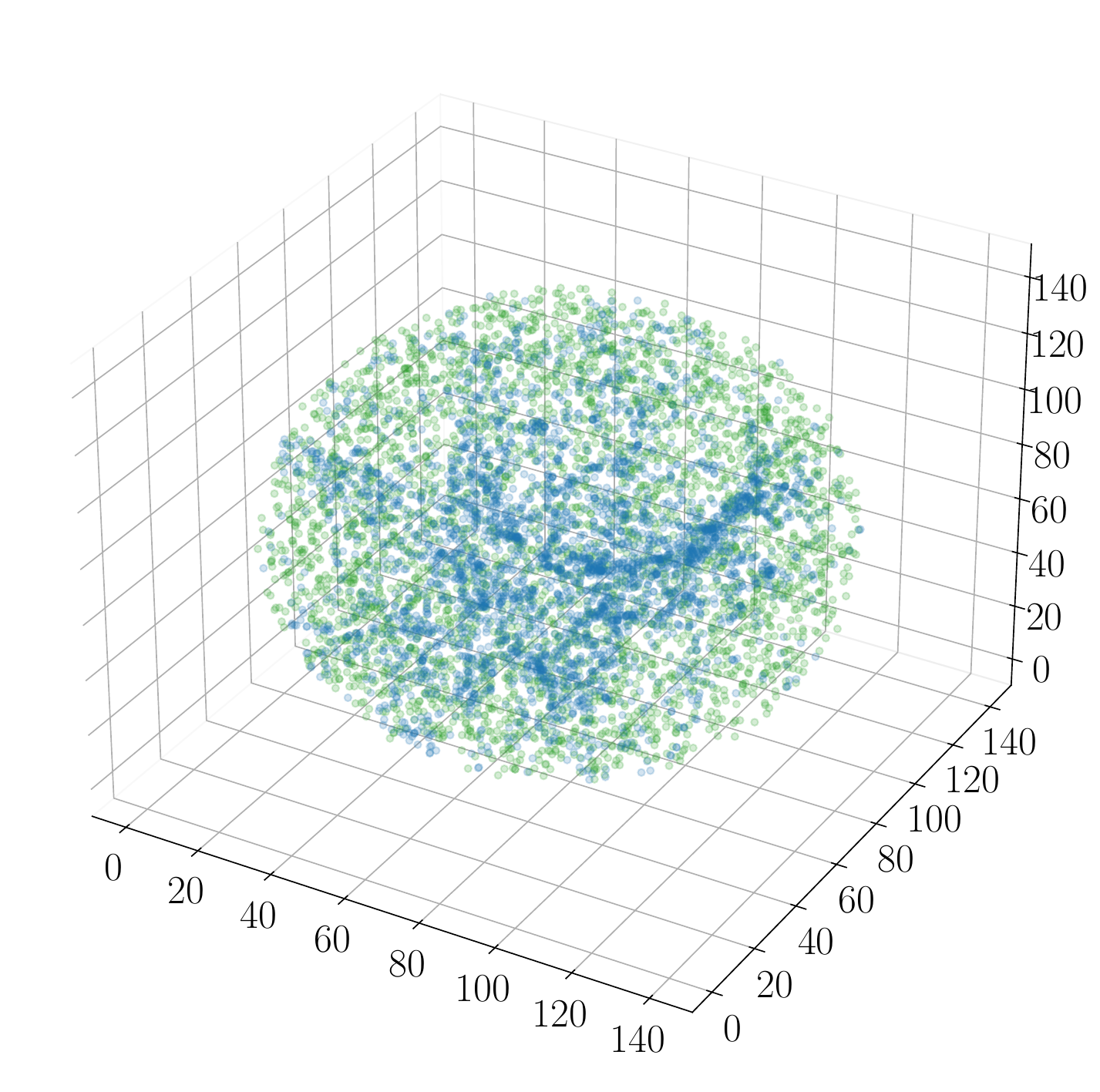}}

     \caption{Example procedure of creating an incomplete catalogue and then completing the missing galaxies, sampling from the uniform sphere. This example describes a $50 \%$ magnitude cut, i.e., throwing away half the galaxies, based on their magnitudes. The $L_{\rm box}=100$ Mpc/h box, with fewer galaxies, has been chosen for visualisation purposes.}
     \label{fig:sphere_completion}
\end{figure}

\subsubsection{How to complete the catalogue?}\label{sec:complete_a_catalogue}

We complete the catalogues, exploiting a `sampling' uniform box, i.e., put galaxies back that are uniformly distributed in comoving volume, as is the standard procedure in the literature \citep{Soares-Santos_et_al_2019}. A method to complete the catalogues retaining clustering information is developed in the complementary work of Barbieri et al., (in prep.).

A better way to understand our method is with a specific example - see also Figure \ref{fig:sphere_completion}. Assume that we want to create a catalogue of $f=50 \%$ completeness and uniformly add galaxies back. We follow the steps below:

\begin{enumerate}
    \item We start with the complete, clustered observable sphere. For the largest box, this corresponds to a sphere with radius $r = L_{\rm box}/2 = 800$ Mpc/h.
    \item We use the `magnitude cut' method, to throw away $50 \%$ of the galaxies, i.e., we remove the $50 \%$ of galaxies with the highest apparent magnitude. 
    \item Now our catalogue is only $50 \%$ complete and we need to fill it again with galaxies to reach the same density of galaxies. The number of galaxies we need is estimated from a comparison to the initial clustered catalogue\footnote{We have checked that the original box used for comparison, either the clustered one or the uniform one, does not introduce any significant change on the results of completion.}: for the same radial bin in the original box, how many galaxies are missing compared to the $50 \%$ cut radial bin? This is $N_{\rm missing}$ per bin.     
    \item From a `uniform box', see section \ref{sec:uniform_box}, we cut again an `observable sphere` of the same size as the initial one, $L_{\rm box}/2$.
    \item From the uniform sphere, we draw randomly $N_{\rm missing}$ galaxies, without replacement (to avoid drawing the same galaxy coordinates more than once), from the distance bins where we have missing galaxies.
    \item We combine the $50 \%$ clustered galaxies and the $50 \%$ uniform galaxies, to have a new ``complete'' catalogue, in our observable sphere.
    \item This sphere is now the basis of our analysis for the $H_0$ posterior of the $f=50 \%$ case. It can be used as input to the analyses described in section \ref{sec:Ho_posteriors}. 
\end{enumerate}

\subsection{Uniform box}\label{sec:uniform_box}
 
In order to quantify the effects of clustering, we aim to compare with previous similar studies \citep{Chen_et_al_2018, Gray_et_al_2020}, but also perform our own baseline analysis for better consistency. For this reason, we create a simulation box with uniformly distributed galaxies to compare the consequences of clustering versus a randomly, uniformly populated catalogue. This dataset includes only information about the galaxy coordinates, without any physical properties.

\begin{figure}
     \centering
     \subfloat[][Clustered Sphere]{\includegraphics[width=0.45\columnwidth]{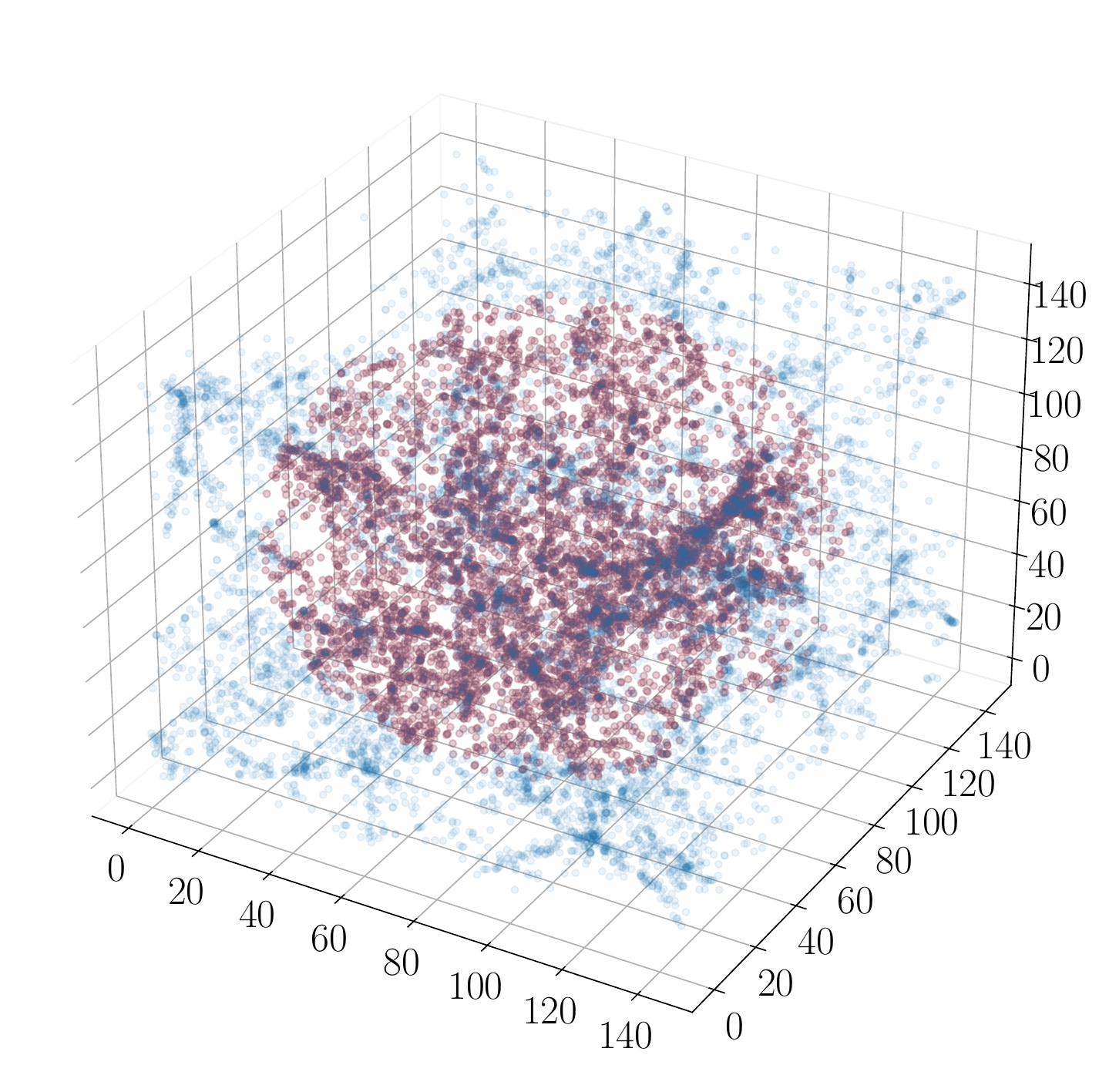}}
     \hspace{0.1cm}
     \subfloat[][Uniform Sphere]{\includegraphics[width=0.45\columnwidth]{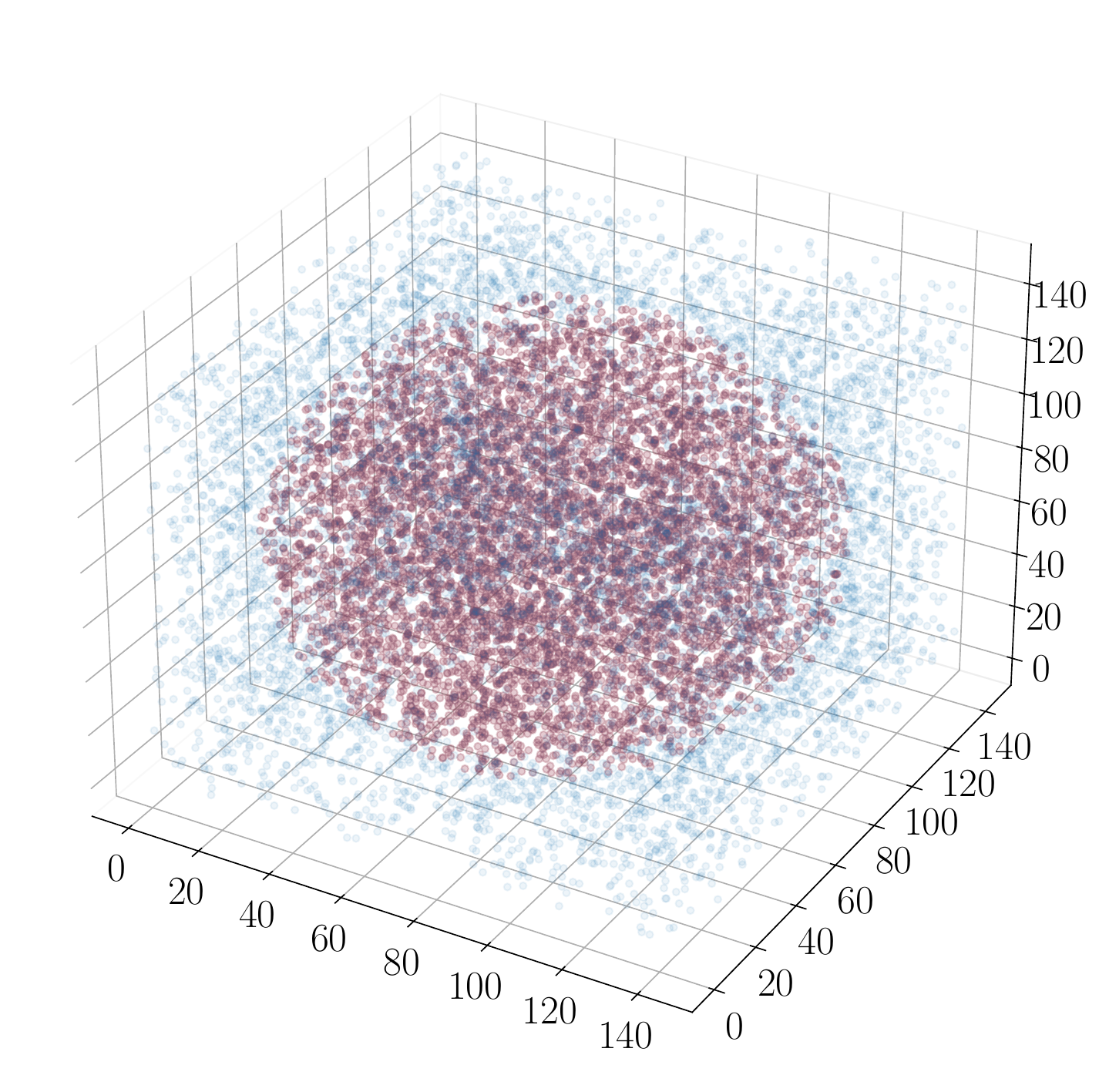}}

     \caption{Construction of ``observable spheres'' in: (a) a clustered catalogue and (b) a uniform catalogue. The $L_{\rm box}=100$ Mpc/h box, with fewer galaxies, box has been chosen for visualisation purposes.}
     \label{fig:sphere_cuts}
\end{figure}

The comparison box has the same number of galaxies as the clustered one. Since we only use an observable spherical region for the clustered case, we need to perform the same cut in the uniform box (Figure \ref{fig:sphere_cuts}). Results will be presented for these two spheres: one with galaxy clustering and one with uniformly distributed galaxies. The procedure to calculate the $H_0$ posterior is the same as described in section \ref{sec:Ho_posteriors}.

\subsection{Multiple realisations}

Finally, to study the robustness of our results and estimate cosmic variance, we run multiple realisations of our analysis. This means that we repeat the analysis for the final $H_0$ posterior for a given number of cones (events) $N_c$, multiple times $N_r$. Each realisation draws different cones and repeats the catalogue completion, i.e., draws new random galaxies from the uniform, sampling catalogue.

To be more explicit, when we investigate the result of $N_c$ cones, a single run will return an $H_0$ posterior, which could be described by a mean $\langle H_0 \rangle$ and error $\sigma_{H_0}$, for these $N_c$ events\footnote{The final posterior, in most cases, is nicely approximated by a Gaussian since $N_c \gg 1$, so the mean and error give a good description of the final distribution. We checked that the results of combining the full posteriors are equivalent. In our analyses, we use the full posteriors.}.

If we repeat the run $N_r$ times, we have:
\begin{itemize}
    \item $\langle H_0 \rangle_1^c$, $\langle H_0 \rangle_2^c$, \ldots, $\langle H_0 \rangle_r^c$, $N_r\ \langle H_0\rangle$ values.
    \item $\sigma_1^c$, $\sigma_2^c$, \ldots, $\sigma_r^c$, $N_r\ \sigma_{H_0}$ values.
\end{itemize}
We then combine the results of the separate realisations as:
\begin{align}
    \bar{H}_0 &= \frac{1}{N_r}\sum^{N_r} \langle H_0 \rangle_i^c \\
    \bar{\sigma}_H &= \frac{1}{N_r}\sum^{N_r} \sigma_i^c \\
    \delta_{\bar{\sigma}_H} &= \sqrt{\frac{1}{N_r}\sum^{N_r} (\sigma_i^c - \bar{\sigma}_H)^2}.
\end{align}
These give respectively an average value of $\langle H_0 \rangle$, an average value of the posterior width over realisations, and the standard deviation of the posterior widths over realisations. We also cross-check our method by adding together the full posteriors of each realisation, which are used for the final results. We use $N_r = 15$ realisations, as the default. Recall that in all these cases, the initial, complete, clustered galaxy catalogue is the same, with each realisation having different cones' positions. For the incomplete catalogues, apart from the new positions of the cones, for each realisation we construct a new $f\ \%$ catalogue, by drawing new galaxies in place of the ones that are missing.

\begin{figure}
     \centering
     \subfloat[][]{\includegraphics[width=\columnwidth]{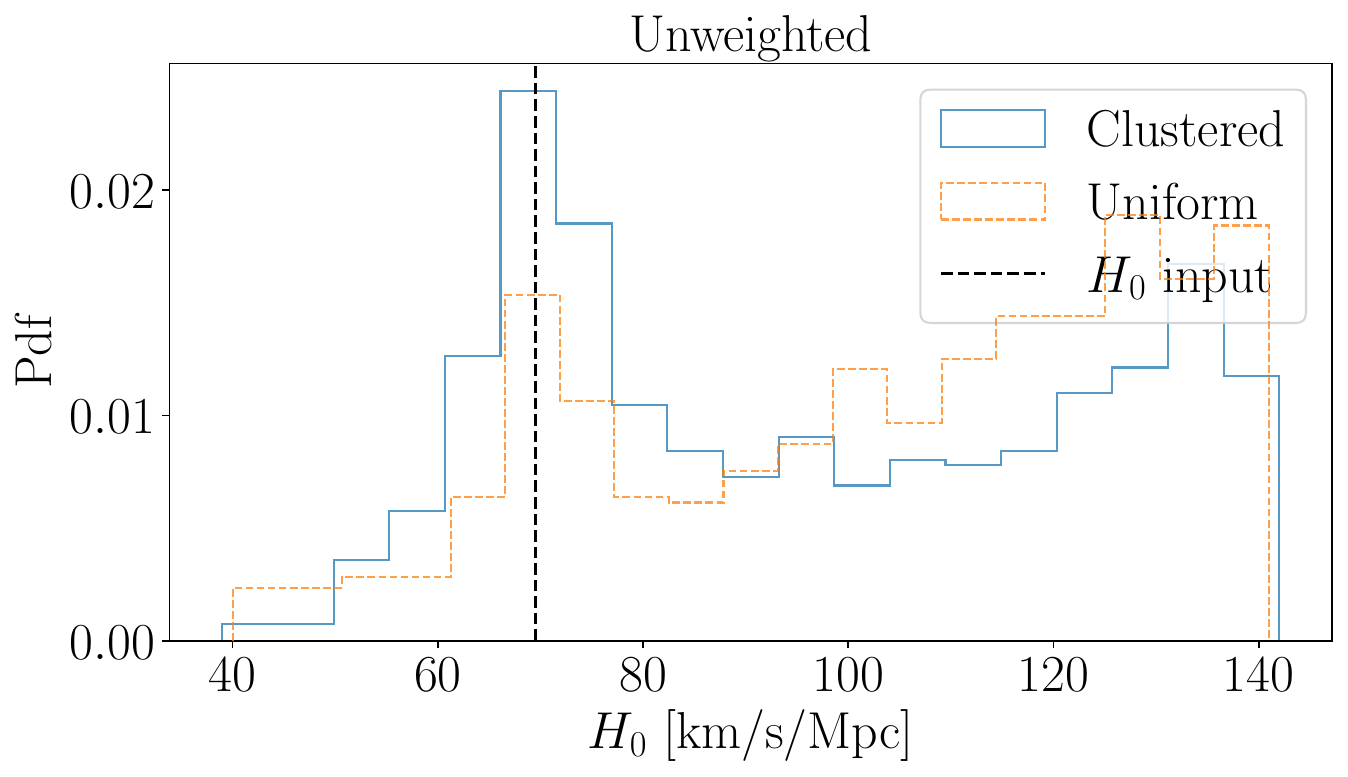}}\\

     \subfloat[][]{\includegraphics[width=\columnwidth]{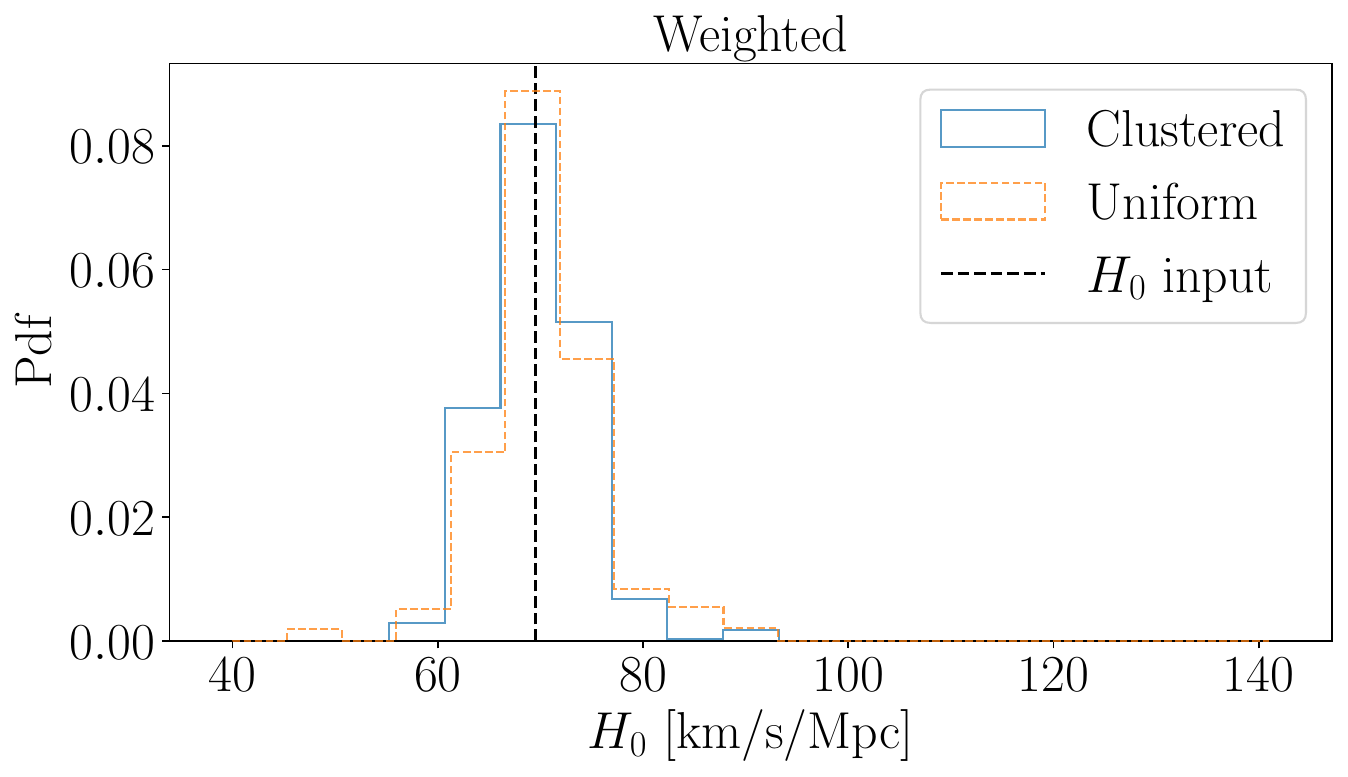}}

     \caption{Stacking histograms for the Hubble parameter for a clustered (blue, solid) and a uniform (orange, dashed) catalogue. The top figure (a) does not involve any weighting and it is biased towards large values of $H_0$, while the bottom figure (b) penalises galaxies based on their position with respect to the cone centre. This correction is needed when realistic detection geometries are used. We have used $N_c=100$ events from the $L_{\rm box}=100$ Mpc/h box.}
     \label{fig:histograms_clustered_uniform}
\end{figure}

\begin{figure}
    \centering
	\includegraphics[width=\columnwidth]{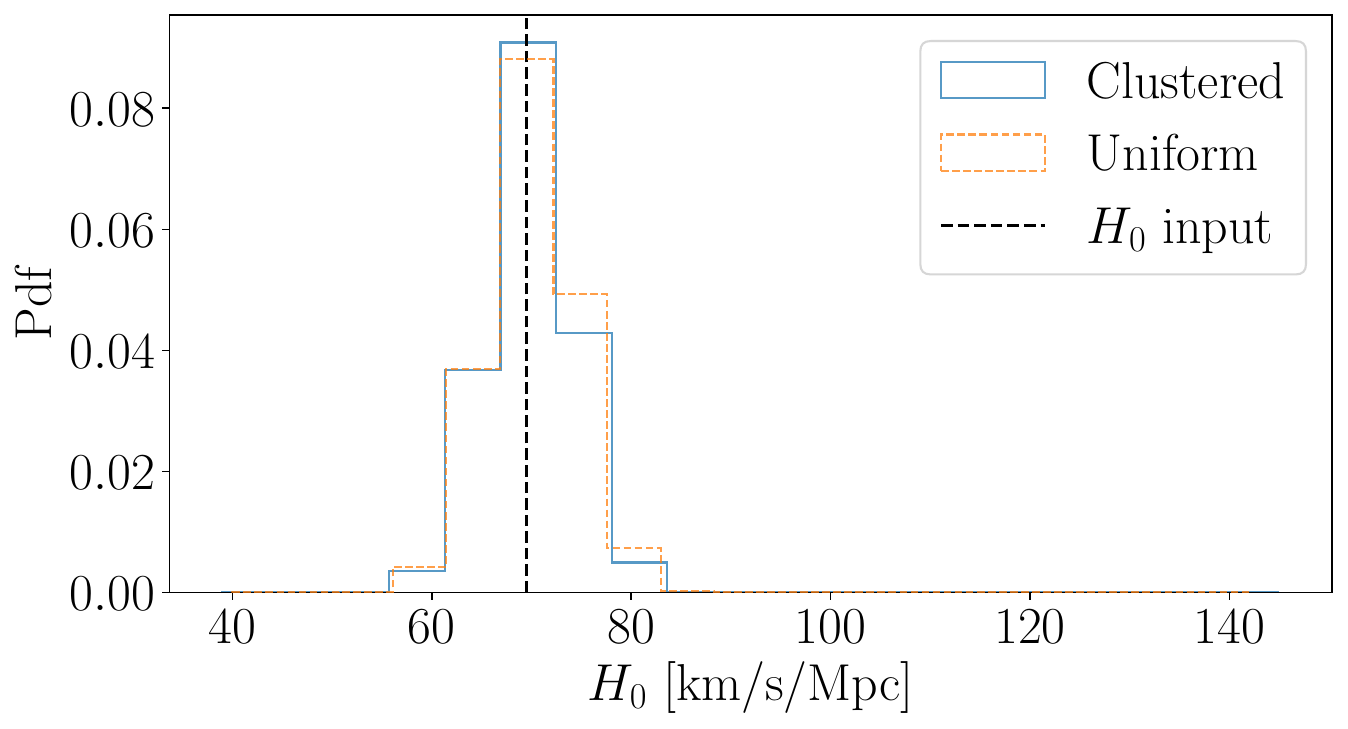}
    \caption{Stacking histograms for the Hubble parameter for a clustered (blue, solid) and a uniform (orange, dashed) catalogue. In both cases, we recover the input value. Both catalogues give equivalent results. We have used $N_c=100$ events from the $L_{\rm box}=1600$ Mpc/h box.}
    \label{fig:histogram_stacking}
\end{figure}

\begin{figure}
    \centering
	\includegraphics[width=0.95\columnwidth]{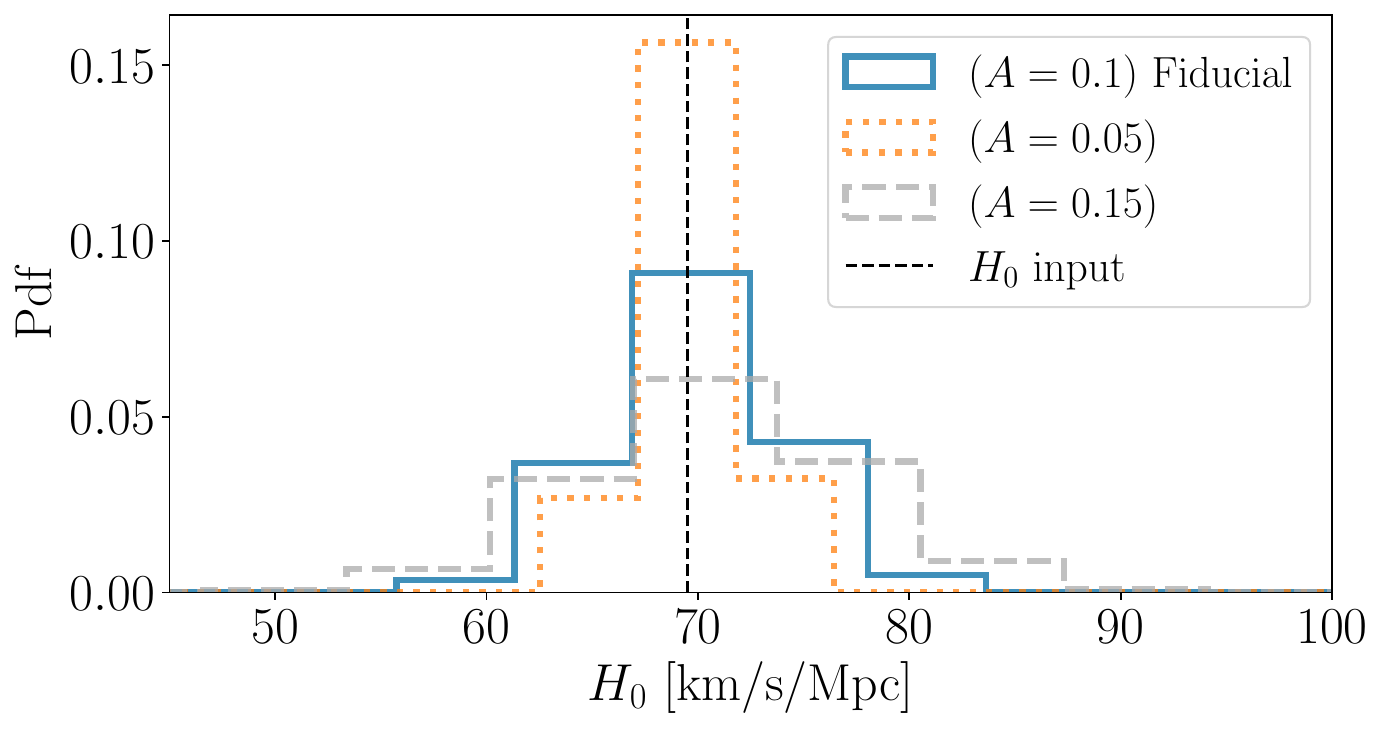}
    \caption{Stacking histograms for the Hubble parameter for different choices of fractional distance error $A$. The histograms follow the expected behaviour: larger errors lead to wider distributions. Note that we have slightly restricted the $H_0$ range only for visualisation purposes. We have used $N_c=100$ events from the $L_{\rm box}=1600$ Mpc/h box.}
    \label{fig:Histogram_comparison}
\end{figure}

\section{Results}\label{sec:results}

We now investigate the $H_0$ inference, considering the main analysis methods of section \ref{sec:Ho_posteriors}. We also discuss the effects of incomplete catalogues and the consequences of the different geometrical or physical weights.

\subsection{Histograms stacking}\label{sec:stacking_results}

The simplest way to estimate the Hubble parameter is to stack the $H_0$ values from all the galaxies, from multiple cones. The results can be seen in Figures \ref{fig:histograms_clustered_uniform} and \ref{fig:histogram_stacking}. For all the plots we use $N = 100$ events, while we compare resolution effects by using the $L_{\rm box}=100$ Mpc/h, small particle number box (Figure \ref{fig:histograms_clustered_uniform} (b)), and $L_{\rm box}=1600$ Mpc/h simulation (Figure \ref{fig:histogram_stacking}).

In Figure \ref{fig:histograms_clustered_uniform} we investigate the importance of penalising the galaxies with geometric weights (recall section \ref{sec:Geometric_physical_weights}). The inclusion of weights is important to correct for the geometric effects of the cone. At the same time, it improves the convergence of the final histogram. In all Figures we plot the histogram of the stacked $H_0$, where the solid line corresponds to the clustered case, while the dashed line to the uniform one.

The power of clustering is not clear in this basic approach, although in the unweighted case, the presence of clustering enhances the peak around the input value. The clustered distribution is similar to the uniform one, for both boxes. To quantify the difference, the calculation of the interquartile gap ($\sigma_q = q_{75} - q_{25}$), gives $\sigma_q^{\rm clu} \approx 11$ vs $\sigma_q^{\rm uni} \approx 16$ for the small box, when comparing the clustered and uniform cases respectively. Similarly, $\sigma_q^{\rm clu} \approx 11$ vs $\sigma_q^{\rm uni} \approx 10$ for the large one (Figure \ref{fig:histogram_stacking}). The small difference between the small and large cases, and clustered and uniform catalogues can be attributed to the analysis method: the events are not considered as independent, and the process of stacking all the $H_0$ values together leads to ``smoothening'' the final histogram, around the input $H_0$ value, which we recover successfully in all cases. A more important parameter than the box size is the observational error of the luminosity distance, given by $A$. In Figure \ref{fig:Histogram_comparison} we compare three different cases of varying $A$, where larger values correspond to larger errors. The smaller the observational errors ($A = 0.05$, a $50 \%$ reduction from the fiducial value of $A=0.1$), the more tight the final histogram after stacking ($\sigma^{\rm clu}_{q, A=0.05} \approx 9$). Although this verifies the validity of the ``stacking method'', its inherent limitations do not allow its use for more robust statistical analyses.

We should emphasize again, that in our analysis we ignore the contributions of the true GWs sources, since the cone centres are perturbed from the true source and are not required to reside on a galaxy.

This makes our results stronger, demonstrating that the knowledge of the true source is not necessary for accurate inference (see also \citep{Bera_et_al_2020_incompleteness}).

\subsection{Bayesian analysis results}\label{sec:bayes_results}

In this section, we investigate how the $H_0$ Pdfs are affected by the number of GWs observations (subsection \ref{sec:multiple_events_results}), the presence of physical weights (subsection \ref{sec:weights_results}), catalogue completeness (subsection \ref{sec:completeness_results}), observational errors (subsection \ref{sec:observational_errors_posteriors}), the inclusion of angular weights (subsection \ref{sec:angle_LHD}), box size (subsection \ref{sec:Box_size_posteriors}), the method of completion (subsection \ref{sec:completion_methods}), and the $H_0$ prior selection (subsection \ref{sec:H0_priors_checks}). Finally, we examine converge issues (subsection \ref{sec:convergence_studies}).

As a reminder, the default settings correspond to:

\begin{itemize}
    \item \emph{Box size}: $L_{\rm box}=1600$ Mpc/h simulation.

    \item \emph{GWs events}: Number of events $N_c=100$ and number of realisations $N_r=15$.

    \item \emph{Observational errors}: $(\Delta \Omega, 100 \cdot A) = (60, 10)$.

    \item \emph{Weights}: Physical weights - None, Geometrical weights - distance and angle weights.

    \item \emph{Pdfs Normalisation}: The normalisation is arbitrary, but consistent among the different analyses.

    \item \emph{Error Calculation}: We report the errors of the Pdfs either calculating a standard deviation (assuming convergence to Gaussians), or/and calculating a highest density interval (HDI) for the $68 \%$ most probable region.
\end{itemize}

\subsubsection{Convergence with number of events}\label{sec:multiple_events_results}

\begin{figure*}
	%\addtocounter{figure}{-1}
	\centering
	\begin{subfigure}[b]{0.48\textwidth}
		\centering
		\includegraphics[width=\textwidth]{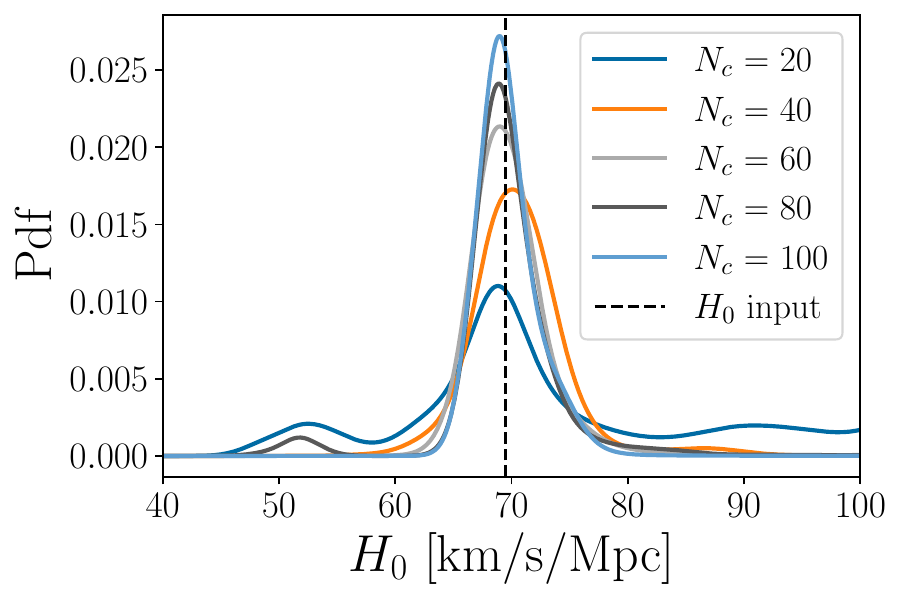}
	\end{subfigure}
	\hspace{0.5cm}
	\begin{subfigure}[b]{0.48\textwidth}
		\centering
		\includegraphics[width=\textwidth]{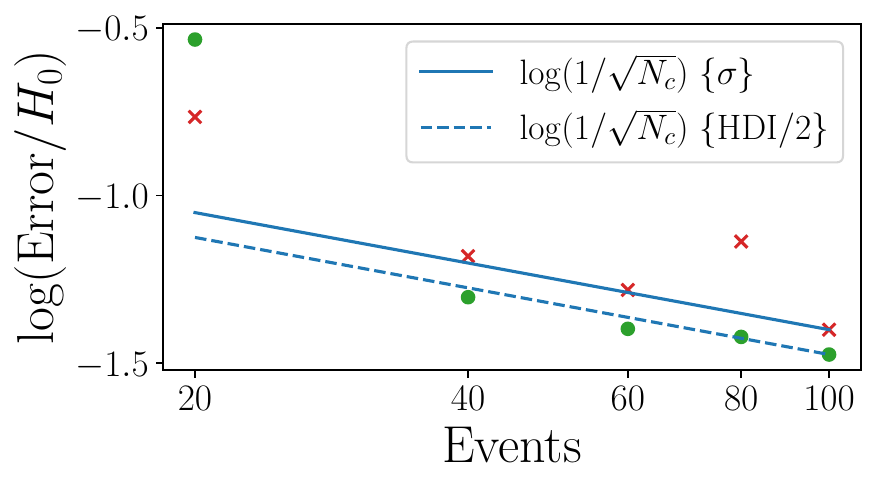}
	\end{subfigure}
	\caption{(Left): $H_0$ posteriors with increasing number of events (from $N_c=20$ to $N_c=100$ GWs events) and $N_r=15$ realisations for each. As expected, more observations lead to tighter constraints. (Right) The mean standard deviation (red crosses) of the final posterior for all realisations for the specific number of events, compared to the input $H_0=69.5$ km/s/Mpc. We compare with the expected scaling of $1/\sqrt{N_c}$ (solid line) in log space, normalised to the most accurate posterior (of $N_c=100$ events). Note that the presence of tails or bumps in the posteriors of $N_c=20, 80$ lead, unsurprisingly, to deviations from the expected behaviour (since the standard deviation assumes a well-shaped Gaussian), which is nevertheless observed for all other cases. We have checked by comparing with other realisations that these are just a statistical fluctuation in the final posterior, while the tightening of the constraints with number of events is observed repeatedly. To further test the robustness of our results we have also calculated the equivalent $68 \%$ HDIs (green dots) and compared with a similar $1/\sqrt{N_c}$ scaling. The errors based on the credible intervals are less sensitive to bumps in the distribution. Results from the $L_{\rm box}=1600$ Mpc/h box.}
	\label{fig:Ho_clustered_Nevents_Nreal}
\end{figure*}

In Figure \ref{fig:Ho_clustered_Nevents_Nreal}, we explore the evolution of $H_0$ posteriors with increasing number of events. These runs use the large, $L_{\rm box}=1600$ Mpc/h box, where different numbers of events are chosen from the complete catalogue and averaged over $N_r=15$ realisations. We confirm the anticipated behaviour: the posteriors become tighter with increasing number of events (from a HDI $(51.4, 92.0)$ km/s/Mpc for $N_c=20$ events, to HDI $(66.8, 71.4)$ km/s/Mpc for $N_c=100$ events). Moreover, the scaling of the pdfs follow the expected $1/\sqrt{N}$ scaling of the standard error of the mean. Finally, we note that when increasing the number of events, we are not adding cones to the previous ones, but doing a completely new $H_0$ inference run, with independent GWs events.

\begin{figure*}
    \centering
	\includegraphics[width=\textwidth]{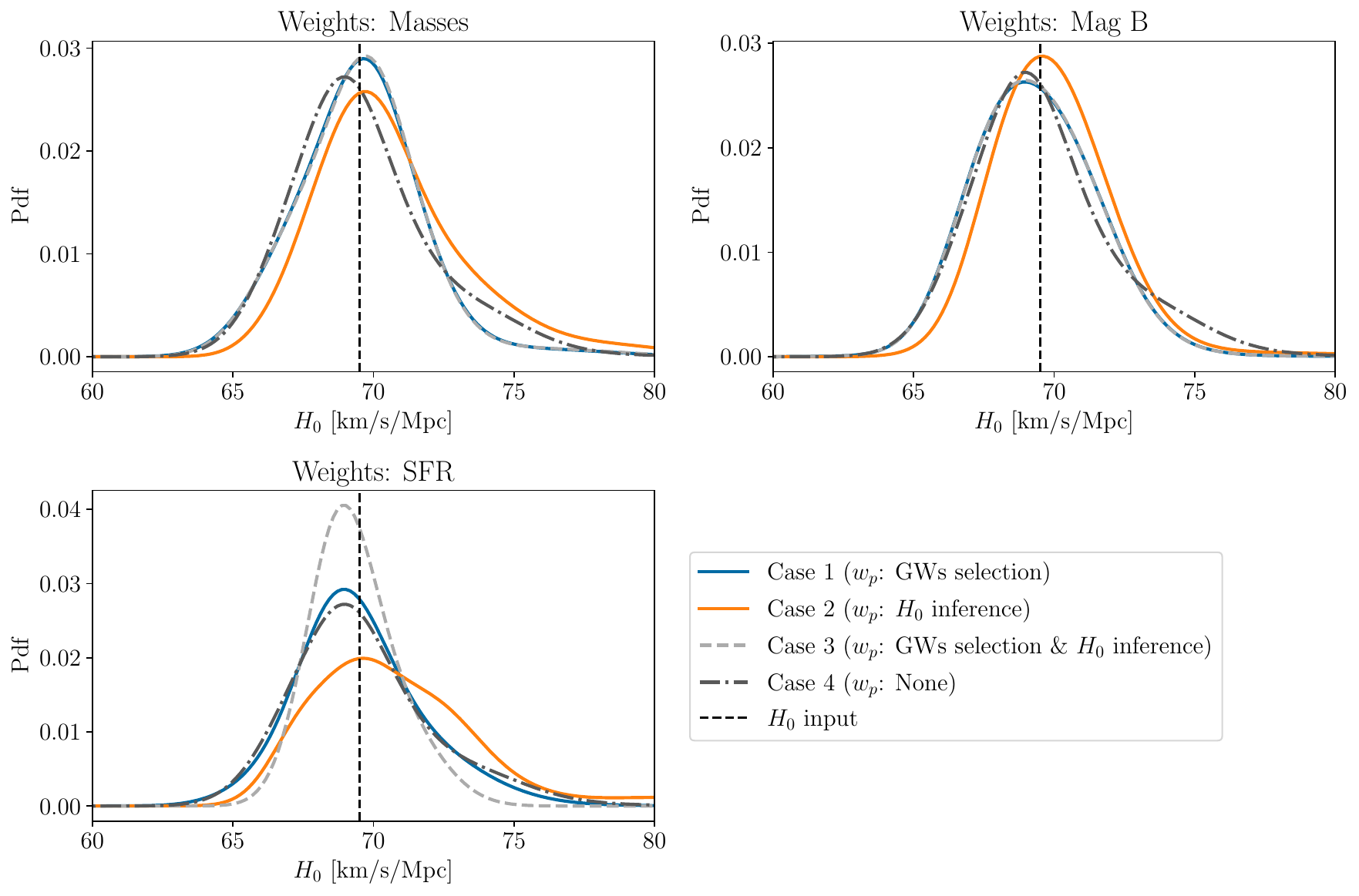}
    \caption{Result of weighted inference for $N_r=15$ realisations and $N_c=100$ events from the $L_{\rm box}=1600$ Mpc/h box. We test the effects of three physical weights (modeled as described in section \ref{sec:Geometric_physical_weights}). For each weight, we compare four cases: 1) use of weights only at the GW host selection phase; 2) use of weights only at the $H_0$ posterior calculation phase; 3) use of weights in both phases; and 4) no use of weights at all, i.e., no extra information on top of clustering. We observe that the presence of weights in the mass and B magnitude cases do not offer any advantage, showing that clustering yields the most important information. Only in the case of SFR, the presence of weights does have a noticeable effect, signifying their potential importance when used in ``statistical dark sirens'' studies. Finally, it seems the presence of clustering contributes to the absence of any bias, when weights are used in a non-consistent way, as in cases 1 and 2. For more details see main text.}
    \label{fig:weights_comparison_multiple_realisations}
\end{figure*}

\begin{figure*}
    \centering
	\includegraphics[width=\textwidth]{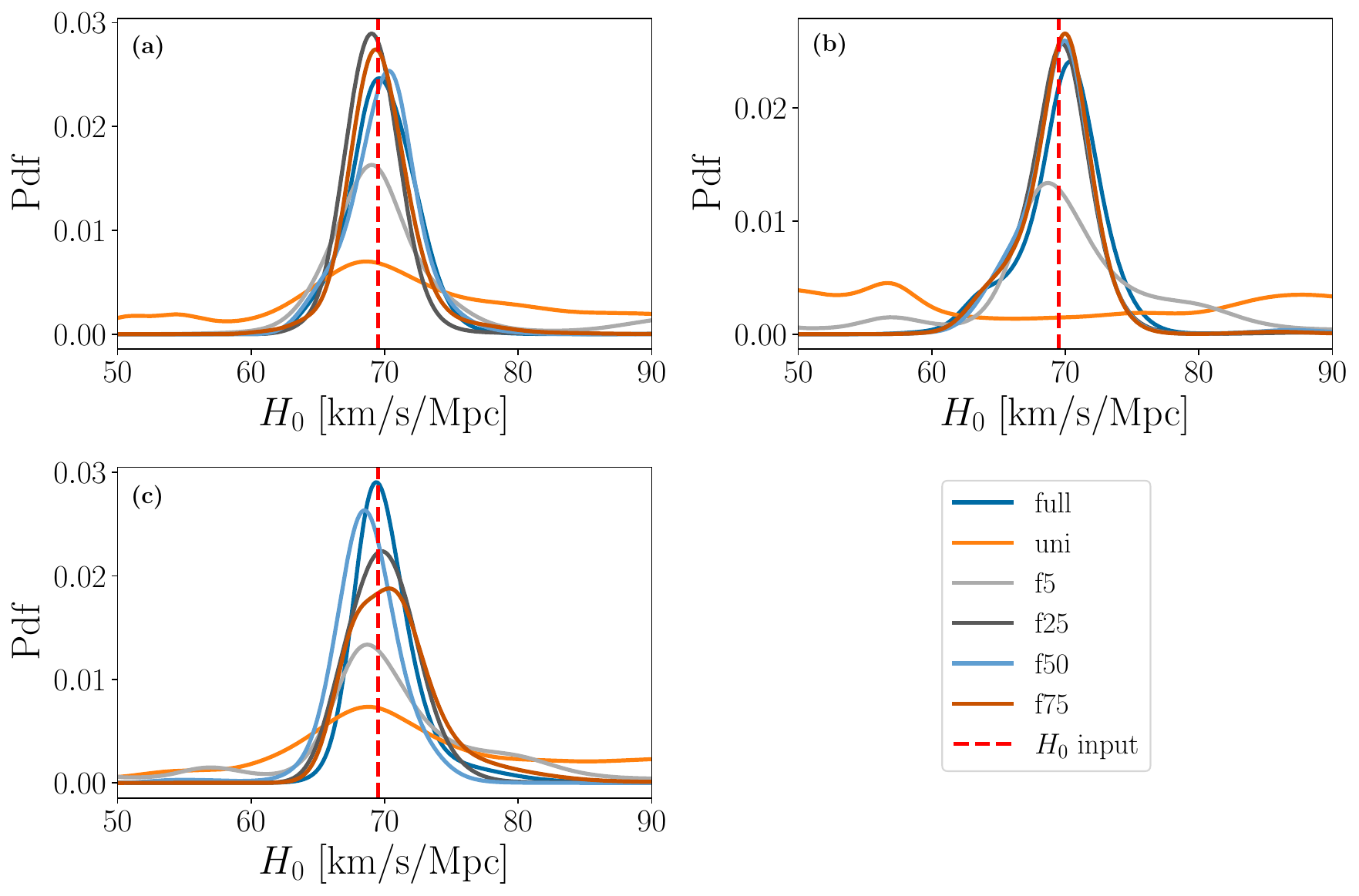}
    \caption{We compare the effects of catalogue incompleteness in three different scenarios: (a) corresponds to the case that new GW events and new catalogues are drawn for each realisation; (b) corresponds to the case that new GW events are drawn for each realisation, but the catalogue is only drawn once and then kept fixed across realisations; and (c) corresponds to the case where the set of GW events are taken from the complete (full) catalogue and used in all the incomplete ones, which are kept fixed. In (a)-(c), we compare the posteriors computed from $N_c=100$ events averaged over $N_r=15$ realisations, between the complete/clustered (full) catalogue, the uniform one and the incomplete catalogues of increasing completeness from $f=5\%$ till $f=75\%$. We observe that the uniform catalogue and the almost uniform catalogue ($f=5\%$) are consistently outperformed by the catalogues with increased clustering presence. We generally observe the expected behaviour, with improved inference for increased completeness. Events are taken from the $L_{\rm box}=1600$ Mpc/h box. For more details see main text.}
    \label{fig:multiple_realisations_catalogues_events_comparison}
\end{figure*}

\subsubsection{Effects of physical weights}\label{sec:weights_results}

Figure \ref{fig:weights_comparison_multiple_realisations} investigates the importance of physical weights, specifically of masses, magnitudes (Mag B) and star formation rates (SFR). The latter are calculated for each halo using the ML techniques developed in \citep{Rob_ML_2021} and are assigned in the analysis as linear weights (section \ref{sec:Geometric_physical_weights}). We emphasise here that since the mapping procedure between the DM haloes and galaxies is done in a consistent way among the different analyses and box sizes, its effect on the final results is negligible. On the other hand, the choice of linear weights, instead of a more complicated weighting function, is expected to influence the results in a more significant fashion, that we leave to investigate in future work. We analyse four cases:

\begin{enumerate}
    \item \emph{Case 1}: We use physical weights in the GW selection procedure, e.g., for mass weights, more massive galaxies would be chosen preferentially as the sources of GW events. However, we ignore weights when estimating $H_0$.

    \item \emph{Case 2}: Opposite of \emph{Case 1}, we use physical weights when calculating $H_0$, but do not use them when actually choosing the host of the GW source.

    \item \emph{Case 3}: Combines \emph{Case 1} and \emph{Case 2} above, and leads to a consistent procedure where the same weights are used for GW source selection and when estimating $H_0$.

    \item \emph{Case 4}: No physical weights. Only geometric information of the galaxies (distances, sky localisation) is used in both GW host selection and $H_0$ inference.
\end{enumerate}

We note that in all cases geometric weights are used, i.e. are part of the GW likelihood (section \ref{sec:bayesian_analysis_method}). Moreover, although the first two cases are not self-consistent, they are used to investigate possible biases in the method. Finally, observe that although ``Case 4'' does not impose any explicit weights, the number of low-mass galaxies is bigger than high-mass galaxies due to their formation (recall Fig. \ref{fig:halo_mass_functions}). As a result, they are preferentially selected in a random draw. However, this is also the case in a realistic scenario with equal prior weights to all galaxies, hence this is not introducing any inconsistency in the ``no physical weights'' analysis.

In general, we see that the presence of weights does not significantly change our $H_0$ posteriors, with the exception of weighting by SFR which seems to lead to tighter constraints in the self-consistent analysis of \emph{Case 3}. Furthermore, we observe that the non-symmetrical presence of weights does not appear to lead to biased results in our case (but see also \citep{Hanselman_et_al_2025, Perna_Mastrogiovanni_Ricciardone_2025}), and this could be attributed to the presence of clustering having the most significant contribution to the analysis. The latter conclusion is reinforced by the work of \citep{VanWyngarden_et_al_2025_cat_completeness}, who studied the effects of physical weights in clustered catalogues and concluded that in some cases even $1\%$ complete catalogues could lead to unbiased $H_0$ estimates. This result qualitatively agrees with our conclusions, with some quantitative variations due to different choices of opening angles and luminosity distance uncertainties.

\begin{figure}
    \centering
	\includegraphics[width=0.95\columnwidth]{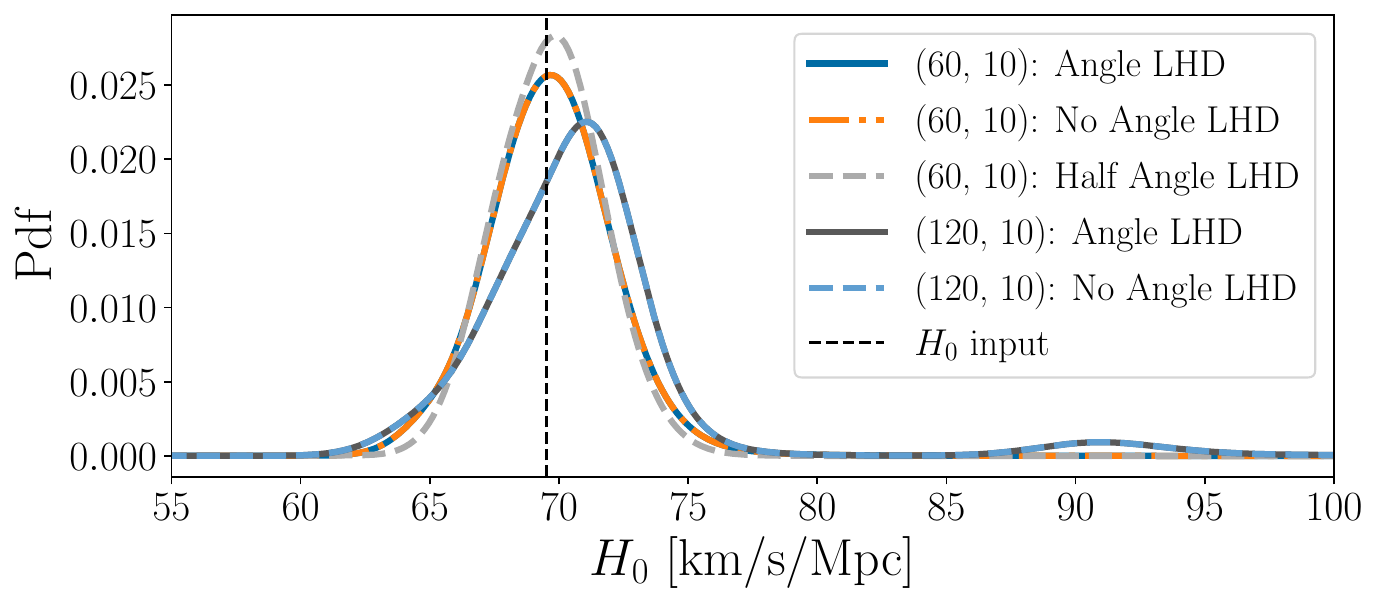}
    \caption{$H_0$ posteriors for complete catalogues, with and without angle weights (section \ref{sec:Geometric_physical_weights}) in the likelihood. We compare two cases of observational errors ($\Delta \Omega, 100 \cdot A$). We find negligible differences in the final pdfs, regardless of the inclusion of weights. This could be a combination of the smallness of the cones' opening angle (relatively small $\Delta \Omega$) and of the importance of clustering. We test this assumption with an ``artificial boost'' of the weight resolution inside the cone - see main text for details, section \ref{sec:angle_LHD} - which indeed leads to a visible improvement (in the grey, dashed line). Results are shown for inference of $N_c=100$ events, averaged over $N_r=15$ realisations, from the $L_{\rm box}=1600$ Mpc/h box.}
    \label{fig:angles_LHD_comparison}
\end{figure}

\begin{figure}
    \centering
	\includegraphics[width=0.98\columnwidth]{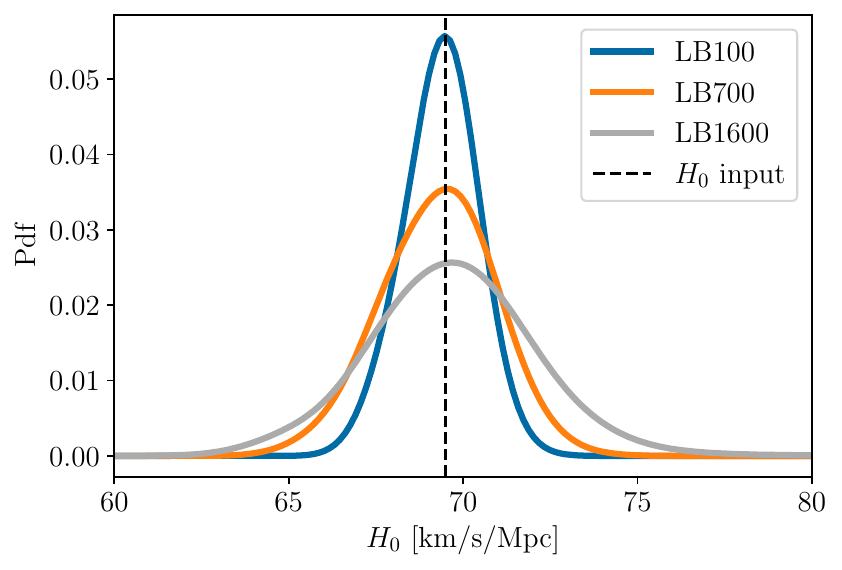}
    \caption{$H_0$ posteriors for complete catalogues for $N_c=100$ events, averaged over $N_r=15$ realisations. We compare runs from the three large particle number boxes $L_{\rm box}=100, 700\ \&\ 1600$ Mpc/h. GW events from much closer provide more than twice the precision on $H_0$ ($\sigma_{\rm LB1600}/\sigma_{\rm LB100} \sim 2.3$).}
    \label{fig:Boxes_clustered_summary}
\end{figure}

\begin{figure}
    \centering
	\includegraphics[width=0.98\columnwidth]{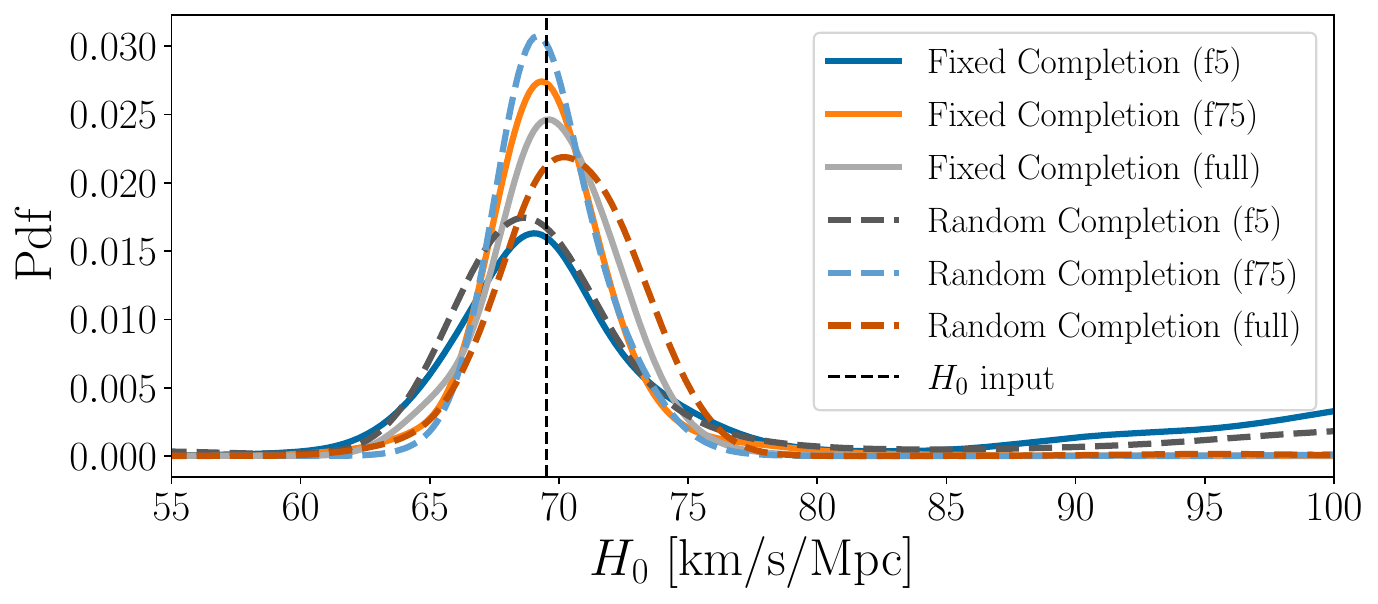}
    \caption{$H_0$ posteriors for two different completion methods - ``fixed'' (fiducial, solid lines) and ``random'' (dashed lines) - for a $f=5\%, 75\%$ and $100\%$ (full) catalogues, for $N_c=100$ events, averaged over $N_r=15$ realisations, from the $L_{\rm box}=1600$ Mpc/h box. For the differences on completion methods, we refer to section \ref{sec:completion_methods}. We observe that the random completion does not lead to any significant bias, but could lead to inflated posteriors of incomplete catalogues, because this sampling method results in a galaxy distribution with increased clustering.}
    \label{fig:completion_comparison}
\end{figure}

\begin{figure*}
	%\addtocounter{figure}{-1}
	\centering
	\begin{subfigure}[b]{0.8\textwidth}
		\centering
		\includegraphics[width=\textwidth]{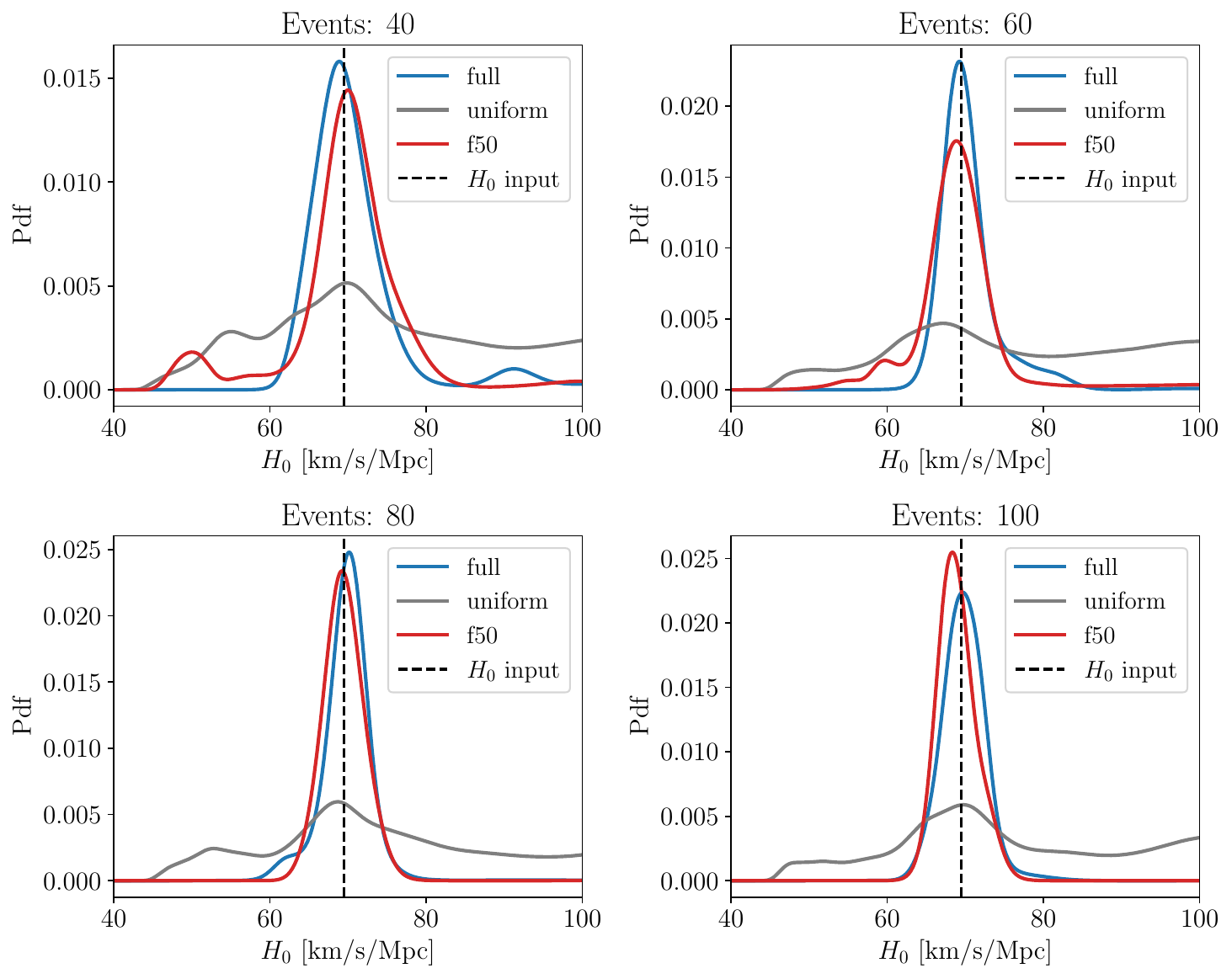}
	\end{subfigure}
	\hspace{0.5cm}
	\begin{subfigure}[b]{0.8\textwidth}
		\centering
		\includegraphics[width=\textwidth]{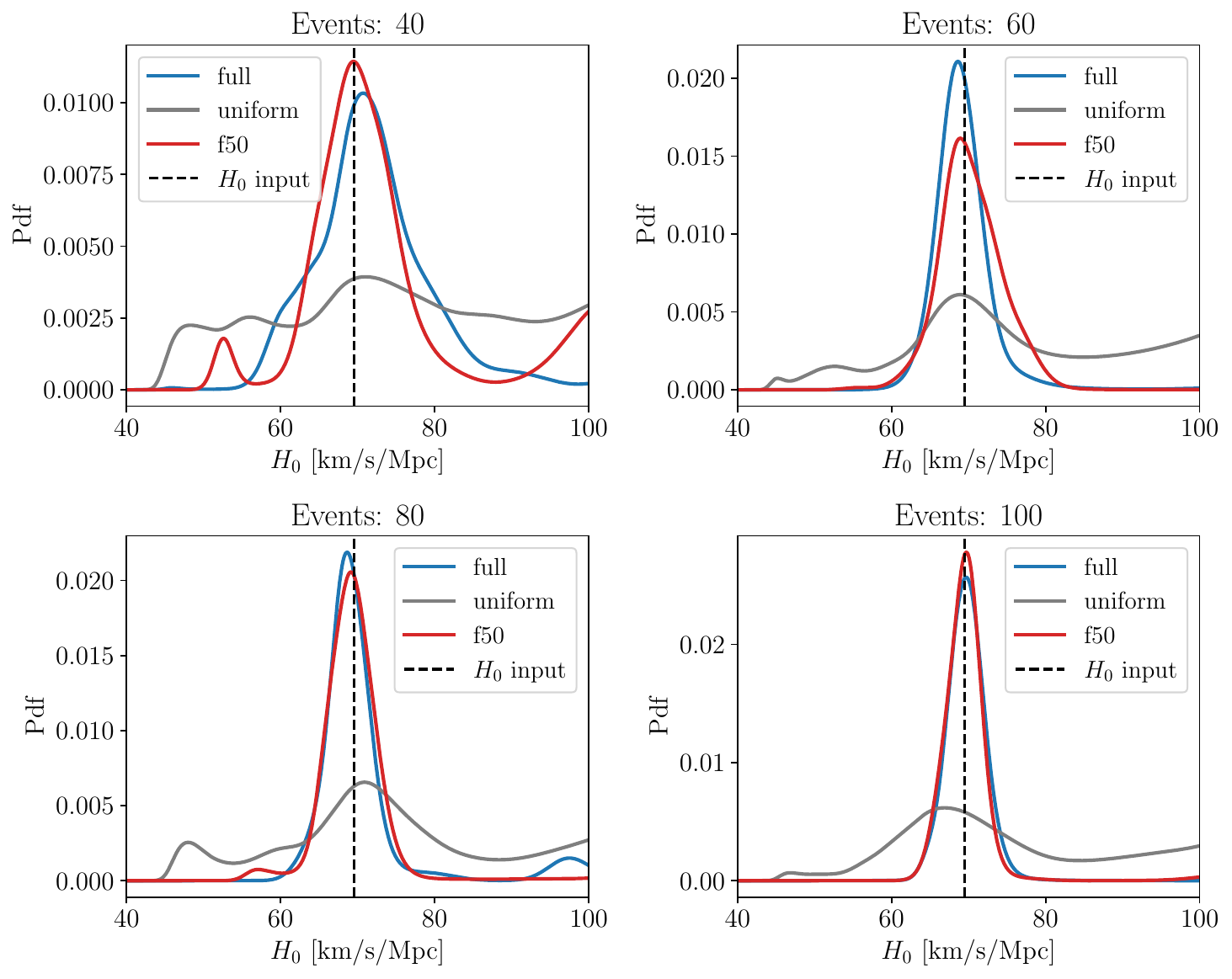}
	\end{subfigure}
	\caption{$H_0$ posteriors comparison, for different number of events, between three catalogues: a complete (full) catalogue, a $f=50 \%$ incomplete catalogue and a uniform catalogue. Top and bottom panels correspond to different realisation scenarios: top $4$ figures to fixed catalogues, bottom $4$ figures to new catalogues. Results of both panels are similar and show that even a half complete catalogue could lead to similar $H_0$ Pdfs, due to the presence of clustering. Pdfs averaged over $N_r=15$ realisations, from the $L_{\rm box}=1600$ Mpc/h box. See main text for more details.}
	\label{fig:clu_uni_f50_pdf_comparison}
\end{figure*}

\subsubsection{Effects of completeness}\label{sec:completeness_results}

We study completeness effects in Figure \ref{fig:multiple_realisations_catalogues_events_comparison}. The Pdfs correspond to $N_c=100$ GW events, averaged over $N_r=15$ realisations from the $L_{\rm box}=1600$ Mpc/h box. The different Figures \ref{fig:multiple_realisations_catalogues_events_comparison} (a)-(c) are produced using different completeness scenarios in order to study potential systematics, and we explain the three different cases below:

\begin{enumerate}
    \item \emph{New GWs, New Catalogues} (Fiducial): In this scenario, for each realisation we create new incomplete catalogues, i.e., we produce a new uniform box, and randomly draw galaxies from there (New Catalogues). Moreover, when running the inference on a specific catalogue, we draw new GW events from the galaxies of each catalogue (New GWs). This is the fiducial technique used when running multiple realisations. Case (a) in Figure \ref{fig:multiple_realisations_catalogues_events_comparison}.
    \item \emph{New GWs, Fixed Catalogues}: Similar to the above, but now the incomplete catalogues are fixed, i.e., we draw galaxies from the uniform box only once to complete the catalogues and then these remain the same for all realisations (Fixed Catalogues). Of course, for each realisation, we draw new (random) GW events, unique for each catalogue (New GWs). Case (b) in Figure \ref{fig:multiple_realisations_catalogues_events_comparison}.
    \item \emph{Fixed GWs, Fixed Catalogues}: In this run, the incomplete catalogues are only created once, exactly as the case above (Fixed Catalogues). In addition, here we also keep the GW fixed, in the sense that their coordinates are selected from the clustered (complete) catalogue and then kept the same for all the other (incomplete) catalogues. In other words, for each realisation, the new GW events are chosen from the complete catalogue, and then the same events, i.e. cone positions/sizes, are used in the incomplete catalogues (Fixed GWs). Case (c) in Figure \ref{fig:multiple_realisations_catalogues_events_comparison}.
    
\end{enumerate}

Irrespective of the method, we draw several general conclusions: 

\begin{itemize}
    \item The uniform catalogue performs consistently worse. This is expected, as it does not include any clustering. In all cases, the HDI covers a region of $\Delta H_0 \geq 38$ km/s/Mpc.

    \item Increasing the catalogue completeness leads to tighter constraints. The most significant improvement occurs on catalogues with completeness levels above $f=25 \%$, where the HDI is of the order of $\Delta H_0 \sim 6$ km/s/Mpc or less.

    \item We observe some variation in terms of which catalogue performs best. A part of this could be explained from the different realisation procedures explored in this section, although the most important factor is the presence of increased clustering in the more complete catalogues, i.e., we have GW events in sky areas where clustering allows tighter $H_0$ constraints, which dominate the final posterior. Hence, we observe that completeness levels as low as $f=25 \%$ can be statistically equivalent to fully completed catalogues.
\end{itemize}

We investigate the last point further in Figure \ref{fig:clu_uni_f50_pdf_comparison}, where we focus on three catalogues and an increasing number of events. For all, we have $N_r=15$ realisations. We use a uniform catalogue, a $f=50 \%$ complete catalogue and a complete catalogue. The four different panels (upper and lower) correspond to the different realisation procedures described in this section. Crucially, this analysis strengthens our results on the importance of clustering, since it demonstrates that even with about half of the galaxies missing from the survey, the posteriors converge strongly and are essentially equivalent to those obtained with the fully complete catalogue. More specifically, as early as $40$ events, both the $f=50\%$ and ``full'' catalogues have a HDI of the order of $\Delta H_0 \sim 10$ km/s/Mpc, reaching $\Delta H_0 \sim 5$ km/s/Mpc for $100$ events.

Moreover, to confirm the robustness of our results, we investigate the galaxy catalogues in terms of their clustering power and the sky positions of the galaxies. We show the correlation function \citep{LS_corr_1993} and their angular distribution in Figure \ref{fig:distribution_checks_big_box}. As expected, we see increased power in the more complete catalogues. This validates their construction. We discuss this in more detail in Appendix \ref{ap:resolution_effects}, where we show that resolution effects are not affecting our main conclusions.

\subsubsection{Effects of angle weights in the likelihood}\label{sec:angle_LHD}

In Figure \ref{fig:angles_LHD_comparison}, we investigate the effect of giving angular weights to the galaxies inside the cones when inferring the Hubble parameter $H_0$. In other words, instead of using eq. (\ref{eq:angles_pdfs}), we use a uniform probability for all angles. We expect the inclusion of this term to improve our inference, however we find that the final Pdfs are very similar: compare the solid lines (with the angle weights) with the dashed lines (without angle weights). We investigated this further, by looking at cones with different opening angles, doubling the observational errors from $\Delta \Omega = 60$ deg. squared to $\Delta \Omega = 120$ deg. squared. We observe no difference between the runs with and without angular weights in either case. 

We believe a combination of reasons lead to this behaviour, and all relate to three various key scales: 1) the GWs angular localisation error, which determines the size of the cone and the width of the exponential weight, 2) the typical size of galaxy clusters, and 3) the typical separation between clusters. We analyse them in turn:

Firstly, if the edge of the cone was only one sigma, then the exponential would be pretty flat across the whole cone. However, our cone edges are more than $1.645 \sigma$ away from the cone centre. At the same time the sky localisation areas we have considered are relatively optimistic, so there are no big differences in sky positions - one can see in Fig. \ref{fig:cone_profiles} the relative small differences between angular weights.

Secondly, if the width of the Gaussian is large compared to the typical cluster size ($\sim$ Mpc) and small compared to the inter-cluster separation ($\gtrsim 50$ Mpc) the behaviour shown in Fig. \ref{fig:angles_LHD_comparison} might be expected, since the weights are approximately constant across a cluster and the cone does not typically contain more than one cluster in any redshift slice. To check if this is the case, we perform a test run, where we ``artificially'' reduce the variance of the exponential, i.e. have higher resolution inside the cones, while keeping cone sizes the same. Indeed, this led to a small improvement of the final posterior - see the gray, dashed line.

\subsubsection{Effects of different box sizes}\label{sec:Box_size_posteriors}

In Figure \ref{fig:Boxes_clustered_summary}, we compare the $H_0$ posteriors for the three large particle number simulations. We draw $N_c=100$ GWs events from the complete catalogue and average their posteriors over $N_r=15$ realisations. The smaller box, with $L_{\rm box}=100$ Mpc/h, yields an $H_0$ Pdf with a width that is half compared to the $L_{\rm box}=1600$ Mpc/h box (with $\Delta H_0 = 2.11$ km/s/Mpc for the $L_{\rm box}=100$ Mpc/h box, $\Delta H_0 = 3.31$ km/s/Mpc for the $L_{\rm box}=700$ Mpc/h box and $\Delta H_0 = 4.66$ km/s/Mpc for the $L_{\rm box}=1600$ Mpc/h box). This emphasises that closer events would be much more impactful in ``statistical dark sirens'' analyses. Finally, we note (see also Appendix \ref{ap:resolution_effects} for additional analyses on resolution effects) that although the $L_{\rm box}=1600$ Mpc/h box has the strongest clustering due to mass effects, the latter plays a secondary role in this type of analysis compared to the detector horizon distance.

\subsubsection{Effects of completion method}\label{sec:completion_methods}

In Figure \ref{fig:completion_comparison}, we compare the effects on $H_0$ posteriors when completing the incomplete catalogues in two different ways: the ``fixed'' completion method, described in detail in section \ref{sec:complete_a_catalogue} and where the main logic is to introduce galaxies per radial bin, and the ``random'' method, where we introduce missing galaxies globally. In other words, in the latter procedure, when we calculate the $N_{\rm missing}$ galaxies, we take into account all the galaxies in the observational sphere. Then we put back $N_{\rm missing}$ galaxies, uniformly over the whole sphere. As a result, this could lead to galaxies being drawn to regions, i.e. radial bins, where no galaxy is missing. We have checked that this procedure changes the expected galaxy distribution: distances further away are not ``completed'' adequately, and there is stronger clustering, since we introduce galaxies to regions that are not missing any. In terms of the $H_0$ posteriors, both methods result in a HDI of $\Delta H_0 \sim 5$ km/s/Mpc for the complete catalogues (this is expected, since for the complete catalogues the two methods are identical as no completion is needed), while the ``random'' vs the ``fixed'' method gives $\Delta H_0^{\rm ran} (f=5\%) \sim 7.5$ km/s/Mpc, $\Delta H_0^{\rm ran} (f=75\%) \sim 3.9$ km/s/Mpc vs $\Delta H_0^{\rm fix} (f=5\%) \sim 9.3$ km/s/Mpc, $\Delta H_0^{\rm ran} (f=75\%) \sim 4.5$ km/s/Mpc respectively. Although a bias is not observed in the ``random'' completion method, due to the above reasons, the ``fixed'' method is the default option throughout.

\begin{figure*}
	%\addtocounter{figure}{-1}
	\centering
	\begin{subfigure}[b]{0.48\textwidth}
		\centering
		\includegraphics[width=\textwidth]{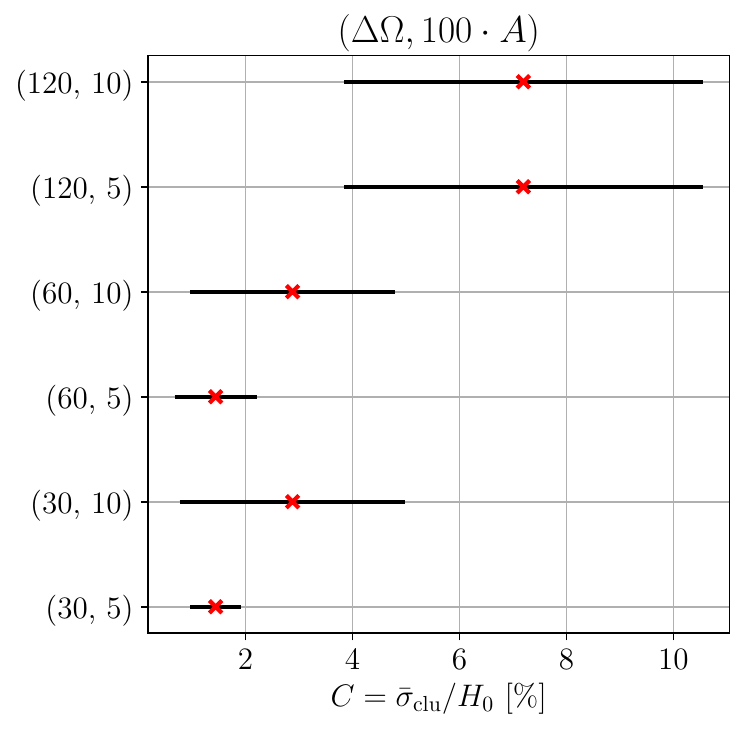}
	\end{subfigure}
	\hspace{0.5cm}
	\begin{subfigure}[b]{0.48\textwidth}
		\centering
		\includegraphics[width=\textwidth]{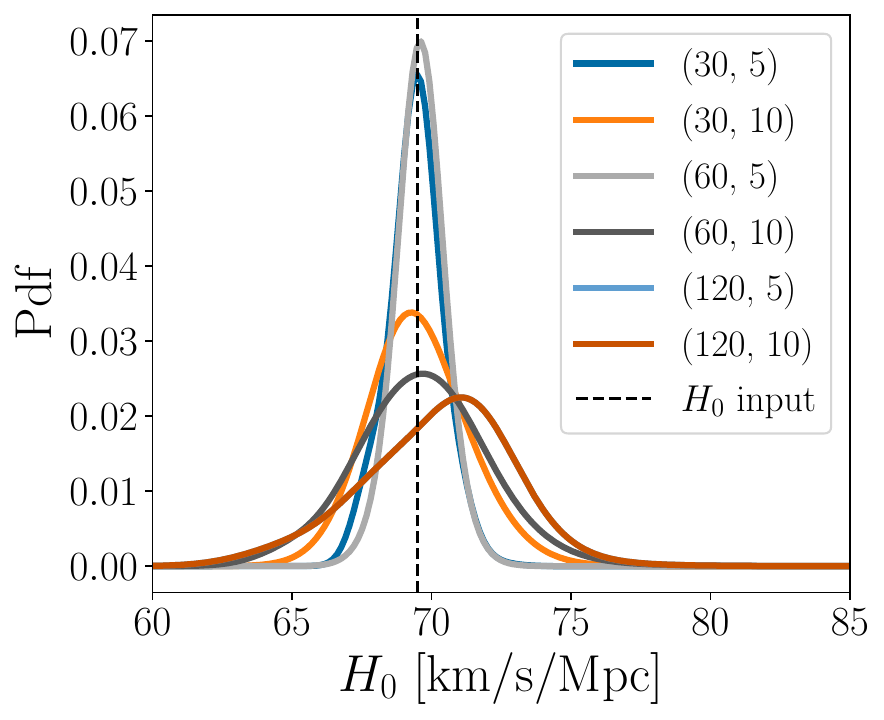}
	\end{subfigure}
	\caption{Study of the impact of observational errors ($\Delta \Omega, 100 \cdot A$) on the $H_0$ Pdfs, for the complete, clustered catalogues. (Left) We show the standard deviation of the Pdfs, normalised to the input $H_0 = 69.5$ km/s/Mpc. For small cone openings, the distance errors play the more important role. For large cone openings, the angular errors dominate. (Right) The full posteriors for the same selection of errors. Note that the ($120, 5)$ posterior is not visible, because it is very similar to the ($120, 10$) run. Recall that the errors are selected at the beginning of each inference simulation and are kept fixed for all events. Results of inference for $N_c=100$ events, averaged over $N_r=15$ realisations, from the $L_{\rm box}=1600$ Mpc/h box.}
	\label{fig:Obs_errors_clustered_summary}
\end{figure*}

\subsubsection{Effects of observational errors}\label{sec:observational_errors_posteriors}

In Figure \ref{fig:Obs_errors_clustered_summary}, we study the impact of observational errors (section \ref{sec:Error_priors_source_selection}) on the final $H_0$ posteriors. For all cases, we consider $N_c=100$ GWs events from the complete catalogue from the big box $L_{\rm box}=1600$ Mpc/h, averaged over $N_r=15$ realisations. As a general conclusion, we establish the expected behaviour: smaller observational errors produce tighter constraints. More specifically, we see that for smaller sky localisation errors, $\Delta \Omega \leq 60$ deg. squared, the distance uncertainty plays the more important role, with the Pdfs of ($\Delta \Omega, 100 \cdot A$) = ($60, 5$) and ($30, 5$) being very similar. In parallel, when the cone openings are large, $\Delta \Omega = 120$ deg. squared, the smaller distance uncertainty is not affecting the $H_0$ Pdf, with ($120, 10$) and ($120, 5$) being almost identical, and similar to the ($60, 10$) posterior). Finally, we note that a $\Delta \Omega = 120$ deg. squared sky localisation would still be an optimistic scenario in current observational data. On the one hand, this is a limitation due to the finite size of the simulation, but on the other, our main interest in this section lies in the relative impact of different errors in ($\Delta \Omega, 100 \cdot A$). Our main conclusion is that above a certain threshold, in our case $\Delta \Omega \geq 120$ deg. squared, the number of galaxies in the region becomes so big, that no significant improvement is expected for the $H_0$ posteriors. We summarise the results in Table \ref{table:obs_errors_HDI_table}.

\begin{table}
\centering
\begin{tabular}{|c c c c c|}
     \hline
     $(\Delta \Omega, 100 \cdot A)$ & $(30, 5)$ & $(30, 10)$ & $(60, 5)$ & $(60, 10)$\\ 
     \hline
     $\Delta H_0$ [km/s/Mpc] & $1.81$  & $3.46$ & $1.66$ & $4.66$ \\ 
     \hline
     \hline
     $(\Delta \Omega, 100 \cdot A)$ & $(120, 5)$ & $(120, 10)$ & & \\
     \hline
     $\Delta H_0$ [km/s/Mpc] & $5.56$ & $5.56$ \\ 
\end{tabular}
\caption{HDI for the pdfs of the realisations with different observational errors. See Figure \ref{fig:Obs_errors_clustered_summary}.}
\label{table:obs_errors_HDI_table}
\end{table}

\subsubsection{Effects of $H_0$ priors}\label{sec:H0_priors_checks}

\begin{figure}
    \centering
	\includegraphics[width=1\columnwidth]{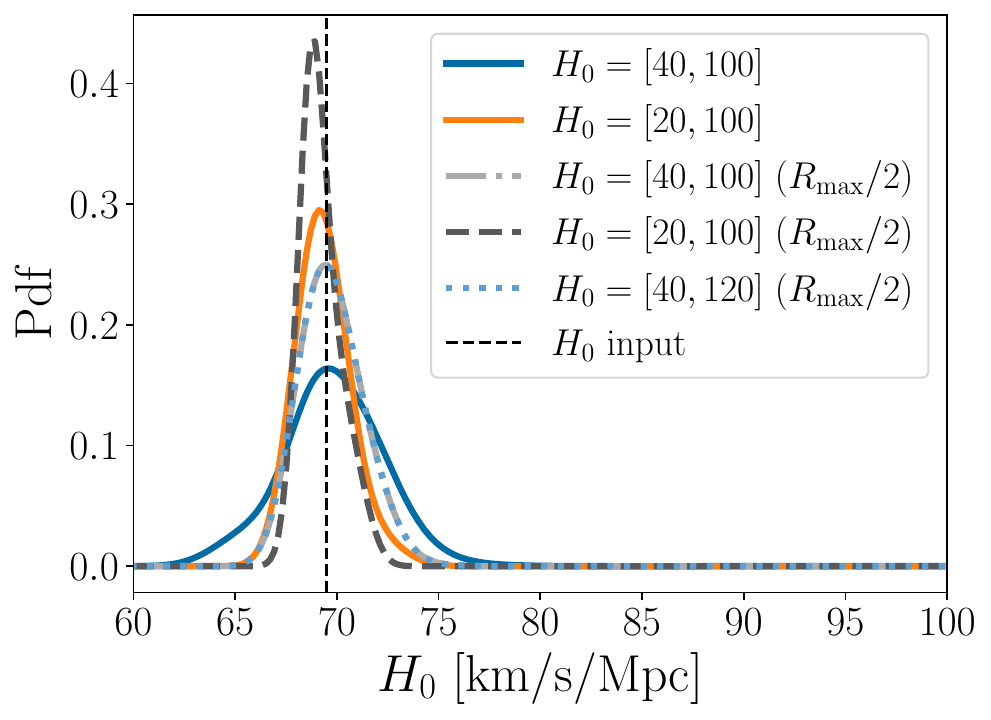}
    \caption{$H_0$ posteriors for three different $H_0$ prior choices: the fiducial one $H_0^{\rm prior} = [40, 100]$ (orange, dotted), the low-asymmetric one $H_0^{\rm prior} = [20, 100]$ (blue, solid) and the high-asymmetric one $H_0^{\rm prior} = [40, 120]$ (grey, dashed). The reason for the extra distance constraints is for consistency between all the runs, since a prior change affects also the construction of the catalogue and of the cones (see main text, section \ref{sec:H0_priors_checks} for details). Taking this into account, we observe that all prior runs are equivalent and there are no biases. Results are shown for inference of $N_c=100$ events, averaged over $N_r=15$ realisations, from the $L_{\rm box}=1600$ Mpc/h box. The normalisation here is defined, so that the Pdfs integrate to $1$.}
    \label{fig:H0priors_comparison}
\end{figure}

In Figure \ref{fig:H0priors_comparison}, we compare the effects on $H_0$ posteriors when assuming different, asymmetric $H_0$ prior choices: $H_0^{\rm prior} = [40, 100]$ (fiducial), $H_0^{\rm prior} = [20, 100]$ (low-asymmetric), and $H_0^{\rm prior} = [40, 120]$ (high-asymmetric). It is important to note that the prior choices are not affecting only the final inference, but also the cone construction and the range of eligible GW sources. This happens because the cone size takes into account the $H_0$ prior range, as described in section \ref{sec:Cone_construction} and similarly for the distance threshold of sources, as seen in eq. (\ref{eq:dcone_centre_threshold}). 

As a result, a smaller lower limit allows a more restricted distance range of GW source, i.e. sources are selected closer to the observer, yielding a more precise measurement. To make a fairer comparison and investigate more thoroughly and consistently the effect of the priors, we impose an extra constrain to the fiducial and high-asymmetric runs: we restrict the distance threshold of the GW events to about $L_{\rm box}/4$, which is roughly the same as the maximum GW source distance recovered in the low-asymmetric case (which uses the fiducial limit $L_{\rm box}/2$, but the maximum GW distance is much closer due to the prior). In this case, we observe that the final posteriors are equivalent between the three cases, and hence confirm that our results are not prior dependent. Finally, note that the prior choice does not induce any bias.

\begin{figure*}
	%\addtocounter{figure}{-1}
	\centering
	\begin{subfigure}[b]{0.48\textwidth}
		\centering
		\includegraphics[width=\textwidth]{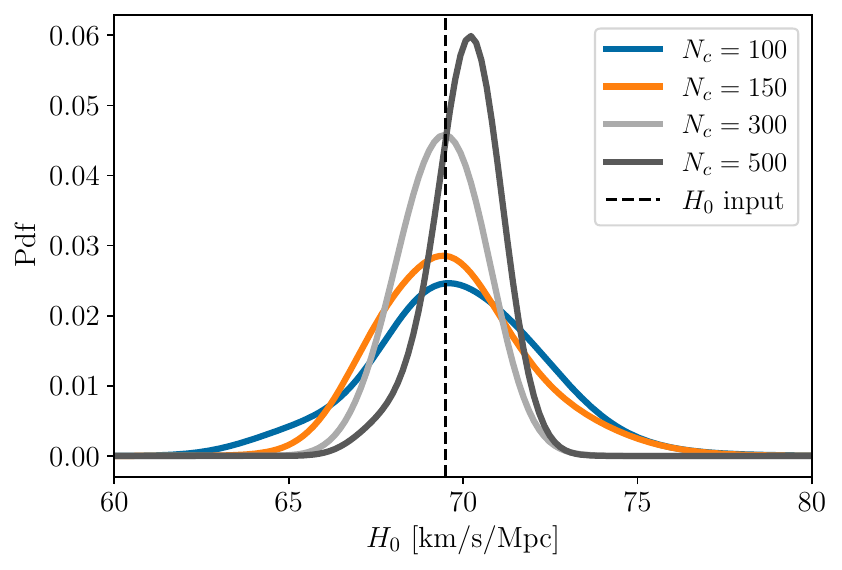}
	\end{subfigure}
	\hspace{0.5cm}
	\begin{subfigure}[b]{0.48\textwidth}
		\centering
		\includegraphics[width=\textwidth]{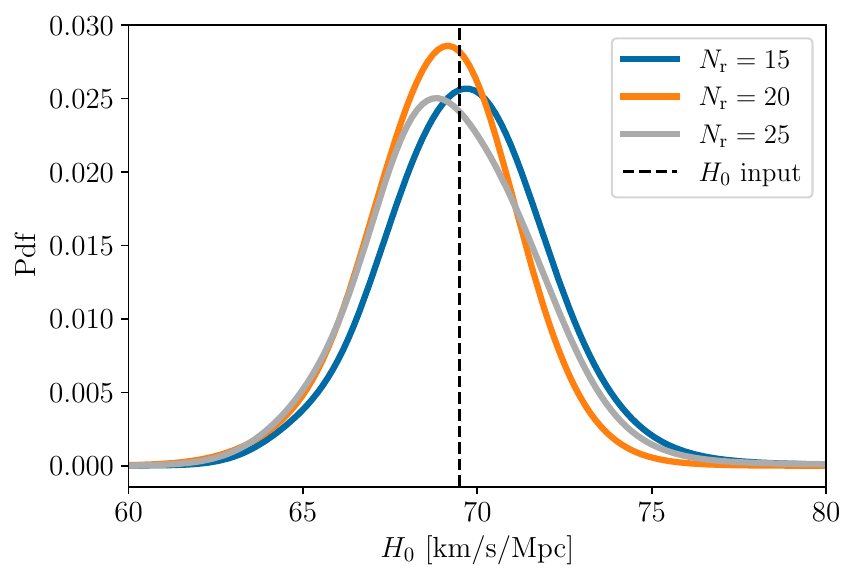}
	\end{subfigure}
	\caption{Checking convergence of our inference procedure. (Left) Similar to Figure \ref{fig:Ho_clustered_Nevents_Nreal}, but reaching $\times 5$ number of GW events. As expected, more observations lead to much more precise results. (Right) We test the impact of the number of realisations $N_r$, for $N_c=100$ GW events. We show that for the default number chosen ($N_r=15$), the pdfs have converged. Results from the $L_{\rm box}=1600$ Mpc/h box.}
	\label{fig:Nevents_Nreal_comparison}
\end{figure*}

\subsubsection{Convergence studies}\label{sec:convergence_studies}

In this last section, we cross-check our results by running multiple realisation scenarios. As a reminder, the default number of realisations is $N_r=15$, and is used throughout unless stated otherwise. This means that for a specific number of events, for example $N_c = 20$, we repeat the analysis $N_r$ times and combine the posteriors. 

Figure \ref{fig:Nevents_Nreal_comparison} illustrates the results of our basic test: we use the complete catalogue of the $L_{\rm box}=1600$ Mpc/h box and compare the $H_0$ posteriors for different numbers of GW events and different realisations. In the first case, we confirm that increasing the number of events, even 5-fold, leads to more precise constraints. In the second case, we show that the default number of realisations chosen for our results $N_r=15$ yields similar $H_0$ Pdfs with $N_r=20$ and $N_r=25$, demonstrating convergence of our analysis.

Moreover, in section \ref{sec:completeness_results} we investigate different completeness realisations to make sure our results are robust with respect to GW event selection and catalogue variability. The results are shown in Figures \ref{fig:multiple_realisations_catalogues_events_comparison} and \ref{fig:clu_uni_f50_pdf_comparison}.

In Figure \ref{fig:multiple_realisations_catalogues_events_comparison}, panels (a)-(c) correspond to the cases (i)-(iii), described in section \ref{sec:completeness_results}, respectively. We find that the completeness and GW selection procedures can quantitatively affect the results, but qualitatively they are not very different. The most important difference appears in Figure (c)/case (iii), where the GW events are chosen from the complete catalogue, and both the GW events and the incomplete catalogues are kept fixed, however the result is not surprising. Keeping the selected GW events and catalogues consistent across completenesses suppresses the differences arising from random fluctuations due to the event selection. As a result the posteriors look much more similar than in the other cases where these random fluctuations are present. 

This panel suggests that $f\geq25 \%$ provides the critical threshold for the completeness levels we have studied, with the quality of the constraint starting to degrade at lower completenesses.

Figure \ref{fig:clu_uni_f50_pdf_comparison} tests cases (i) (top four panels) and (ii) (bottom four panels), demonstrating the qualitative equivalency of the two methods: in both cases, catalogues with clustering perform much better than the uniform one, and even heavily incomplete catalogues could offer similar constraints as complete ones.

\begin{figure}
     \centering
     \subfloat[][RA distribution]{\includegraphics[width=\columnwidth]{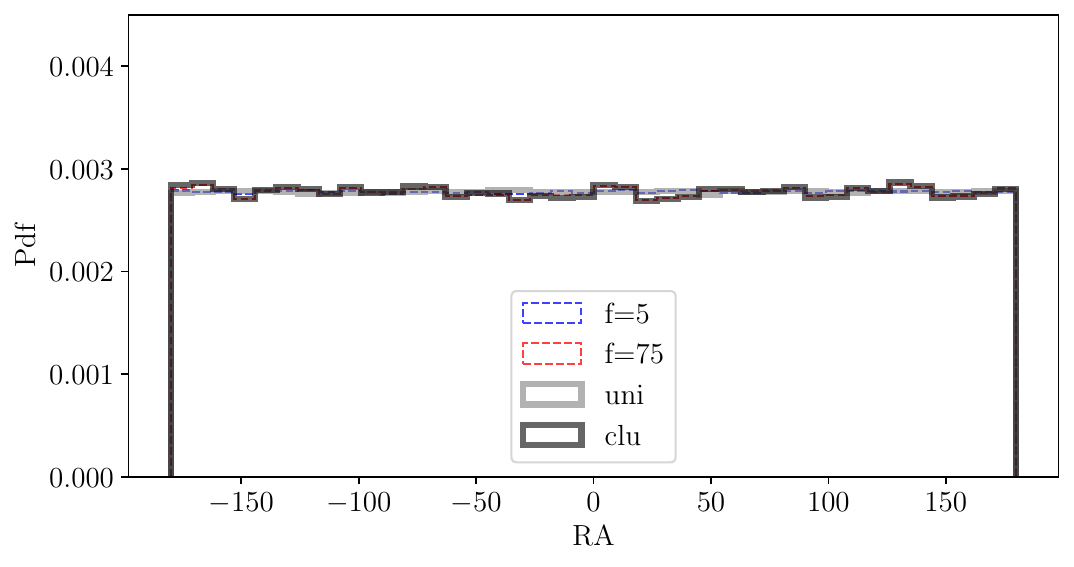}}\\
     \subfloat[][DEC distribution]{\includegraphics[width=\columnwidth]{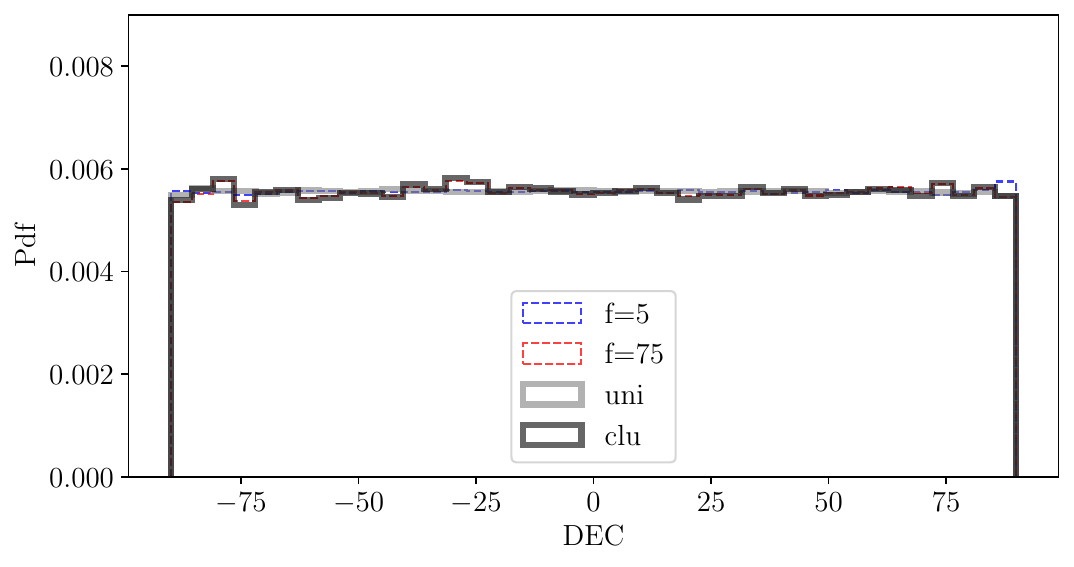}}\\
     \subfloat[][Correlation function]{\includegraphics[width=\columnwidth]{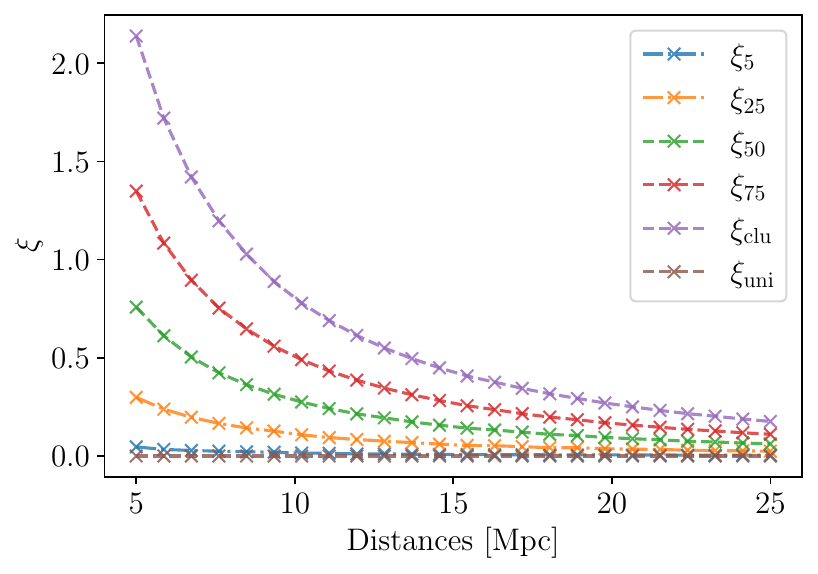}}

     \caption{\textbf{L = $1600$ Mpc/h}: Sanity checks for our completion procedure: (a)-(b) the angular distribution of the galaxies in each catalogue - the more clustered the catalogue, the more features we see: the uniform catalogue here is almost flat, with the $5 \%$ (blue, dashed) following it very closely; (c) the correlation function of each catalogue. As expected, the clustering increases as we increase the completeness of the catalogue. Note that the $5 \%$ catalogue is almost identical to the uniform one. For DEC, the cosine distribution has been transformed to uniform, for easier comparison.}
     \label{fig:distribution_checks_big_box}
\end{figure}

\section{Conclusions}\label{sec:conclusions}

In this paper, we have studied the effects of clustering and incompleteness of galaxy surveys, when constraining the Hubble-Lema\^itre parameter with dark GW sirens. For this, we exploited high-resolution cosmological simulations with realistic structure formation and physical weights. We find that:

\begin{itemize}
    \item Care is needed to avoid biases in the final $H_0$ posterior. Based on our simulated analyses, the most important effects come from the geometric characteristics of the problem, i.e., the cone geometry and positions of galaxies. These can be counterbalanced by galaxy weights based on their positions and by selection effects, as well as considering a consistent approach on simulation studies of GW ``statistical dark sirens''. Our results confirm the potential problems with biased $H_0$ inference \citep{Trott_Huterer_2021}, and solutions discussed in \citep{Hitchhiker_2022, Alfradique_Bom_Castro_2025}. As a result, we find that different methods - histogram ``stacking'' of all $H_0$ values (section \ref{sec:Ho_histogram_stacking}), and the full Bayesian analysis (section \ref{sec:bayesian_analysis_method}) of calculating the $H_0$ distribution give consistent results, and converge to the input value as the number of observations increases. We do not observe any clear bias in our results.
    
    \item The inclusion of physical weights (masses, magnitudes, star formation rates) does not always lead to a significant improvement of the final posterior: weighting by star formation rate seems to have the most significant impact. We caution however that this can be implementation dependent. On the other hand, geometric weights are important and necessary. This reinforces the importance of clustering, which in addition allows for unbiased results even when weights are not applied self-consistently (Figure \ref{fig:weights_comparison_multiple_realisations}).
    
    \item Clustered catalogues improve convergence compared to uniform ones (Figures \ref{fig:multiple_realisations_catalogues_events_comparison} and \ref{fig:clu_uni_f50_pdf_comparison}). This is not true for all methods. In the simplest ``stacking'' method, the interquartile gap is similar between clustered and uniform catalogues, irrespective of box size and particle numbers. On the other hand, the ``Bayesian'' analysis shows a clear improvement on the $H_0$ determination.
    
    \item Incomplete catalogues, based on magnitude cuts, perform better than complete uniform ones (Figure \ref{fig:multiple_realisations_catalogues_events_comparison}). This is a direct result of clustering. In some cases, even catalogues only $f=25 \%$ complete, can be competitive with full catalogues. And already from $f=5 \%$ completeness, we see a clear peak in the posterior. We find that these results are robust against different realisation scenarios.  
    
    \item In terms of observational errors, we find that the angle weights in the likelihood play a secondary role, at least for the cone geometries we consider (Figure \ref{fig:angles_LHD_comparison}). Moreover, for small cone openings ($\Delta \Omega < 120$ deg. squared), we showed that the distance errors $A$, influence strongly the tightness of our posteriors, while for larger openings ($\Delta \Omega \geq 120$ deg. squared) the posteriors are dominated by the angular errors (Figure \ref{fig:Obs_errors_clustered_summary}). This behaviour is not unexpected, since in bigger cones the distribution of galaxies ``smoothens out''.

    \item Considering catalogues of different sizes, i.e., different simulation boxes, we confirm that good quality, close-by events are more important than more distant ones. More specifically, the same number of events lead to almost double the Pdf width for the larger observing sphere ($\sigma_{\rm LB1600}/\sigma_{\rm LB100} \sim 2.3$) (Figure \ref{fig:Boxes_clustered_summary}).
    
    \item The most important result of this work demonstrates the power of clustering in real galaxy surveys. Catalogues that globally retain even $f=5 \%$ of the true number of galaxies in the Universe lead to unbiased and informative constraints (Figure \ref{fig:multiple_realisations_catalogues_events_comparison}). Completeness levels of $f=50 \%$ already perform statistically similar to fully complete catalogues (Figure \ref{fig:clu_uni_f50_pdf_comparison}). This indicates that a strategy that focusses on improving GW detector sensitivity and minimizing detector downtime, so that the number of GW events is maximized, is more important for ``statistical dark siren'' cosmology than obtaining more complete galaxy catalogues.

    \item Our results are consistent and complement different ``completion'' procedures in the literature. Most approaches complete the catalogues in a statistical way, either by assuming a uniform in comoving volume distribution of galaxies \citep{Gray_et_al_2020}, assuming a multiplicative factor based on inhomogeneities \citep{Finke_et_al_DS_2021, Borghi_et_al_2026}, assuming large-scale structure clustering information (a factor based on the variance of the distribution of galaxies in the catalogue \citep{Dalang_Baker_2023, Dalang_Fiorini_Baker_2024} or using a Bayesian reconstruction based on the power spectrum \citep{Cosmic_cartography_2024, Cosmic_cartography_2025}), or alternatively leaving the catalogue incomplete and examining the cosmological power of the brightest galaxies of the region \citep{VanWyngarden_et_al_2025_cat_completeness}. Our methodology here is different in the following way: we complete the catalogues by putting galaxies back, and not in a statistical way, in the likelihood construction. Our completion procedure in this work assumes a uniform distribution, while in our upcoming paper (Barbieri et al., in prep.), we consider clustering information when adding galaxies in an incomplete catalogue. Similarly to previous works, our incomplete catalogues are constructed by setting a magnitude threshold, with the distinction lying in the galaxy catalogues employed - in our case, we exploit galaxies in high-resolution 3D cosmological simulations, with properties assigned through a ML framework (see section \ref{sec:simulations} for details). Finally, even though our results are qualitatively and quantitatively similar ($\sigma_{H_0}/H_0 \sim 4 \%$) to other studies, the different configurations used, i.e., different GW events and catalogue details, do not allow for a direct comparison. At present, different ``completion analyses'' are essential to examine possible systematic biases and leverage the full power of ``statistical dark sirens'' analyses.    
    
    \item Finally, we tested the robustness of our analysis with a number of sanity-checks, which it passes successfully: catalogue construction and clustering behaviour (Figure \ref{fig:distribution_checks_big_box}, convergence with number of GW events follows an $1/\sqrt{N}$ scaling (Figures \ref{fig:Ho_clustered_Nevents_Nreal} and \ref{fig:Nevents_Nreal_comparison}), convergence of posteriors with number of realisations (Figure \ref{fig:Nevents_Nreal_comparison}) and resolution effects (section \ref{ap:resolution_effects}).
\end{itemize}

\section*{Acknowledgements}

MK would like to thank Konstantinos Migkas for useful discussions during the 15th Hellenic Astronomical Conference, Andrea Incatasciato $\&$ Jose Onorbe for helpful discussions concerning the LEGACY simulations, and Tessa Baker for detailed comments in a very early draft of this paper. The authors also thank Konstantin Leyde for the very helpful internal LVK review, and Simone Mastrogiovanni and Maciej Bilicki for useful comments, as well as the reviewer for their careful report that improved the quality of this work. SK acknowledges funding via STFC Small Grant ST/Y001133/1. For the purpose of open access, the authors have applied a Creative Commons Attribution (CC BY) license to any Author Accepted Manuscript version arising from this submission.

\emph{Software}: For the analysis, we used \verb#Matplotlib# \citep{Hunter:2007}, \verb#NumPy# \citep{oliphant2006guide, van2011numpy}, and \verb#SciPy# \citep{Scipy}. 

\section*{Data Availability}

The data underlying this article will be shared on reasonable request to the corresponding author. The codes to reproduce the results of this paper, will appear in \url{https://github.com/MariosNT/GWs_DarkSirens_completeness}.

%%%%%%%%%%%%%%%%%%%%%%%%%%%%%%%%%%%%%%%%%%%%%%%%%%



\bibliographystyle{mnras}
\bibliography{biblioX} % if your bibtex file is called example.bib

%\begin{thebibliography}{99}
%\end{thebibliography}

%%%%%%%%%%%%%%%%%%%%%%%%%%%%%%%%%%%%%%%%%%%%%%%%%%

%%%%%%%%%%%%%%%%% APPENDICES %%%%%%%%%%%%%%%%%%%%%

\appendix

\section{Number of galaxies in simulations}\label{ap:volume_cones}

Resolution limitations restrict the number of galaxies in the region of interest compared to the real Universe. We find about an order of magnitude fewer galaxies than in a real survey. This would lead to individual posteriors to be more optimistic than real observations, however the effects of clustering should hold both qualitatively and quantitatively, since they are based on the relative comparison with an identically populated uniform box. This is confirmed by the investigation of resolution effects in Appendix \ref{ap:resolution_effects}.

\section{Consistency checks for physical parameters}

\begin{figure}
	\includegraphics[width=\columnwidth]{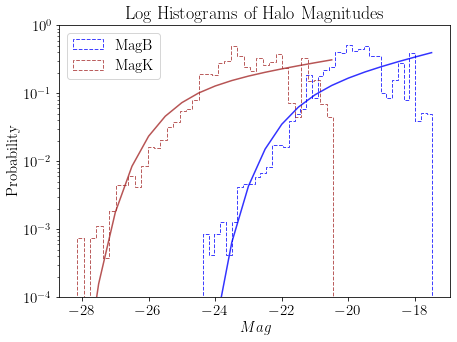}
    \caption{Absolute magnitudes in $B$ and $K$ bands, approximated by a Schechter magnitude function.}
    \label{fig:Schechter_mag}
\end{figure}

Before using the ML generated magnitudes and the stellar masses from eq. (\ref{eq:SHM}), we performed sanity checks to assess their characteristics. We removed the least massive haloes, which cannot be resolved due to resolution limitations and we checked that the maximum stellar mass of the fiducial $L_{\rm box}=1600$ Mpc/h box is smaller than $M_* < 10^{12} M_{\odot}$, which is reasonable for galaxies scales. For the SFR and apparent magnitudes, we retained only galaxies which had physical values assigned, e.g. not negative SFRs. Finally, we confirmed that the distributions of K and B bands (note though that only the latter are used in the main text) of the remaining galaxies follow a Schechter magnitude function \citep{Schechter_1976}. In Figure \ref{fig:Schechter_mag}, we plot such a function, based on \citep[eq. 5]{Dunlop_2012}, with $(M_*, \alpha) = (-25.5, -1.2)$ and $(M_*, \alpha) = (-22, -1.3)$ for $K$ and $B$ band respectively.  

\section{Resolution effects}\label{ap:resolution_effects}

To develop our analysis, we have used a low particle number box from the LEGACY suite of cosmological simulations. This test box has $100 \rm{Mpc}/h$ length per side, with a resolution of $256^3$ particles. Together with the large particle number boxes of different sizes, as described in the main text, this box can be used to investigate any resolution issues of our results. Note that between the high-res runs,  the largest box ($1600$ Mpc/h), can be compared with the $100$ Mpc/h one which has the same number of particles (comparing size effects). At the same time, the $100$ Mpc/h high-res box can be directly compared with the $100$ Mpc/h ``test'' box, which includes many fewer particles (comparing particle number effects), while the $700$ Mpc/h box has the same mass resolution to the $100$ Mpc/h ``test'' box (comparing mass effects). We find that our main results still hold qualitatively in all the boxes.

\begin{figure}
     \centering
     \subfloat[][RA distribution]{\includegraphics[width=\columnwidth]{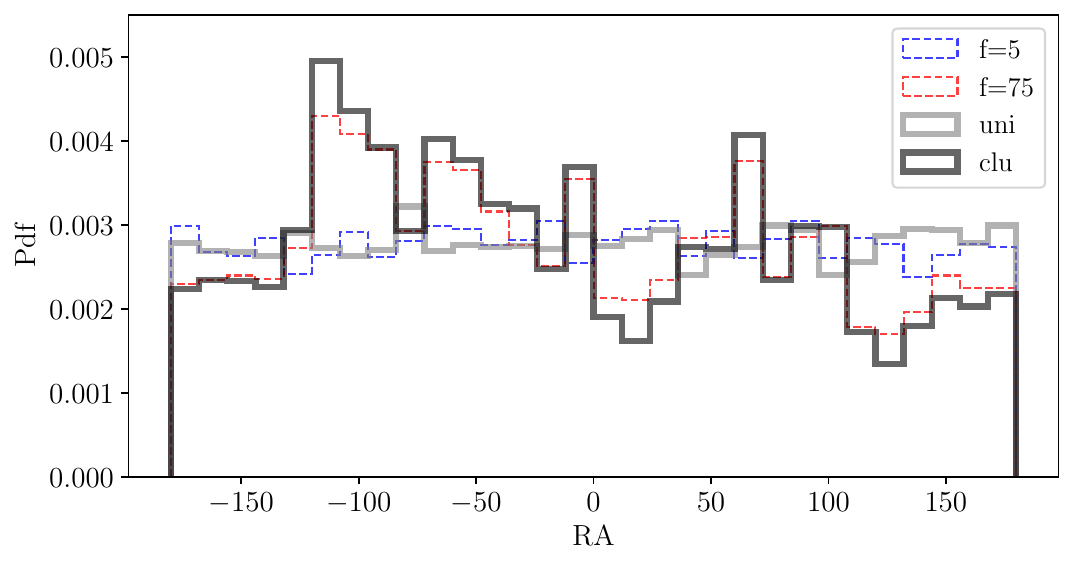}}\\
     \subfloat[][DEC distribution]{\includegraphics[width=\columnwidth]{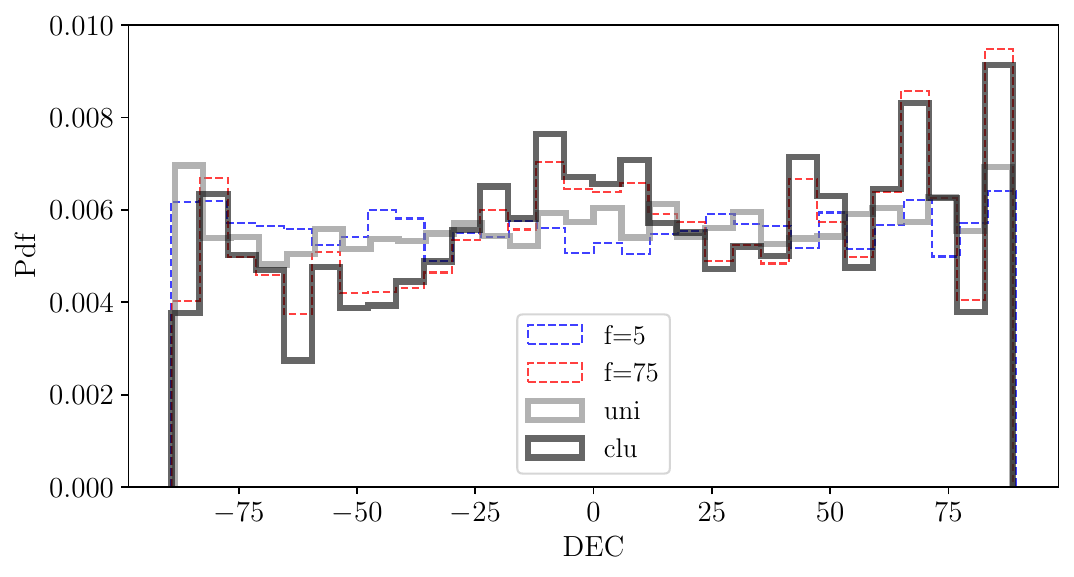}}\\
     \subfloat[][Correlation function]{\includegraphics[width=\columnwidth]{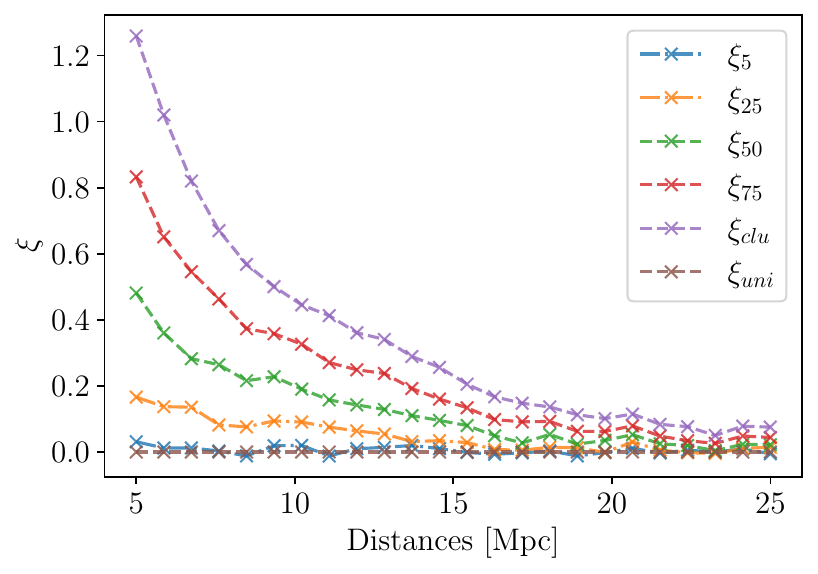}\label{fig:correlation_mini_box}}

     \caption{\textbf{L = $100$ Mpc/h - small particle number}: Sanity checks for our completion procedure: (a)-(b) the angular distribution of the galaxies in each catalogue - the more clustered the catalogue, the more features we see; and (c) the correlation function of each catalogue. As expected, the clustering increases as we increase the completeness of the catalogue. For DEC, the cosine distribution has been transformed to uniform, for easier comparison.}
     \label{fig:distribution_checks_low_res}
\end{figure}

In Figure \ref{fig:distribution_checks_low_res}, we perform a number of sanity checks to validate our completion procedure: we calculate the angular distributions on the sky of the galaxies for four catalogues (complete, uniform, $f=5 \%$ and $f=75 \%$ completeness) and their correlation function \citep{LS_corr_1993}. In all cases, the qualitative behaviour is as expected: the more complete the catalogue, i.e., the higher the clustering, the more features in the angular distribution of galaxies and stronger correlation power (compare with Figure \ref{fig:distribution_checks_big_box}).

\begin{figure}
     \centering
     \subfloat[][Investigating convergence with number of events]{\includegraphics[width=0.9\columnwidth]{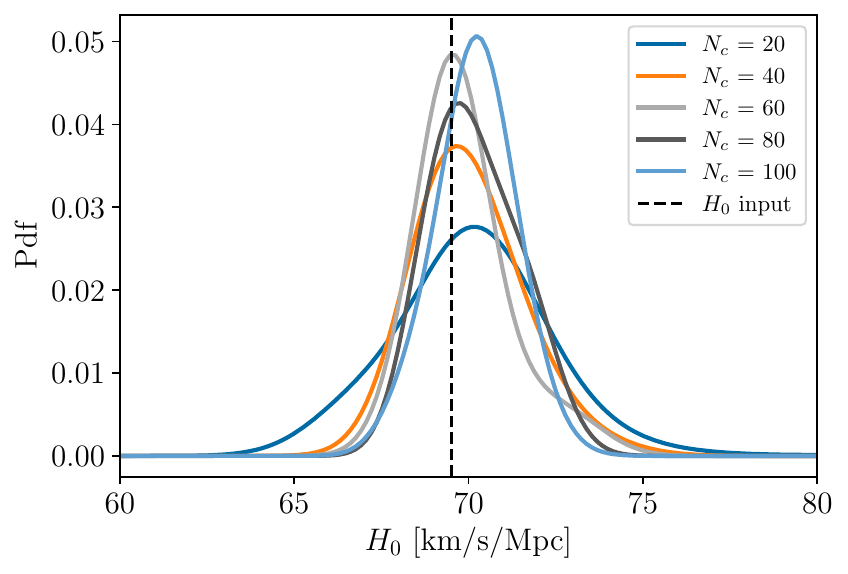}\label{fig:Ho_clustered_Nevents_Nreal_mini_box}}\\
     \subfloat[][Investigating completeness]{\includegraphics[width=\columnwidth]{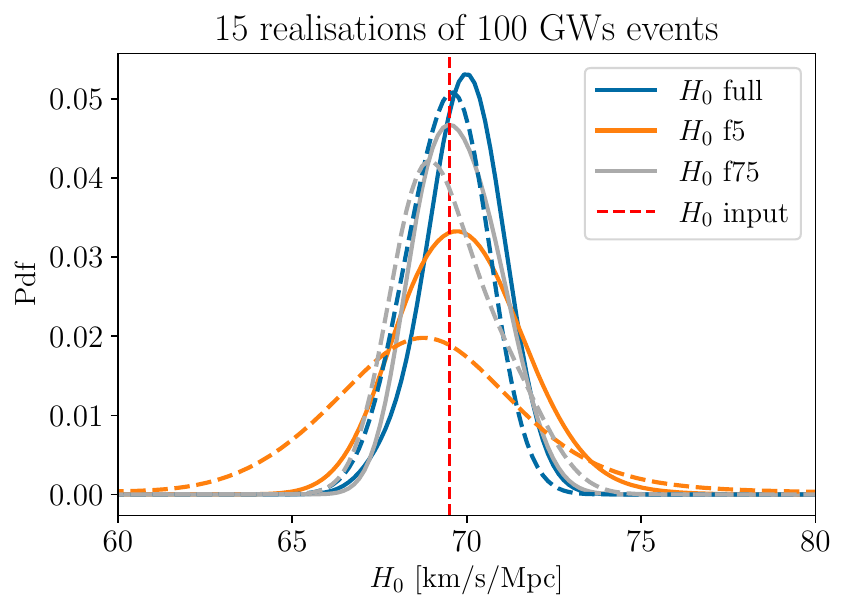}\label{fig:completeness_NewEvents_NewCatalogues_resolution}}\\

     \caption{$H_0$ posteriors in the small, low particle number, test box: (a) shows the convergence of the posteriors with the number of events; and (b) compares the posterior for $N_c=100$ events from a number of catalogues of different completeness, and different resolution (the dashed lines correspond to the posteriors of the high-res `LB100' for the same level of completeness. For both resolutions, the qualitative behaviour is similar: more events and more complete catalogues lead to tighter constraints. Note also that quantitatively, for the low-res box (solid lines) the posteriors are tighter for the same completeness level of the high-res box. This is due to the increased clustering in the low-res box (see Figure \ref{fig:correlation_function_comparison}), which reinforces our conclusions on the power of clustering. For both figures the results are the average over $N_r=15$ realisations. For visualisation purposes, we have plotted only a few representative catalogues.}
     \label{fig:Ho_summary_low_res}
\end{figure}

\begin{figure}
	\includegraphics[width=\columnwidth]{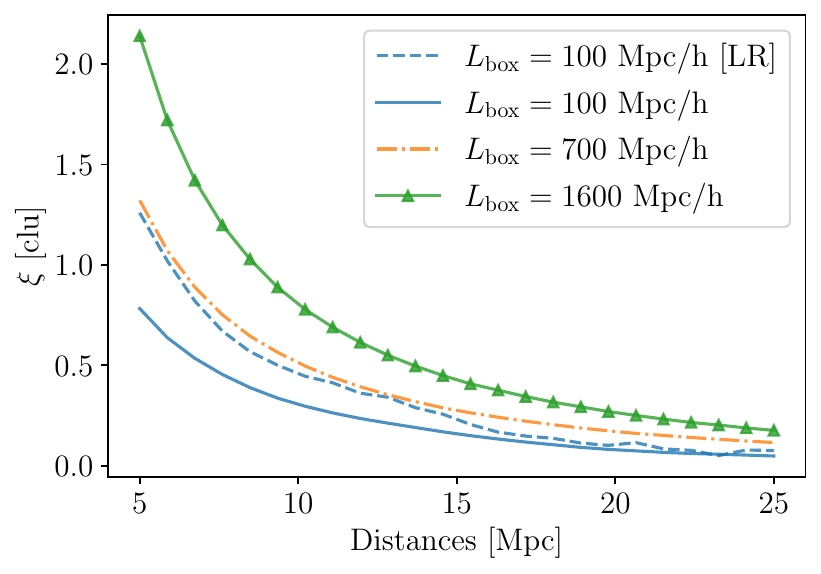}
    \caption{Comparison of the correlation functions of the three high-res simulations. The smaller box resolves better both smaller scales and smaller masses. Although clustering is stronger at smaller scales (this can be seen for a fixed box size), the most important feature when comparing clustering between the boxes is the halo mass. Higher-mass haloes, tend to cluster more strongly. Larger boxes are able to resolve haloes of higher masses and indeed show a larger correlation function. The differences though are not important enough to outweigh other effects in the $H_0$ posteriors (see section \ref{sec:Box_size_posteriors} for further discussion).}
    \label{fig:correlation_function_comparison}
\end{figure}

In Figures \ref{fig:Ho_clustered_Nevents_Nreal_mini_box} and \ref{fig:completeness_NewEvents_NewCatalogues_resolution}, we present $H_0$ posteriors for increasing numbers of observations and increasing catalogue completeness in the small, low-res box. They behave qualitatively as expected, with increasing number of events and increasing completeness leading to tighter constraints. Also, quantitively we compare the posteriors of complete and incomplete catalogues with the ones of the $100$ Mpc/h high-res box (dashed lines). The low-res box, which has higher clustering (see Figure \ref{fig:correlation_function_comparison}), has tighter constraints for the same completeness percentage, which reinforces the role of clustering.

Finally, in Figure \ref{fig:correlation_function_comparison} we compare the correlation functions of the three high-res simulation boxes and the low-red box. First, we compare with the $100$ Mpc/h ``test'' box (Figure \ref{fig:correlation_mini_box}, clustered catalogue). We see that the correlation functions are similar, and the low-res box resembles more the $700$ Mpc/h box, which has the same mass resolution. Second, the same effect is visible in the comparison of the high-res boxes, where the largest $1600$ Mpc/h box exhibit the strongest clustering, since it is able to resolve better the most massive haloes, which cluster more. This is seen in spite of the fact that the $100$ Mpc/h box would be able to resolve better smaller scales. We note however, that the small differences in clustering seem to play a secondary role in the $H_0$ posteriors, where the size of the box, and as a consequence the horizon distance of detection, dominates (refer again to Figure \ref{fig:Boxes_clustered_summary}). This demonstrates that box size is the most important feature in this kind of analysis, followed by clustering.

We conclude that resolution is not impacting our results for the complete or incomplete catalogues.

\begin{figure}
	\includegraphics[width=\columnwidth]{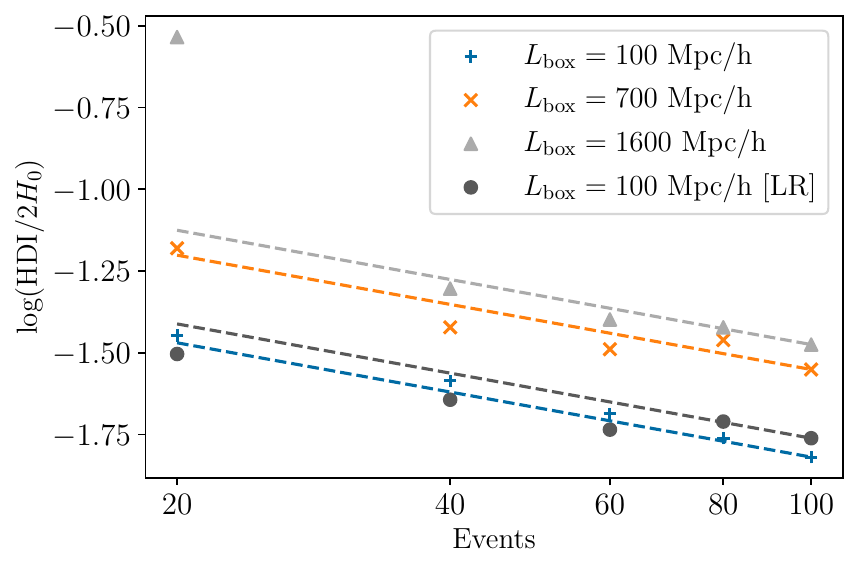}
    \caption{Posterior $68 \%$ HDI for complete catalogues for all simulation boxes (compare with right panel of Figure \ref{fig:Ho_clustered_Nevents_Nreal}, where we plot the `LB1600' results). We observe that the convergence with the number of events and the error scalings are not affected by resolution. Moreover, similarly to the main results of section \ref{sec:Box_size_posteriors}, the smaller the box size, i.e., the closer the GWs events, the tighter the posterior is still true irrespective of the resolution (recall Figure \ref{fig:Boxes_clustered_summary}.)}
    \label{fig:errors_scaling_multiple_realisations_resolution}
\end{figure}

%%%%%%%%%%%%%%%%%%%%%%%%%%%%%%%%%%%%%%%%%%%%%%%%%%

% Don't change these lines
\bsp	% typesetting comment
\label{lastpage}
\end{document}